\documentclass[11pt]{article}
\pdfoutput=1
\usepackage{jcappub,natbib}
\input{colordvi.tex}

\def\ga{\mathrel{\raise.3ex\hbox{$>$\kern-.75em\lower1ex\hbox{$\sim$}}}}
\def\la{\mathrel{\raise.3ex\hbox{$<$\kern-.75em\lower1ex\hbox{$\sim$}}}}

\title{Scale-dependent gravitational waves  from a rolling axion}

\author[a]{Ryo Namba,}
\author[b]{Marco Peloso,}
\author[a]{Maresuke Shiraishi,}
\author[c]{Lorenzo Sorbo,}
\author[b]{and Caner Unal}

\affiliation[a]{ Kavli Institute for the Physics and Mathematics of the Universe (WPI),
The University of Tokyo Institutes for Advanced Study,
The University of Tokyo, Kashiwa, Chiba 277-8583, Japan}
\affiliation[b]{ School of Physics and Astronomy, and Minnesota Institute for Astrophysics, 
University of Minnesota, Minneapolis, 55455 (USA)}
\affiliation[c]{Amherst Center for Fundamental Interactions, Department of Physics, University of Massachusetts, Amherst, MA 01003 (USA)}

\abstract{We consider a model in which a pseudo-scalar field $\sigma$ rolls for some e-folds during inflation, sourcing one helicity of a gauge field. These fields are only gravitationally coupled to the inflaton, and therefore produce scalar and tensor primordial perturbations only through gravitational interactions. These sourced signals are localized on modes that exit the horizon while the roll of $\sigma$ is significant. We focus our study on cases in which the model can simultaneously produce {\em (i)} a large gravitational wave signal, resulting in observable  B-modes of the CMB polarizations,  and {\em (ii)} sufficiently small scalar perturbations, so to be in agreement with  the current limits from temperature anisotropies.  Different choice of parameters can instead lead to a localized and visible departure from gaussianity in the scalar sector, either at CMB or LSS scales.
}

\begin{document}

\begin{flushright}  ACFI-T15-10, IPMU15-0130, UMN--TH--3448/15  \end{flushright}

\maketitle
\flushbottom

\section{Introduction}
\label{sec:intro}

There is currently  strong experimental effort to detect the gravitational waves (GW)  generated during inflation. This signal is conventionally parameterized by the tensor-to-scalar ratio $r$, which is the ratio between the GW power spectrum (summed over both polarizations) and the density power spectrum at a given large scale. A  joint BICEP2/Keck Array and {\it Planck} analysis has recently reported the $95\%$ CL bound $r< 0.12$ at  the scale $0.05 \, {\rm Mpc}^{-1}$ \cite{Ade:2015tva}.  It is expected that this limit will be  improved by several ongoing and forthcoming CMB polarization measurements. For example, a statistical uncertainty  $\sigma \left( r \right) = 0.001$ or below is quoted in the  proposed stage $4$ of ground-based CMB experiments \cite{Abazajian:2013vfg}. 

It is commonly stated that observing the GW generated during inflation provides both {\em (i)} the energy scale of inflation, and {\em (ii)} a lower bound on the  excursion of the inflaton field during inflation (the so-called Lyth  bound \cite{Lyth:1996im}, which holds in the case of single field inflation). In terms of $r$, these results read 
\begin{equation}
V^{1/4} \simeq 10^{16} \, {\rm GeV} \, \left( \frac{r}{0.01} \right)^{1/4} \;\;,\;\; 
\frac{\Delta \phi}{M_p} \ga \left( \frac{r}{0.01} \right)^{1/2} \,. 
\label{standard-GW}
\end{equation}
The observation of an inflationary GW signal does not guarantee that the expressions (\ref{standard-GW}) are valid, as they rely on the assumption that the observed GW are  vacuum modes of the metric, amplified by the inflationary expansion, while it is possible that the observed GW are sourced by some other field during inflation. Several recent studies  have shown that it is rather nontrivial to realize such mechanisms.  Whatever the GW source is, it couples at least gravitationally not only to the tensor metric perturbations, but also to the scalar ones, and so it unavoidably affects also the scalar perturbations to some degree. The source typically leads to a statistics that is significantly less gaussian than that of the vacuum modes.  Therefore, respecting the stringent limits imposed by the gaussianity of the scalar perturbations often restricts the source term to a too low level to lead to visible GW. 

This problem manifest itself for example~%
\footnote{Other mechanisms for the generation of GW  which are alternative to the standard vacuum production, but are still embedded in the inflationary scenario include the use of  spectator fields with low sound speed \cite{Biagetti:2013kwa,Biagetti:2014asa,Fujita:2014oba}, the modification of the dispersion relation of the tensor modes in the effective-field-theory approach \cite{Cannone:2014uqa,Cannone:2015rra}, varying sound speed of the tensor \cite{Cai:2015dta},  the strong tachyonic growth of chiral tensor modes in the chromo-natural inflation \cite{Dimastrogiovanni:2012ew,Adshead:2013nka,Obata:2014loa} and in the gauge-flation model \cite{Namba:2013kia}, and  preheating \cite{Khlebnikov:1997di,Easther:2006gt,GarciaBellido:2007dg,Dufaux:2007pt,Dufaux:2008dn,Dufaux:2009wn,Dufaux:2010cf,Figueroa:2013vif} (in this last case the produced GW signal is typically at scales much shorter than the CMB ones).  It has also been investigated~\cite{Antoniadis:2014xva,Kleban:2015daa}  whether the presence of many light degrees of freedom could modify the standard relation~(\ref{standard-GW}) between the scale of inflation and $r$; as shown in~\cite{Kleban:2015daa}, this does not appear to be the case.}
in what is probably the most studied mechanism of GW generation \cite{Cook:2011hg,Senatore:2011sp,Barnaby:2012xt,Carney:2012pk}. In this mechanism, the mass of a scalar field $\chi$ depends on the value of the inflaton, and it is arranged in such a way that it vanishes when the inflaton reaches a given value $\phi_*$ during inflation. As the inflaton crosses $\phi_*$, a burst of quanta of the field $\chi$ is generated non-perturbatively \cite{Chung:1999ve} and can source GW. In all the implementations of this mechanism that have been worked out in some details the GW production is however negligible;  as soon as the inflaton moves past $\phi_*$, the quanta of $\chi$ rapidly become non-relativistic. This suppresses their quadrupole moment, so that the sourced GW signal is below the observational level \cite{Barnaby:2012xt}.  On the contrary, the production of scalar perturbations from these quanta is not suppressed. The non observation of these non-gaussian scalar modes puts strong bounds on this mechanism.

To have a successful mechanism for the production of a visible GW during inflation it is therefore crucial to maximize the ratio between the sourced GW and the sourced scalar modes \cite{Barnaby:2012xt}.~\footnote{ We note that also in  warm inflation \cite{Berera:1995ie} the tensor-to-scalar ratio is  reduced compared to the standard case due to the dissipative effects and the much greater production of scalar with respect to tensor perturbations due to the coupled inflaton \cite{BasteroGil:2009ec}.} For example, one could imagine that the mass of $\chi$ is controlled by a field $\sigma$ different from the inflaton.  The absence of a direct coupling between the inflaton and $\chi$ decreases the amount of scalar perturbations produced by the latter. Even in this case, however, the gravitational coupling between $\chi$ and the inflaton leads to a much greater production of scalar than tensor perturbations, due to the non-relativistic nature of $\chi$ \cite{Barnaby:2012xt}. This led ref.~\cite{Barnaby:2012xt} to study the case in which the source of GW is both {\em (i)} only gravitationally coupled to the inflaton, and {\em (ii)} a relativistic gauge field $A_\mu$. 

This is a modification of the mechanism studied in \cite{Sorbo:2011rz}, in which the vector field is produced by a pseudo-scalar inflaton $\phi$. It was shown in~\cite{Anber:2006xt} that the pseudo-scalar coupling $\frac{\phi}{f} F \, {\tilde F}$~\footnote{In this expression,  $f$ is a mass scale, often denoted as the ``axion scale'', $F_{\mu \nu}$ is the field strength of the vector, and ${\tilde F}^{\mu \nu} \equiv \frac{\epsilon^{\mu \nu \alpha \beta}}{2 \sqrt{-g}} F_{\alpha \beta}$ its dual. The quantity $\epsilon^{\mu \nu \alpha \beta}$ is totally anti-symmetric, and normalized to $\epsilon^{0123} = 1$.} amplifies one helicity of the vector field. This helicity mostly sources the GW  of one definite chirality  \cite{Sorbo:2011rz}, through the two-body interaction $\delta A + \delta A \rightarrow \delta g$. The chiral nature of the signal is particularly interesting,~\footnote{Moreover, the $\phi F {\tilde F}$ interaction does not lead to such a ghost instability as the mechanism of chiral GW production from a gravitational Chern-Simons term coupled to the inflaton \cite{Lyth:2005jf,Satoh:2010ep}.}  since it leads to violation of parity in the CMB that can \cite{Lue:1998mq,Saito:2007kt,Gluscevic:2010vv,Ferte:2014gja} allow to discriminate the GW produced by this mechanism from the vacuum ones.~\footnote{The phenomenological constraints on the production of chiral magnetic fields and their subsequent sourcing of chiral gravitational waves have been studied in \cite{Caprini:2003vc,Caprini:2014mja}.} However, the vector quanta also inverse decay to inflaton perturbations,~\footnote{At the technical level, in the presence of a direct coupling the sourced scalar power is enhanced with respect to the sourced tensor power by a factor of $\frac{1}{\epsilon_\phi^2}$, cf. eqs. (8) and (10) of \cite{Barnaby:2010vf}  and the discussion in section 4.1 of \cite{Sorbo:2011rz}.} $\delta A + \delta A \rightarrow \delta \phi$. This again would lead to an exceedingly large CMB non-gaussianity, so that, once the CMB limits are respected, the produced GW are undetectable \cite{Barnaby:2010vf,Barnaby:2011vw}. The scalar non-gaussianity can be reduced if the inflation sources a large number $N$ of vector species, so that the inflaton perturbations sourced by these vectors are gaussian due to the central limit theorem (obtaining a visible GW signal however requires $N \sim {\rm few \; hundreds} - {\rm thousands}$  \cite{Sorbo:2011rz}). An observable GW signal may be also achieved if the CMB anisotropies are due to a curvaton   \cite{Sorbo:2011rz}. Ref.  \cite{Barnaby:2012xt} studied instead the case in which the rolling pseudo-scalar is a field $\sigma$ decoupled as much as possible from the inflaton (thus realizing the conditions (i) and (ii) mentioned above~\footnote{A similar mechanism that also satisfies these conditions, and that also leads to observable GW, has been  recently studied in \cite{Choi:2015wva}. In this work, the sourcing vector field is produced by a rolling dilation field $\sigma$ (only gravitationally coupled to the inflaton) through the Ratra $f \left( \sigma \right) \, F^2$ mechanism \cite{Ratra:1991bn}.}). It was shown there that, in this case, the $\delta A + \delta A \rightarrow \delta \phi$ gravitational interaction produces a negligible amount of inflaton perturbations, so that the mechanism can indeed lead to a visible GW signal. 

The mechanism of   \cite{Barnaby:2012xt} was further studied in \cite{Mukohyama:2014gba}, as a possible way to generate a blue GW spectrum, as required to reconcile the original inflationary BICEP2 interpretation  \cite{Ade:2014xna} of their signal, and the upper bounds on $r$ at larger scales. Some specific potentials for $\sigma$ were considered in  \cite{Barnaby:2012xt}, and it was noted there that for the mechanism to be successful, one should also ensure that $\sigma$ does not contribute to the observed CMB perturbations. It was mentioned that the simplest way to achieve this is to impose that $\sigma$ becomes massive short after the CMB modes leave the horizon. This issue was further studied in  \cite{Ferreira:2014zia} and in \cite{Mirbabayi:2014jqa}, where it was shown that the $\delta \phi$ production is dominated by their linear interaction with the $\delta \sigma$ quanta that are sourced by the gauge field and that the amplitude of the $\delta \phi$ fluctuations sourced this way is proportional to the number of e-folds during which $\sigma$ is rolling.  In this work we note that the $\delta \phi$ fluctuations are sourced only for modes that leave the horizon when $\dot\sigma\neq 0$ and that the constraints from non-observation of a nonvanishing scalar bispectrum are relatively weak for modes with $\ell\lesssim 100$ or so. As a consequence, if $\dot\sigma \simeq 0$ when modes with $\ell\ga {\cal O}(10^2)$ leave the horizon, then the constraints from non-gaussianity can be more easily evaded.

In the computations of  \cite{Barnaby:2012xt} and  \cite{Ferreira:2014zia}, $\dot{\sigma}$ was assumed to be constant. Here we study the generation of primordial perturbations in a more realistic setup, namely in a model in which the field $\sigma$ evolves in a given potential. The simplest potential for an axion field, 
\begin{equation}
V_\sigma \left( \sigma \right) = \frac{\Lambda^4}{2} \left[ \cos \left( \frac{\sigma}{f} \right) + 1 \right]\,,
\label{V-sigma}
\end{equation}
allows for  a very small $\dot{\sigma}$ both at early and late times, when $\sigma$ is, respectively, close to the maximum ($\sigma =0$) and the minimum of the potential ($\sigma =  f \pi$). The speed has a peak at an intermediate time, when $\sigma$ is between the maximum and the minimum (we assume that $V_\sigma$ is subdominant with respect to the inflaton potential; moreover, we assume that the inflaton potential is very flat, so that the Hubble rate $H$ can be treated as constant; this is the most interesting regime for our discussion, since a very flat inflaton potential corresponds to unobservable vacuum GW). The peak lasts for a number of e-folds roughly of ${\rm O } \left( \frac{H^2}{m^2} \right)$, where $m$ is the curvature of (\ref{V-sigma}). So, remarkably, the simplest axion potential is a perfect candidate for generating a visible GW signal, while keeping the $\delta \phi$ production under control. In this work we show that this is indeed the case through explicit computations. 

We present some specific examples (namely, some choice of parameters in the model) for which the sourced tensor mode strongly dominates over the vacuum one at large scales, leading to observable B modes of the CMB polarization. The GW signal also leads to a marginally observable TB correlation (as a consequence of the broken parity invariance of the mechanism) and to a well observable (high signal-to-noise ratio) BBB correlation. At the same time,  in such examples we find no statistically significant signatures in the TT, and TTT temperature correlators. 

The plan of the  work is the following. In Section \ref{sec:model} we present the model, the background evolution, and the vector field production. In Section \ref{sec:sourced} we study the cosmological perturbations (scalar and tensor modes) sourced by the vector field.  In Section \ref{sec:pheno} we summarize our results for the two- and three-point scalar and tensor perturbations, and we discuss their phenomenology. In Section \ref{sec:conclusions} we present our conclusions. The work is complemented by six appendices. In Appendix \ref{app:WKB} we compute (in WKB approximation) the gauge field produced in the case of nonconstant 
$\dot{\sigma}$. In Appendix \ref{app:constant}  we review the computation of scalar modes produced in the case of constant $\dot{\sigma}$. In Appendices \ref{app:scalar} and \ref{app:tensor} we give details of, respectively, the scalar and tensor mode computation. In Appendix \ref{app:BS} we present some properties of the bispectra produced in the model. In Appendix \ref{app:NG} we  estimate the departure from gaussianity of the statistics of the sourced modes. 

\section{Model, background evolution, and vector field production}
\label{sec:model}

We will consider a system containing the inflaton $\phi$ and a second rolling field $\sigma$ which interacts with the $U(1)$ gauge field $A_\mu$ via an axionic coupling, so that the lagrangian reads 
\begin{equation}\label{lagr}
{\cal L} = - \frac{1}{2} \left( \partial \phi \right)^2  - \frac{1}{2} \left( \partial \sigma \right)^2 - V(\phi,\,\sigma) - \frac{1}{4} F^2 - 
\alpha\frac{\sigma}{4 f} F {\tilde F}\,.
\end{equation}

The rolling of $\sigma$ provides a time dependent background for the gauge field and amplifies its vacuum fluctuations. Such a phenomenon, on a de Sitter Universe with expansion rate $H$, is controlled by the dimensionless quantity 
\begin{equation}
\xi\equiv\frac{\alpha\,\dot\sigma}{2\,f\,H}\,\,,
\end{equation}
which must be larger than unity or so to give a significant effect~\cite{Anber:2006xt}.
 
In order to decouple as much as possible the fluctuations in the gauge field from the scalar perturbations that are strongly constrained by CMB measurements, we will assume as in~\cite{Barnaby:2012xt} that there is no direct coupling between the inflaton and the field $\sigma$, and we will write $V(\phi,\,\sigma)=V_\phi(\phi)+V_\sigma(\sigma)$.

As we have discussed in the Introduction, if $\dot\sigma$ is approximately constant throughout inflation, then an exceedingly large non-gaussian component of the metric perturbations is generated. However,  a large weight in the constraints on the primordial bispectrum from cosmic microwave background radiation measurements is carried by  relatively high multipoles, $\ell={\cal O}(100-1000)$, whereas a possible observation of primordial tensor modes relies on measurements at larger scales, corresponding to $\ell\sim {\cal O}(10 - 100)$. As a consequence, we will focus on the case where $\dot\sigma$ is significantly different from zero only during the epoch when scales of the order of $10^{-2} - 10^{-1}$ times the current size of the observable Universe have left the inflationary horizon.

Remarkably, the most natural potential for an axion-like field such as $\sigma$, namely eq. (\ref{V-sigma}),  does the job:  assuming $V_\sigma\ll V_\phi$, one finds  the slow roll solution
\begin{equation}\label{solsigma}
\sigma = 2 f \, \arctan \left[  {\rm e}^{\delta H \left( t - t_* \right) } \right] \,, 
\end{equation} 
where $\delta\equiv \Lambda^4/(6\,H^2\,f^2)$ and where we have  denoted by $t_*$ the time at which $\sigma = \frac{\pi}{2}\,f$. As a consequence,
\begin{equation}
\dot{\sigma}  = \frac{f\, H\, \delta}{\cosh \left[ \delta H \left( t - t_* \right) \right]} \,. 
\label{dot-sigma}
\end{equation} 
The slow roll condition gives 
\begin{equation}
\frac{\ddot{\sigma}}{3 H \dot{\sigma}} = - \frac{\delta}{3} \, \tanh \left[  \delta H \left( t - t_* \right)  \right] \,, 
\end{equation}
so, to ensure that the slow-roll solution is valid, we require that $\delta \ll 3$. Finally, the parameter $\xi$ is given by 
\begin{equation}
\xi \equiv \frac{\alpha \, \dot{\sigma}}{2 H f} 
\equiv \frac{\xi_*}{\cosh \left[ \delta H \left( t - t_* \right) \right]} 
= \frac{2 \, \xi_*}{\left( \frac{a}{a_*} \right)^\delta + \left( \frac{a_*}{a} \right)^\delta } \,, 
\label{xi-def}
\end{equation} 
where $\xi_*$ and by $a_*$ are, respectively, the value of $\xi$ and of $a$ at $t=t_*$; we note that  $\xi_*\equiv \alpha\,\delta/2$. The value of $\xi$ is significantly different from zero and of the order of $\xi_*$ only for $(2\,\xi_*)^{-1/\delta}\lesssim a/a_*\lesssim (2\,\xi_*)^{1/\delta}$. 

\subsection{Amplification of the fluctuations of the gauge field}
\label{gaugemodes}

The last term in the lagrangian~(\ref{lagr}) is responsible for the amplification of the vacuum fluctuations of the gauge field. To see this, we choose the Coulomb gauge and we define
\begin{eqnarray}\label{a16}
{\hat A}_i(\tau,\,{\vec x})=\int\frac{d^3k}{\left(2\pi \right)^{3/2}} \, {\rm e}^{i{\vec k\cdot \vec x}}{\hat A}_i(\tau,{\vec k})=\sum_{\lambda=\pm}\int \frac{d^3k}{\left(2\pi \right)^{3/2}}\left[\epsilon_i^{(\lambda)}(\vec k)\,A_\lambda(\tau,\,\vec k)\,{\hat a}_\lambda \left( \vec k \right) \, {\rm e}^{i{\vec k\cdot \vec x}}+{\mathrm {h.c.}}\right] \,, \nonumber\\ 
\end{eqnarray}
where the helicity vectors $\vec{\epsilon} \,{}^{(\pm)}$ satisfy $\vec{k}\cdot \vec{\epsilon} \,{}^{(\pm)}=0$, $i\,\vec{k}\times\vec{\epsilon} \,{}^{(\pm)}=\pm k\,\vec{\epsilon} \,{}^{(\pm)}$, $\vec{\epsilon} \,{}^{(\pm)} \cdot\vec{\epsilon} \,{}^{(\mp)}=1$ and $\vec{\epsilon} \,{}^{(\pm)} \cdot\vec{\epsilon} \,{}^{(\pm)}=0$. Explicitly, if ${\hat k} = \left( \sin \theta  \cos \phi ,\,  \sin \theta  \sin \phi ,\, \cos \theta \right)$, then
\begin{eqnarray}
\vec{\epsilon} \,{}^{(\pm)} \left( {\hat k } \right) &=& \frac{1}{\sqrt{2}} \left( \cos \theta \, \cos \phi \mp i \, \sin \phi ,\, 
 \cos \theta \, \sin \phi \pm i \, \cos \phi ,\, - \sin \theta \right)\,.
\end{eqnarray} 

The mode functions $A_\pm$ satisfy the equation 
\begin{align}\label{eom-apm}
A_{\pm}''+\left(k^2 \mp k\,\frac{\alpha \, \sigma'}{f}\right)A_{\pm}=0\,,
\end{align}
where the prime denotes differentiation with respect to the conformal time $\tau$. In the case of   spatially flat, inflating Universe with Hubble parameter $H$ and scale factor $a(\tau)=-1/(H\,\tau)$,\footnote{In this work, we disregard all terms that are subdominant in a slow-roll expansion.} the second term in parentheses in eq.~(\ref{eom-apm}) reads $k\,\alpha\,\sigma'/f=-2\,k\,\xi/\tau$. 

For constant values of $\xi$, eq.~(\ref{eom-apm}) can be solved exactly~\cite{Anber:2006xt}.  In our case, we must solve the following equation
\begin{equation}
A''_\pm + \left( k^2 \pm \frac{4\,k\,\xi_*}{\tau \left[ \left( \tau / \tau_* \right)^\delta + \left( \tau_* / \tau \right)^\delta \right]} \right) A_\pm = 0\,,
\label{eom-Apm-app}
\end{equation}
where $\xi_*$ is defined in \eqref{xi-def}, and $\tau_* \equiv - 1 / (a_* H)$. Without loss of generality, we set $\xi_*>0$. As a consequence, only positive helicity modes are amplified, and we can neglect the  $A_-$ mode.

The solution of eq.~(\ref{eom-Apm-app}) cannot be written in closed form. However we can obtain a good analytical approximation of the time dependence of $A_+(\tau,k)$ using the WKB approximation, finding (see Appendix \ref{app:WKB}) 
\begin{equation}
A_+\left(\tau , k\right) \simeq \left[ \frac{-\tau}{8\,k\,\xi(\tau)} \right]^{1/4} \tilde A \left( \tau , k \right) \; , \quad
A_+'\left(\tau , k\right) \simeq \left[ \frac{k\,\xi(\tau)}{-2\,\tau} \right]^{1/4} \tilde A \left(\tau , k \right) \; ,
\label{approxaplus}
\end{equation}
where we have defined
\begin{equation}\label{defatilde}
\tilde A \left(\tau,k\right) \equiv N\left[ \xi_* ,\, x_* ,\, \delta \right] \, \exp \left[ - \frac{4 \, \xi_*^{1/2}}{1+\delta} \left( \frac{\tau}{\tau_*} \right)^{\delta/2} \left( - k \tau \right)^{1/2} \right] \; ,
\end{equation}
with $x_* \equiv - k \tau_* = k / (a_*H)$. We determine the time-independent normalization factor $N\left[ \xi_* ,\, x_* ,\, \delta \right]$ by matching $A_+$ at late time $-k\tau \ll 1$ to the full numerical solution of \eqref{eom-Apm-app}. We choose the arbitrary initial phase of $A_+$ such that $N\left[ \xi_* ,\, x_* ,\, \delta \right]$ is real and positive.
As a result, in what follows we will use the following expression for the gauge field 
\begin{eqnarray}
{\hat A}_0 =0\,,\qquad 
{\hat A}_i \left( \tau , \vec k \right) = \int \frac{d^3 k}{\left( 2 \pi \right)^{3/2}} \, {\rm e}^{i \vec{k} \cdot \vec{x}} \, \epsilon_i^{(+)} \left( {\hat k} \right) A_+ \left( \tau ,\, k \right) \left[ {\hat a}_+ \left( \vec{k} \right) + {\hat a}_+^\dagger \left( - \vec{k} \right) \right]\,,
\end{eqnarray}
with the function $A_+(\tau,\,k)$ given by eq.~(\ref{approxaplus}).

For future reference we give here also the ``electric'' and ``magnetic'' field,~\footnote{We do not need to identify the gauge field considered here with the Standard Model photon. Nonetheless, we sometimes use electromagnetic notation for convenience.}  which are related to the vector potential by 
\begin{eqnarray}\label{def_EB}
{\hat E}_i = - \frac{1}{a^2} \,{\hat A}_i'\,,\qquad
{\hat B}_i = \frac{1}{a^2}\, \epsilon_{ijk}\, \partial_j {\hat A}_k \,, 
\end{eqnarray} 
and therefore read
\begin{eqnarray}\label{EB_approx}
{\hat E}_i \left( \tau , \vec k \right) &=& - \int \frac{d^3 k}{\left( 2 \pi \right)^{3/2}} \, {\rm e}^{i \vec{k} \cdot \vec{x}} \, \epsilon_i^{(+)} \left( {\hat k} \right) 
H^2\tau^2 \left[ \frac{k\,\xi(\tau)}{-2\,\tau} \right]^{1/4} \tilde A \left(\tau , k \right)
 \left[ {\hat a}_+ \left( \vec{k} \right) + {\hat a}_+^\dagger \left( - \vec{k} \right) \right] \,, \nonumber\\ 
{\hat B}_i \left( \tau , \vec k \right) &=&  \int \frac{d^3 k}{\left( 2 \pi \right)^{3/2}} \, {\rm e}^{i \vec{k} \cdot \vec{x}} \, \epsilon_i^{(+)} \left( {\hat k} \right) 
H^2\tau^2
\left[ \frac{-k^3\,\tau}{8\,\xi(\tau)} \right]^{1/4} \tilde A \left( \tau , k \right)
 \left[ {\hat a}_+ \left( \vec{k} \right) + {\hat a}_+^\dagger \left( - \vec{k} \right) \right] \; , \nonumber\\ 
\end{eqnarray} 
using the approximate solution \eqref{approxaplus}.

\section{Sourced primordial perturbations}
\label{sec:sourced}

This section is divided in two subsections. In Subsection \ref{scalarmodes} we study the scalar modes generated by the vector modes computed in the previous section. In Subsection \ref{tensormodes} we study the sourced tensor modes. 

\subsection{Generation of scalar perturbations}
\label{scalarmodes}

The action \eqref{lagr} contains two scalar and two tensor degrees of freedom as dynamical modes. After taking the spatially flat gauge,
the scalar sector metric reads
\begin{equation}
d s^2 =  a^2 \left( \tau \right) \left[ - \left( 1 + 2 \phi \right) d \tau^2 + 2 \partial_i B \, dx^i d \tau + \delta_{ij} d x^i dx^j \right] \,, 
\end{equation}
and, after solving for the non-dynamical variables $\phi$ and $B$, we decompose the remaining physical modes $\hat\phi(\tau,\vec x)$ and $\hat\sigma(\tau,\vec x)$ as
\begin{eqnarray}
{\hat \phi} (\tau,\vec x) =  \phi \left( \tau \right) + \int \frac{d^3 k}{\left( 2 \pi \right)^{3/2}} \, {\rm e}^{i  \vec{k} \cdot \vec{x}} \, \frac{{\hat Q}_\phi \left( \vec{k} \right)}{a \left( \tau \right)} \; , \nonumber\\ 
{\hat \sigma} (\tau,\vec x) =  \sigma \left( \tau \right) + \int \frac{d^3 k}{\left( 2 \pi \right)^{3/2}} \, {\rm e}^{i  \vec{k} \cdot \vec{x}} \, \frac{{\hat Q}_\sigma \left( \vec{k} \right)}{a \left( \tau \right)}  \; ,
\label{dec_phisig}
\end{eqnarray}
and denote $\left( \phi_1 , \phi_2 \right) \equiv \left( \phi , \sigma \right)$ and $\left( {\hat Q}_1 , {\hat Q}_2 \right) \equiv \left( {\hat Q}_\phi , {\hat Q}_\sigma \right)$. 
The free part of the action for $\hat Q_i$ then reads
\begin{equation}
S^{(2)}_{\rm free} \big[\hat Q_i \big] = \frac{1}{2} \int d\tau d^3k
\left[ \hat Q_i'{}^\dagger \, \hat Q_i' - \hat Q_i^\dagger \left( k^2 \delta_{ij} + \tilde M_{ij}^2 \right) \hat Q_j \right] \; ,
\label{act_freescalar}
\end{equation}
where prime denotes derivative with respect to the conformal time $\tau$, and
\begin{equation}
\tilde M_{ij}^2 = - \frac{a''}{a} \delta_{ij} + a^2 V_{,ij} + \left( 3 - \frac{\phi_k' \phi_k'}{2 M_p^2} \, \frac{a^2}{a'^2} \right) \frac{\phi_i' \phi_j'}{M_p^2} + \frac{a^3}{M_p^2 a'} \left( \phi_i' V_{,j} + \phi_j' V_{,i} \right) \; ,
\end{equation}
with $V_{,i} \equiv \partial V / \partial \phi_i$. The interaction term in \eqref{lagr} gives
\begin{equation}
S_{\rm int} = - \int d^4x \sqrt{-g} \,\alpha\, \frac{\sigma}{4f} F_{\mu\nu} \tilde F^{\mu\nu} 
= \int d^4x \, a^4 \,\alpha\, \frac{\sigma}{f} \, \vec{\hat E} \cdot \vec{\hat B} \; ,
\label{act_intscalar}
\end{equation}
where the ``electric'' and ``magnetic'' fields are defined in \eqref{def_EB}. The equations of motion for $\hat Q_\phi$ and $\hat Q_\sigma$, derived from \eqref{act_freescalar} and \eqref{act_intscalar}, are~%
\footnote{
In principle the inflaton also couples to the gauge sector through gravity and receives contributions of the type $\delta A + \delta A \rightarrow \delta\phi$. This coupling term is a Planck suppressed operator, and parametrically one may expect its effects to be as large as the ones from $\delta A + \delta A \rightarrow \delta\sigma \rightarrow \delta\phi$, which is of our main interest in this work. However, $\delta\sigma$ is produced near horizon exit and keeps sourcing $\delta\phi$ while being outside the horizon, whereas the direct production of $\delta\phi$ from the gauge field occurs only near horizon crossing and, as shown in~\cite{Barnaby:2012xt}, is negligible. Therefore the latter mechanism, mediated by $\delta\sigma$, provides the leading contribution from the gauge field to the curvature perturbations. 
}
\begin{align}
& \left( \frac{\partial^2}{\partial \tau^2} + k^2 + \tilde M_{\phi\phi}^2 \right) \hat Q_\phi + \tilde M_{\phi\sigma}^2 \hat Q_\sigma = 0 \; ,
\nonumber\\
& \left( \frac{\partial^2}{\partial \tau^2} + k^2 + \tilde M_{\sigma\sigma}^2 \right) \hat Q_\sigma + \tilde M_{\sigma\phi}^2 \hat Q_\phi =\alpha \frac{a^3}{f} \int \frac{d^3x}{(2\pi)^{3/2}} \, {\rm e}^{-i \vec k \cdot \vec x} \, \vec{\hat E} \cdot \vec{\hat B} \; .
\label{eom_Qi_exact}
\end{align}
Expanding $\tilde M_{ij}$ to first order in the slow roll parameters, we have 
\begin{equation}
\tilde M_{ij}^2 \simeq - \frac{1}{\tau^2} \left(
\begin{array}{cc}
2 + 9 \, \epsilon_\phi + 3 \, \epsilon_\sigma - 3 \, \eta_\phi &
6 \sqrt{\epsilon_\phi \epsilon_\sigma} \\
6 \sqrt{\epsilon_\phi \epsilon_\sigma} &
2 + 9 \, \epsilon_\sigma + 3 \, \epsilon_\phi - 3 \, \eta_\sigma
\end{array}
\right) \; ,
\label{Mij_app}
\end{equation}
where we have defined the slow-roll parameters as
\begin{equation}
\epsilon_{\phi_i} \equiv \frac{\dot{\phi_i^2}}{2 M_p^2 H^2} \; , \quad
\eta_{\phi_i} \equiv M_p^2 \frac{V_{,ii}}{V} \; ,
\end{equation}
with no summation for the repeated $i$ indices. In the following computation, we focus on the production of $\delta\sigma$ from the gauge field $\delta A$ and its subsequent sourcing of $\delta\phi$, namely the process $\delta A + \delta A \rightarrow \delta\sigma \rightarrow \delta\phi$. In this regard, we only take the dominant terms in \eqref{Mij_app} and approximate \eqref{eom_Qi_exact} as
\begin{align}
& \left( \frac{\partial^2}{\partial \tau^2} + k^2 - \frac{2}{\tau^2} \right) \hat Q_\phi \simeq \frac{6}{\tau^2} \sqrt{\epsilon_\phi \epsilon_\sigma} \, \hat Q_\sigma  \; ,
\label{eom_Qi_app_phi}\\
& \left( \frac{\partial^2}{\partial \tau^2} + k^2 - \frac{2}{\tau^2} \right) \hat  Q_\sigma 
\simeq \alpha\frac{a^3}{f} \int \frac{d^3x}{(2\pi)^{3/2}} \, {\rm e}^{-i \vec k \cdot \vec x} \, \vec{\hat E} \cdot \vec{\hat B} 
\equiv \hat {\cal S}_\sigma \left(\tau, \vec k \right) \; .
\label{eom_Qi_app_sigma}
\end{align}
We note here that we are disregarding the backreaction of the $\ \hat Q_\phi $ quanta sourced by $ \hat Q_\sigma $ on the sourcing $ \hat Q_\sigma $ modes; this effect is of higher order in slow-roll.  Let us also emphasize that since we are interested in the feature due to the change in $\dot{\sigma}$ during the trajectory, we do not neglect the time dependence of $\epsilon_\sigma$, while we assume constant $\epsilon_\phi$. 

In the spatially flat gauge, the scalar curvature perturbations are related to the inflaton perturbations by~\footnote{In a general two scalar field model,  the curvature $\zeta$ is a linear combination $\zeta \left( t \right) = c_1 \left( t \right) \delta \phi \left( t \right) + c_2 \left( t \right) \delta \sigma \left( t \right)$, where the coefficients $c_{1,2}$ depend on the background (the orthogonal combination being an isocurvature mode). With the choice of potential (\ref{V-sigma}), and with our assumptions on the parameters, the field $\sigma$ becomes massive short after the CMB modes are produced. As $\sigma$ becomes a massive field in an inflationary universe, its energy density and pressure very rapidly drop to zero, and so does $c_2$. The only potentially observable effect of $\delta \sigma$ is through its couplings to the inflaton and metric perturbations, which are effective only as long as $\dot{\sigma} \neq 0$. The dominant among these interactions is the linear coupling to $\delta \phi$ that we are computing here (all the other couplings are nonlinear in $\delta \sigma$, and highly subdominant; they can be disregarded, once we impose that the effects of the linear coupling are below the observational level).}  
\begin{equation}
\hat\zeta\left(\tau,\vec k \right) \simeq - \frac{H}{\dot{\phi}} \, \delta \hat{\phi} \left(\tau , \vec k \right) 
\simeq \frac{H \tau}{\sqrt{2 \, \epsilon_\phi} \, M_p} \, \hat Q_\phi \left(\tau , \vec k \right) \; ,
\label{zeta-Qphi}
\end{equation}
assuming $\dot{\phi} >0$.
To find $\hat Q_\phi \left(\tau , \vec k \right)$, we solve eq.~\eqref{eom_Qi_app_phi} by separating $\hat Q_\phi$ into two parts,
\begin{equation}
\hat Q_\phi = \hat Q_\phi^{(0)} + \hat Q_\phi^{(1)} \; ,
\label{split_Qphi}
\end{equation}
where $\hat Q_\phi^{(0)}$ is the homogeneous solution of eq.~\eqref{eom_Qi_app_phi} and $\hat Q_\phi^{(1)}$ is its particular solution. 
 These two contributions are statistically uncorrelated.
We decompose the operator $\hat Q_\phi^{(0)}$, corresponding to the vacuum fluctuations, as 
\begin{equation}
\hat Q_\phi^{(0)} \big(\tau , \vec k \big) = Q_\phi^{(0)} (\tau , k) \, a \big( \vec k \big) + Q_\phi^{(0) *} (\tau,k) \, a^\dagger \big( -\vec k \big)  \; ,
\label{Q0}
\end{equation}
where $a^\dagger$ and $a$ are the creation and annihilation operators for $\hat Q_\phi^{(0)}$, respectively, and the mode function $Q_\phi^{(0)}$ is solved to be, with the Bunch-Davies initial conditions,~\footnote{We neglect the scale dependence of the homogeneous solutions of eq.~(\ref{eom_Qi_exact}) that can be induced by relatively large values of $\epsilon_\sigma$ and $\eta_\sigma$; we have checked that the mixing angle, see eq.~(\ref{diago_mat}), between $\hat{Q}_\phi$ and $\hat{Q}_\sigma$ can be made small enough that such a scale dependence will affect only the unobservable $\hat{Q}_\sigma$ mode and will not leak into the metric perturbations, that are associated to $\hat{Q}_\phi$.} 
\begin{equation}
Q_i^{(0)}(\tau, k) = 
\frac{{\rm e}^{-i k \tau}}{\sqrt{2k}} \left( 1 - \frac{i}{k\tau} \right) \; .
\label{Q0-mode}
\end{equation}

The particular solution $\hat Q_\phi^{(1)}$ is obtained by solving eq.~\eqref{eom_Qi_app_sigma} and plugging its solution into~\eqref{eom_Qi_app_phi}:
\begin{equation}
\hat Q_\phi^{(1)} = 6 \sqrt{\epsilon_\phi} \int d\tau' \, G_k(\tau, \tau') \, \frac{\sqrt{\epsilon_\sigma(\tau')}}{\tau'^2} \int d\tau'' \, G_k(\tau' , \tau'') \, \hat{\cal S}_\sigma \big(\tau'' , \vec k \big) \; ,
\label{sol_Qphi1}
\end{equation}
where the retarded Green function $G_k(\tau , \tau')$ reads
\begin{equation}
G_k(\tau , \tau') =
\Theta(\tau-\tau') \, \frac{\pi}{2} \sqrt{\tau \tau'}
\left[ J_{3/2}(-k\tau) \, Y_{3/2}(-k \tau') - Y_{3/2}(-k\tau) \, J_{3/2}(-k \tau') \right]\,,
\label{Green_JY}
\end{equation}
where $J$ and $Y$ denote the Bessel function of real argument. In this work for the first time we evaluate the sourced solution (\ref{sol_Qphi1}) in the case in which $\dot{\sigma}$ is time-dependent. In Appendix \ref{app:constant} we show that the solution reproduces that of \cite{Ferreira:2014zia} (up to the correction of a typo) in the case of constant $\dot{\sigma}$. 

\subsection{Generation of tensor perturbations}
\label{tensormodes}

Let us next focus on the tensor perturbations, considering a metric of the form 
\begin{equation}\label{lagr_tens}
d s^2 = a^2 \left( \tau \right) \left[ - d \tau^2 + \left( \delta_{ij} + \hat{h}_{ij} \left( \tau ,\, \vec{x} \right) \right) dx^i dx^j \right] \,, 
\end{equation}
where $\hat{h}_{ij}$ is transverse and traceless. The leading expression for $\hat{h}_{ij}$ is obtained by expanding the action to second order in $\hat{h}_{ij}$, including the first order interaction term with the gauge field: ~\footnote{Interaction terms of the form $hhFF$ give contributions to the two-point function of the tensor perturbations that are of the same order in $M_P^{-2}$ as those of the form $hFF$. However the terms of the form $hFF$ give a contribution to the tensor power spectrum that is proportional to $e^{4\pi\xi}$, much larger than  those of the form $hhFF$, whose contribution is proportional to $e^{2\pi\xi}$, which therefore will be  neglected \cite{Eccles:2015ipa}.}
\begin{equation}
S_{\rm GW} = \int d^4 x \left[ \frac{M_p^2 a^2}{8} \left( \hat{h}_{ij}' \hat{h}_{ij}' - \hat{h}_{ij,k} \hat{h}_{ij,k} \right) - \frac{a^4}{2} \, \hat{h}_{ij}  \left(  \hat{E}_i \, \hat{E}_j +  \hat{B}_i \, \hat{B}_j \right) \right] \,. 
\end{equation} 

To obtain a canonically normalized field describing tensor modes in Fourier space we decompose 
\begin{equation}
\hat{h}_{ij} \left( \tau , \vec k \right) = \frac{2}{M_p \, a(\tau)} \int \frac{d^3 k}{\left( 2 \pi \right)^{3/2}} \, {\rm e}^{i \vec{k} \cdot \vec{x}} \sum_{\lambda = +,-} \Pi_{ij,\lambda}^* \left( {\hat k} \right) \, 
\hat{Q}_\lambda \left( \tau,\, \vec{k} \right) \,, 
\label{deco-hij}
\end{equation}
where the polarization operators are 
\begin{eqnarray} 
\Pi_{ij,\pm}^* \left( {\hat k} \right) \equiv \epsilon_i^{(\pm)} \left( {\hat k} \right)  \epsilon_j^{(\pm)} \left( {\hat k} \right) \,. 
\label{pol_tensor}
\end{eqnarray}

The equations of motion for $\hat{Q}_\lambda$
\begin{equation}
\left( \frac{\partial^2}{\partial \tau^2} + k^2 -\frac{2}{\tau^2}\right) \hat{Q}_\lambda \left( \vec{k},\,\tau \right) = - \frac{a^3}{M_p} \, \Pi_{ij,\lambda} \left( {\hat k} \right)\int \frac{d^3x}{(2\pi)^{3/2}} \, {\rm e}^{-i \vec k \cdot \vec x}  \left[\hat{E}_i \, \hat{E}_j +  \hat{B}_i \, \hat{B}_j  \right] \equiv \hat{\cal S}_\lambda(\tau,\,\vec{k}) \,, 
\label{eom_Qlambda}
\end{equation} 
are solved, as in the scalar case considered in the previous subsection, by separating $\hat{Q}_\lambda$ into a vacuum mode $\hat{Q}^{(0)}_\lambda$, solution of the homogeneous equation, and a sourced mode $\hat{Q}^{(1)}_\lambda$ (also in this case the two modes are statistically uncorrelated). The vacuum mode is given by
\begin{eqnarray}
\hat{Q}^{(0)}_\lambda \left( \vec{k} \right) &=&  h_\lambda \left( \tau,\, k \right) \, \hat{a}_\lambda \left( \vec{k} \right) +  h_\lambda^* \left( \tau,\, k \right) \, \hat{a}_\lambda^\dagger \left( - \vec{k} \right) \,, \nonumber\\ 
h_\lambda \left( \tau ,\, k \right) &=& \frac{{\rm e}^{-i k \tau}}{\sqrt{2 k} } \left( 1 - \frac{i}{k \, \tau} \right)\,,
\label{sol_tensorvac}
\end{eqnarray} 
where  $\hat{a}_\lambda^\dagger$ creates gravitons of helicity $2\,\lambda$.

For the sourced mode, we have the formal solution
\begin{equation}
\hat{Q}^{(1)}_\lambda \left( \tau ,\, \vec{k} \right) = \int^\tau d \tau' \, G_k \left( \tau ,\, \tau' \right)\, \hat{\cal S}_\lambda(\tau,\,\vec{k})\,,
\label{Q1lambda_formal}
\end{equation} 
where the retarded propagator is the same as that for the scalar perturbations, eq.~(\ref{Green_JY}). 

\section{Results and phenomenology}
\label{sec:pheno}

The gauge quanta produced by the rolling pseudo-scalar $\sigma$ source quanta of $\sigma$ through inverse decays $\delta A + \delta A \rightarrow \delta \sigma$. They also 
 source inflaton perturbations and gravity waves through the $2 \rightarrow 1$ processes $\delta A + \delta A \rightarrow \delta \phi$ and  $\delta A + \delta A \rightarrow h_\lambda$.  Under the assumption of no direct coupling between the inflaton and $\sigma$ in the potential, the  $\delta A + \delta A \rightarrow \delta \phi$ interaction is of gravitational strength, and negligible with respect to the gravity wave production \cite{Barnaby:2012xt}. For this reason we disregard it in this paper. On the other hand, the inverse decay into quanta of $\sigma$ can produce a large signal, which has a much stronger departure from gaussianity than the vacuum mode \cite{Barnaby:2010vf}. In the model under consideration the field $\sigma$ has a completely negligible energy density at the end of inflation / beginning of reheating (both at the background and at the perturbation level), so that the inflation perturbations can be identified with the observed curvature perturbations $\zeta$, see eq. (\ref{zeta-Qphi}).  However, $\delta \sigma$ is not completely irrelevant for observations: as long as $\sigma$ is rolling (see eq. (\ref{Mij_app})), there is a linear gravitational coupling between perturbations of $\sigma$ and of the inflaton, leading to conversion of some  $\delta \sigma$ into $\delta \phi$ quanta \cite{Ferreira:2014zia}. 

In the previous Section (see also appendices \ref{app:scalar} and \ref{app:tensor}) we computed the curvature perturbations $\zeta$ and the gravity waves produced in the model. The results are summarized in Subsection \ref{subsec:correlators}. In the following Subsection \ref{subsec:phenoresults} we study the phenomenological implications of these results. 

\subsection{Scalar and tensor correlators}
\label{subsec:correlators}

We are interested in the power spectrum and bispectrum of the scalar curvature $\zeta$ and of the gravity wave polarizations $h_\pm$. They are defined as 
\begin{eqnarray}
&& \left\langle {\hat \zeta} \left( \vec{k} \right) \,  {\hat \zeta} \left( \vec{k}' \right) \right\rangle \equiv \frac{2 \pi^2}{k^3} \, {\cal P}_\zeta \left( k \right) \delta^{(3)} \left( \vec{k} + \vec{k}' \right) \,, \nonumber\\ 
&& \left\langle {\hat \zeta} \left( \vec{k}_1 \right) \,  {\hat \zeta} \left( \vec{k}_2 \right)  \,  {\hat \zeta} \left( \vec{k}_3 \right) \right\rangle \equiv  {\cal B}_\zeta \left( k_1 ,\, k_2 ,\, k_3 \right) \delta^{(3)} \left( \vec{k}_1 + \vec{k}_2 +   \vec{k}_3 \right) \,, 
\label{def-P-B}
\end{eqnarray}
and analogously for $h_\pm$. The scalar curvature $\zeta$ and gravity waves $h_\pm$ produced in the model (\ref{lagr}) are the sum of a vacuum component and a component sourced by the gauge quanta. The two components, which we denote, respectively,  as 
\begin{equation}
\zeta = \zeta^{(0)} + \zeta^{(1)} \;\;,\;\; 
h_\pm = h_\pm^{(0)} + h_\pm^{(1)}  \,,  
\label{dec_zetahpm}
\end{equation}
are uncorrelated among each other, and therefore the power spectra and bispectra of these modes are the sum of the vacuum and of the sourced spectra: 
\begin{equation}
{\cal P}_\zeta \left( k \right) = {\cal P}_\zeta^{(0)} \left( k \right) +  {\cal P}_\zeta^{(1)} \left( k \right) \;\;\;,\;\;\; 
{\cal B}_\zeta \left( k \right) = {\cal B}_\zeta^{(0)} \left( k \right) +  {\cal B}_\zeta^{(1)} \left( k \right) \,,  
\end{equation}
and analogously for $h_\pm$. 

The vacuum power spectra are  given by the standard relations \cite{Riotto:2002yw}
\begin{equation}
{\cal P}_\zeta^{(0)} \left( k \right) = \frac{H^2}{8 \pi^2 \epsilon_\phi M_p^2} \;\;,\;\; 
{\cal P}_+^{(0)} \left( k \right) = {\cal P}_-^{(0)} \left( k \right) = \frac{H^2}{\pi^2 M_p^2} \,, 
\label{standard-P}
\end{equation}
where we disregard subleading corrections in slow-roll. This gives the  vacuum relation
\begin{equation}
r_{\rm vac} \equiv \frac{{\cal P}_+^{(0)} + {\cal P}_-^{(0)} }{{\cal P}_\zeta^{(0)}} \simeq 16 \epsilon_\phi \,, 
\label{r-vac}
\end{equation}
for the tensor-to-scalar ratio. The vacuum bispectra are known to be well below the current observational limits, and we disregard them in this work. 

The sourced terms $\zeta^{(1)}$ and $h_\pm^{(1)}$ present a localized bump in momentum space. This corresponds to modes that leave the horizon when $\dot{\sigma}$ is close to its maximum value. The gauge field production is exponentially proportional to the parameter $\xi$, given in eq. (\ref{xi-def}) \cite{Anber:2006xt}.  Therefore the height of the bump is exponentially proportional to the maximum value $\xi_*$ acquired by $\xi$. The width of the peak decreases with increasing values of the parameter $\delta$. This is because, as we have see in  Section \ref{sec:model}, the field $\sigma$ has a significant evolution only for a number of e-folds $\Delta N \simeq \frac{1}{\delta}$, and, therefore, the larger $\delta$ is the fewer are the modes produced while $\sigma$ has a substantial rolling.

As we show in Appendices  \ref{app:scalar} and \ref{app:tensor}, we can factorize the following parametric and functional dependence from the scalar and tensor spectra 
\begin{eqnarray}
{\cal P}_\zeta^{(1)} \left( k \right) &=& \left[ \epsilon_\phi \, {\cal P}_\zeta^{(0)} \left( k \right) \right]^2 \; f_{2,\zeta} \left( \frac{k}{k_*} ,\, \xi_* ,\, \delta \right) \,, \nonumber\\ 
{\cal P}_\lambda^{(1)} \left( k \right) &=& \left[   \epsilon_\phi \, {\cal P}_\zeta^{(0)} \left( k \right) \right]^2  \; f_{2,\lambda} \left( \frac{k}{k_*} ,\, \xi_* ,\, \delta \right) \;\;\;,\;\;\; \lambda = +,- \,, \nonumber\\ 
{\cal B}_\zeta \left( k_1 ,\, k_2 ,\, k_3 \right) &=& \frac{ \left[ \epsilon_\phi \, {\cal P}_\zeta^{(0)} \left( k \right) \right]^3 }{ k_1^2 \, k_2^2 \, k_3^2 } \; f_{3,\zeta} \left( \frac{k_1}{k_*} ,\,  \frac{k_2}{k_*} ,\,  \frac{k_3}{k_*} ,\, \xi_* ,\, \delta \right) \,, \nonumber\\ 
{\cal B}_\lambda \left( k_1 ,\, k_2 ,\, k_3 \right) &=& \frac{ \left[ \epsilon_\phi \, {\cal P}_\zeta^{(0)} \left( k \right) \right]^3 }{ k_1^2 \, k_2^2 \, k_3^2 } \; f_{3,\lambda} \left( \frac{k_1}{k_*} ,\,  \frac{k_2}{k_*} ,\,  \frac{k_3}{k_*} ,\, \xi_* ,\, \delta \right) \;\;\;,\;\;\; \lambda = +,- \,.  
\label{f23-def}
\end{eqnarray} 
For brevity, we denote the functions at the right hand side as  $f_{i,j}$, where $i \in \left\{ 2 ,\, 3 \right\}$ and $j \in \left\{ \zeta ,\, + ,\, - \right\}$. These functions are  dimensionless.  The functions $f_{3,j}$  encode the full dependence of the bispectrum on the momenta $k_i$.  In eq. (\ref{shape})  we provide an approximate relation for the shape dependence,  written in terms only of the two point function and of the three point function in the exact equilateral case. The expression (\ref{shape}) is exact by construction on equilateral triangles, and it is very accurate where the signal is maximum (see Figure \ref{fig:isosceles}). Therefore, to study the phenomenology of the scalar and tensor two- and three-point correlation function we simply need to provide the functions, $f_{2,j}$, as well as  $f_{3,j}$ for equal momenta. 

\begin{figure}[tbp]
\centering 
\includegraphics[width=.45\textwidth,clip]{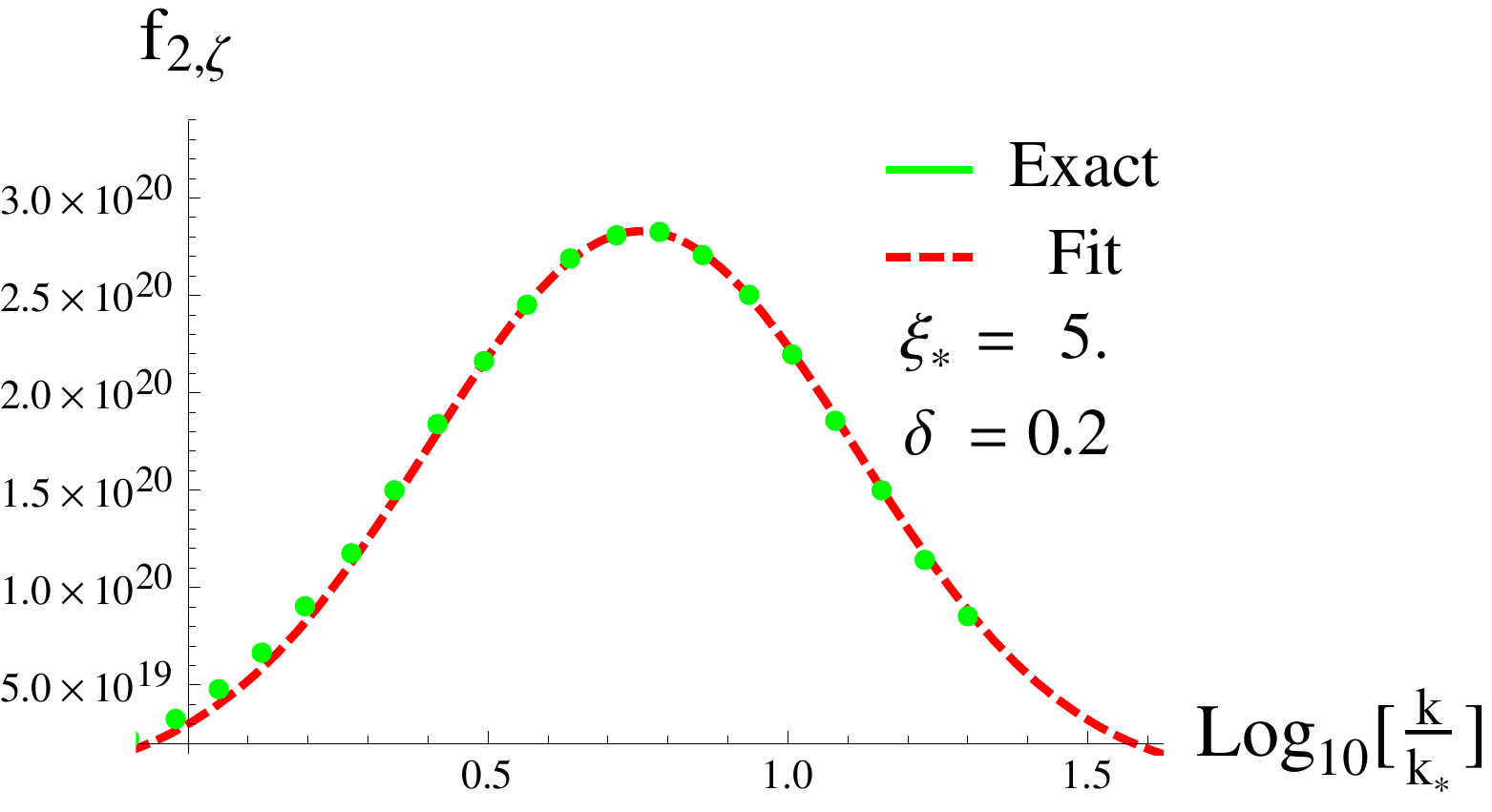}
\includegraphics[width=.45\textwidth,clip]{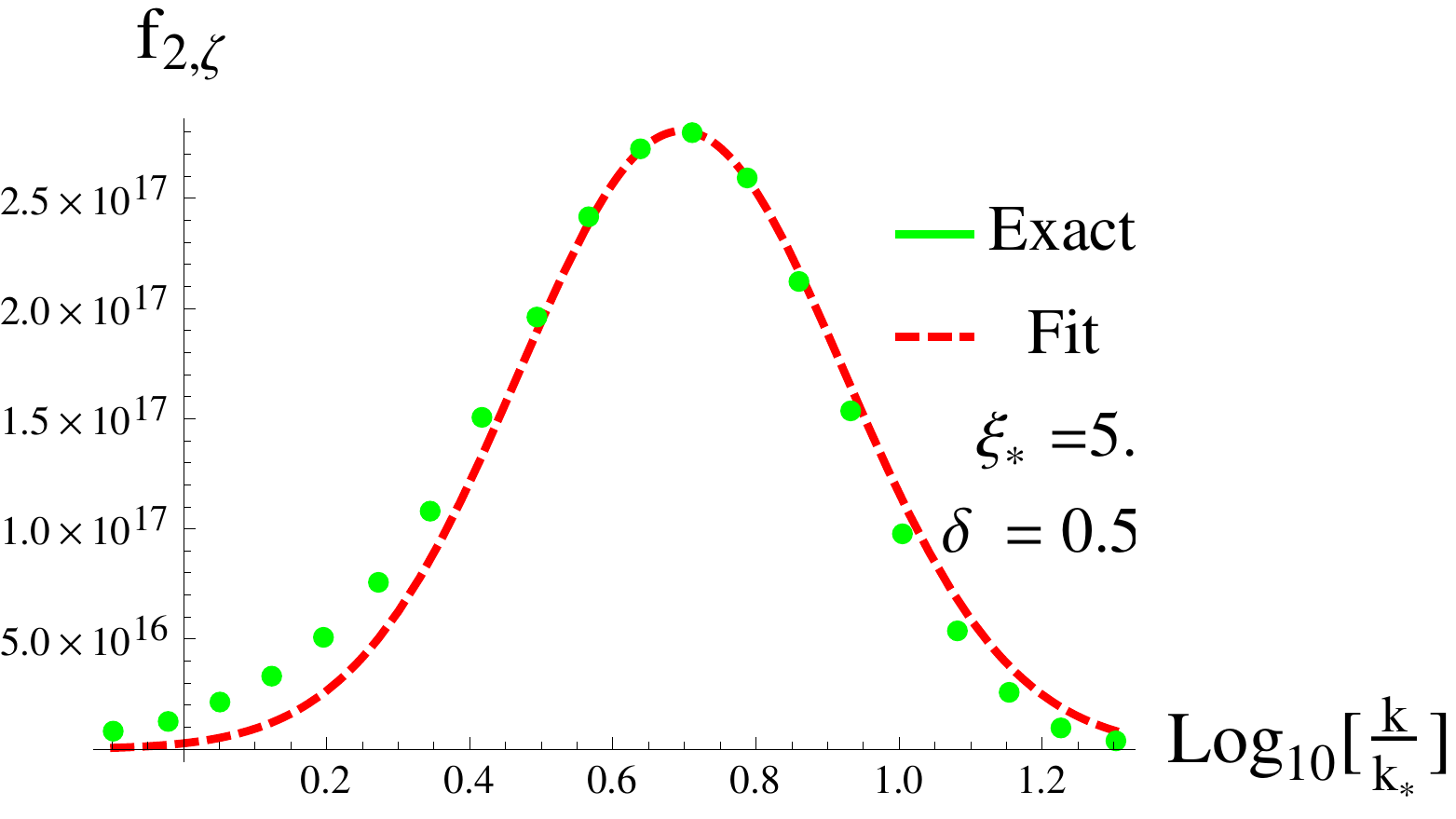}
\hfill
\caption{Comparison of the exact function $f_{2,\,\zeta}$ (green dots) and of the approximate form (red dashed line) given in eq. (\ref{f23-fit}), for the two values of $\delta$ that we have studied and for one illustrative choice of $\xi_*$.  The function $f_{2,\zeta}$, defined in \eqref{f23-def}, depicts a bump in the sourced scalar power spectrum, featuring a short period of a relatively fast rolling of $\sigma$ and the production of the gauge field during this stage.  A bump feature is also present in the tensor sector. The phenomenological consequences of these localized signal are studied in Subsection \ref{subsec:phenoresults}.  
}
\label{fig:fitbump}
\end{figure}

We first studied the momentum dependence of these functions for fixed values of $\xi_*$ and $\delta$. We found that they are well approximated by 
\begin{equation}
f_{i,j}  \left( \frac{k}{k_*} ,\, \xi_* ,\, \delta \right) \simeq f_{i,j}^c \left[ \xi_* ,\, \delta \right] \, {\rm exp} \left[ - \frac{1}{2 \sigma_{i,j}^2  \left[ \xi_* ,\, \delta \right] } \,  
\ln^2 \left( \frac{k}{k_* \, x_{i,j}^c  \left[ \xi_* ,\, \delta \right] }  \right) \right] \,. 
\label{f23-fit}
\end{equation} 
Namely, the momentum dependence~\footnote{For the three point functions ($i=3$), the approximation (\ref{f23-fit}) refers to the exact equilateral case, $k_1=k_2=k_3=k$.} is encoded by the three functions $f^c ,\, \sigma ,\, x^c$, which in turn depend on the evolution of $\dot{\sigma}$ through the model parameters $\xi_*$ and $\delta$. 
The function (\ref{f23-fit}) has a bump at $k = k_* \, x^c$, where it evaluates to $f^c$. The parameter $\sigma^2$ controls the width of the bump. For each choice of $\xi_*$ and $\delta$, the values of $f_{i,j}^c ,\, x_{i,j}^c ,\,$ and $\sigma_{i,j}$ are obtained by fitting the right hand side of ({\ref{f23-fit}}) to reproduce the position, height, and curvature of the bump. In Figure \ref{fig:fitbump} we verify the accuracy of the approximate relation (\ref{f23-fit}) by showing the comparison between the exact and approximate form of $f_{2,\zeta}$ for two choices of parameters (the chosen values have no particular relevance, and have been chosen for illustrative purpose only). An analogous level of accuracy is obtained for the other $f_{i,j}$ functions and for the other choices of $\xi_*$   that we have made.

\begin{figure}[tbp]
\centering 
\includegraphics[width=0.32\textwidth,angle=0]{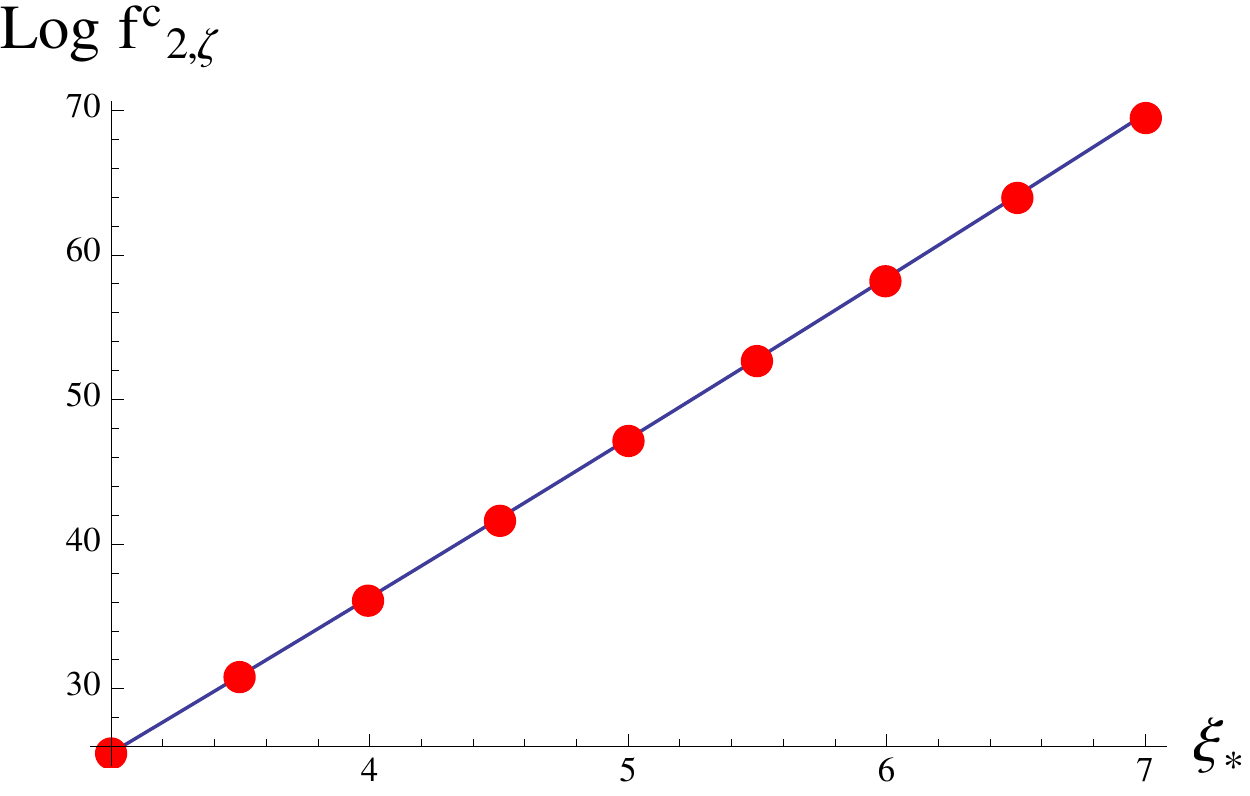}
\includegraphics[width=0.32\textwidth,angle=0]{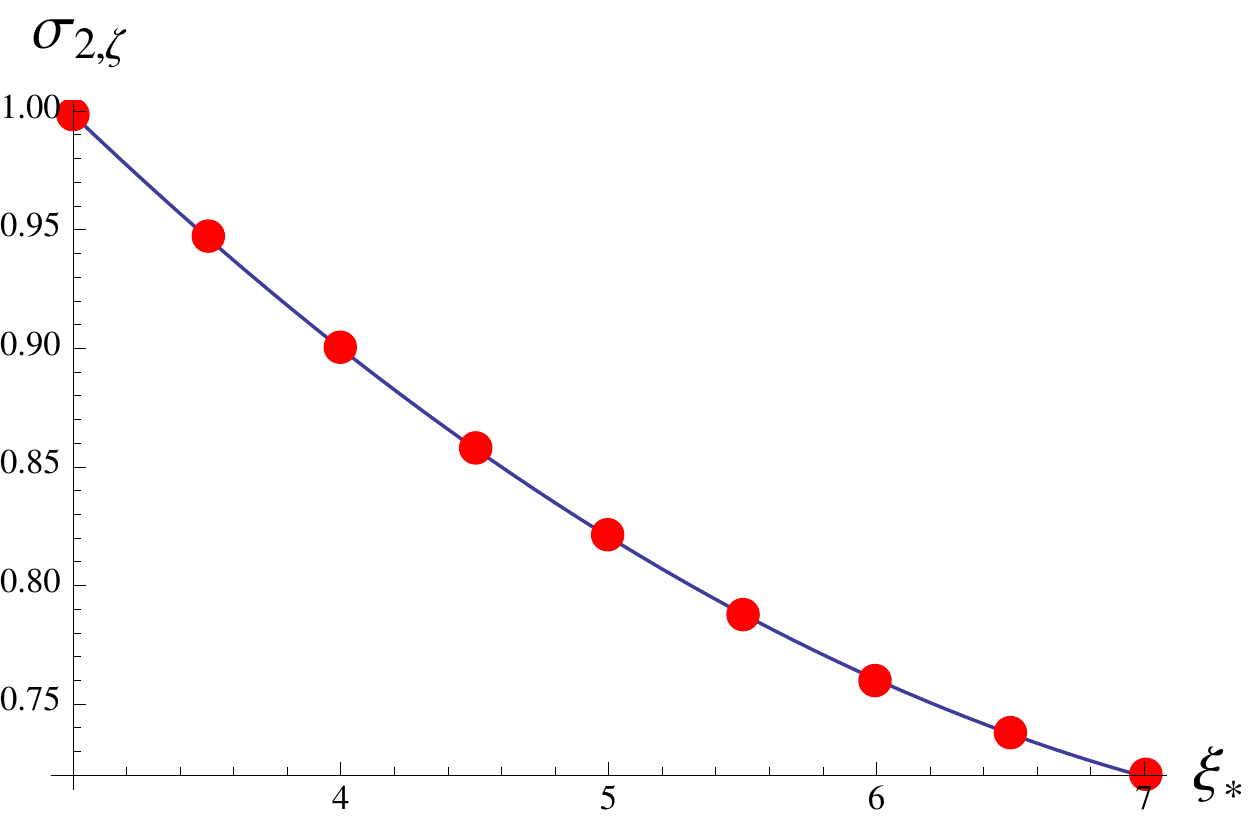}
\includegraphics[width=0.32\textwidth,angle=0]{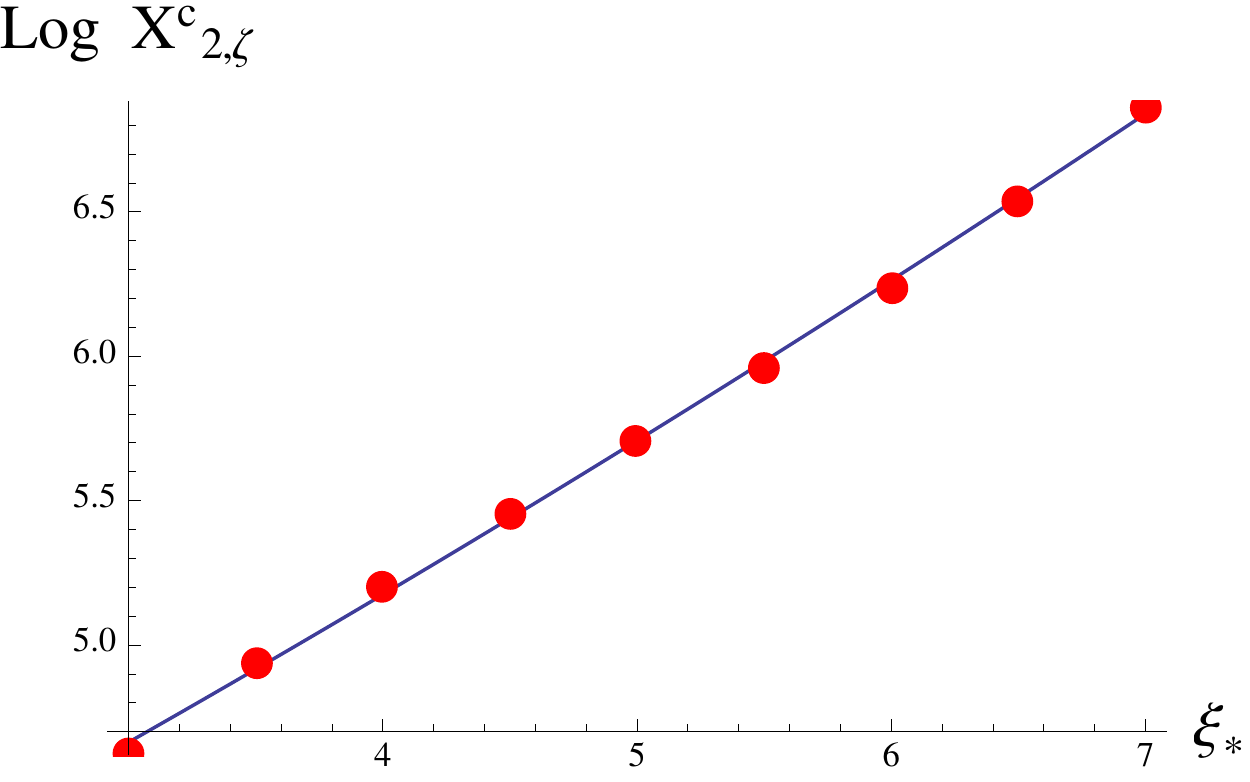}
\hfill
\caption{$\xi_*$ dependence of the fitting $f_{2,\zeta}$, $x^c_{2,\zeta}$ and $\sigma_{2,\zeta}$ entering in (\ref{f23-fit}), for $\delta = 0.2$ (an analogous agreement is obtained in the $\delta = 0.5$ case, and for th other $\left\{ i,\, j \right\}$ functions).  The red dots denotes the values obtained by evaluating the power spectrum at $\xi_* =  \left\{ 3 ,\, 3.5 ,\, 4 ,\, 4.5 ,\, 5 ,\, 5.5 ,\, 6 ,\, 6.5,\,7 \right\}$. The solid lines are the polynomial fits reported in the first row  of Table \ref{tab:fij-d02}. 
}
\label{fig:fitbump2}
\end{figure}

Secondly, we need to specify the dependence of $f_{i,j}^c ,\, x_{i,j}^c ,\, \sigma_{i,j}^2$ on the model parameters $\xi_*$ and $\delta$. 
We evaluated these functions for $\delta = 0.2$ and $\delta = 0.5$, and for  $\xi_* = \left\{ 2.5 ,\, 3 ,\, 3.5 ,\, 4 ,\, 4.5 ,\, 5 ,\, 5.5 ,\, 6 ,\, 7 \right\}$. We found that the dependence on $\xi_*$ is rather smooth, and can be well described by a second degree polynomial. We provide the fitting functions in Tables \ref{tab:fij-d02} and  \ref{tab:fij-d05}. As we see from the examples shown in Figure \ref{fig:fitbump2}, the fitting functions are very accurate. 

\begin{table}[t]
\begin{center}
  \begin{tabular}{|c||c|c|c|c|c|c|c|c|} \hline
$\left\{ i,\,j \right\}$ & $\ln \,  \vert f_{i,j}^c  \vert \simeq $  & $x_{i,j}^c \simeq $ & $\sigma_{i,j} \simeq $ \\ \hline
$\left\{ 2,\zeta \right\} $ & 
$\!\!\!-5.60  + 10.1 \, \xi_*  + 0.0947  \, \xi_*^2$ & 
$\!\!\!3.26 + 0.435 \, \xi_* + 0.0109  \, \xi_*^2$ & 
$\!\!\!1.41 - 0.166 \, \xi_* + 0.00962 \, \xi_*^2$  \\ \hline 
$\left\{ 2,+ \right\} $ & 
$\!\!\!-7.98 + 10.0 \, \xi_* + 0.0979 \, \xi_*^2$ & 
$\!\!\!5.45 + 0.455 \, \xi_* + 0.0316 \, \xi_*^2$ & 
$\!\!\!1.38 - 0.178 \, \xi_* + 0.0103 \, \xi_*^2$  \\ \hline 
$\left\{ 2,- \right\} $ & 
$\!\!\!-13.8 + 9.96 \, \xi_* + 0.104 \, \xi_*^2$ & 
$\!\!\!2.39 + 0.129 \, \xi_* + 0.0214 \, \xi_*^2$ & 
$\!\!\!1.44 - 0.169 \, \xi_* + 0.0102 \, \xi_*^2$  \\ \hline 
$\left\{ 3,\zeta \right\} $ & 
$\!\!\!-4.72 + 15.1 \, \xi_* + 0.142 \, \xi_*^2$ & 
$\!\!\!3.37 + 0.353 \, \xi_* + 0.0142 \, \xi_*^2$ & 
$\!\!\!1.13 - 0.143 \, \xi_* + 0.00830 \, \xi_*^2$  \\ \hline 
$\left\{ 3,+ \right\} $ & 
$\!\!\!-7.77 + 15.1 \, \xi_* + 0.147 \, \xi_*^2$ & 
$\!\!\!5.24 + 0.361 \, \xi_* + 0.0360 \, \xi_*^2$ & 
$\!\!\!1.11 - 0.150 \, \xi_* + 0.00880 \, \xi_*^2$  \\ \hline 
  \end{tabular}
\end{center}
\caption{$\xi_*$ dependence of the functions entering in (\ref{f23-fit}), for $\delta = 0.2$. Among the entries in the first column, only $f_{3,+}^c$ is negative, while the other $f_{i,j}^c$ are positive. 
}\label{tab:fij-d02}
\end{table}

\begin{table}[t]
\begin{center}
  \begin{tabular}{|c||c|c|c|c|c|c|c|c|} \hline
$\left\{ i,\,j \right\}$ & $\ln \,  \vert f_{i,j}^c  \vert \simeq $  & $x_{i,j}^c \simeq $ & $\sigma_{i,j} \simeq $ \\ \hline
$\left\{ 2,\zeta \right\} $ & 
$\!\!\!-6.47 + 9.04 \, \xi_* + 0.0586 \, \xi_*^2$ & 
$\!\!\!1.64 + 0.630 \, \xi_* + 0.00738 \, \xi_*^2$  & 
$\!\!\!0.823 - 0.0872 \, \xi_* + 0.00558 \, \xi_*^2$  \\ \hline 
$\left\{ 2,+ \right\} $ & 
$\!\!\!-6.85 + 9.05 \, \xi_* + 0.0596 \, \xi_*^2$ & 
$\!\!\!2.70 + 0.896 \, \xi_* + 0.0187 \, \xi_*^2$  & 
$\!\!\!0.768 - 0.0993 \, \xi_* + 0.00608 \, \xi_*^2$  \\ \hline 
$\left\{ 2,- \right\} $ & 
$\!\!\!-12.5 + 8.97 \, \xi_* + 0.0656 \, \xi_*^2$ & 
$\!\!\!1.22 + 0.396 \, \xi_* + 0.00976 \, \xi_*^2$  & 
$\!\!\!0.858 - 0.0813 \, \xi_* + 0.00530 \, \xi_*^2$  \\ \hline 
$\left\{ 3,\zeta \right\} $ & 
$\!\!\!-5.98 + 13.6 \, \xi_* + 0.0861 \, \xi_*^2$ & 
$\!\!\!1.71 + 0.569 \, \xi_* + 0.00972 \, \xi_*^2$  & 
$\!\!\!0.641 - 0.0792 \, \xi_* + 0.00495 \, \xi_*^2$  \\ \hline 
$\left\{ 3,+ \right\} $ & 
$\!\!\!-6.03 + 13.6 \, \xi_* + 0.0870 \, \xi_*^2$ & 
$\!\!\!2.63 + 0.841 \, \xi_* + 0.0193 \, \xi_*^2$  & 
$\!\!\!0.606 - 0.0844 \, \xi_* + 0.00520 \, \xi_*^2$  \\ \hline 
  \end{tabular}
\end{center}
\caption{$\xi_*$ dependence of the functions entering in (\ref{f23-fit}), for $\delta = 0.5$. Among the entries in the first column, only $f_{3,+}^c$ is negative, while the other $f_{i,j}^c$ are positive. 
}\label{tab:fij-d05}
\end{table}

\subsection{Phenomenology}
\label{subsec:phenoresults}

In this subsection we study the phenomenology of the model (\ref{lagr}).  The total scalar and tensor modes produced during inflation are the sum of a nearly scale invariant vacuum mode, plus a peaked sourced signal (see Figure \ref{fig:fitbump} for the scale dependence of the latter). The sourced scalar and tensor mode modify the tensor-to-scalar ratio from its vacuum value (\ref{r-vac}), $r_{\rm vac} \simeq 16 \, \epsilon_\phi$,  to
\begin{equation}
r \left( k \right) \simeq  \frac{{\cal P}_+^{(0)} \left( k \right) + {\cal P}_-^{(0)} \left( k \right) + {\cal P}_+^{(1)} \left( k \right)}{{\cal P}_\zeta^{(0)} \left( k \right) + {\cal P}_\zeta^{(1)} \left( k \right)} = 
r_{\rm vac} \: \frac{1+ \frac{ \epsilon_\phi }{ 16 } \, {\cal P}_\zeta^{(0)} \left( k \right) \, f_{2,+} \left( k \right)}{1+ \epsilon_\phi^2 \, {\cal P}_\zeta^{(0)} \left( k \right) \, f_{2,\zeta} \left( k \right)} \;,  
\label{r-tot}
\end{equation}
where we have disregarded the sourced $h_-^{(1)}$ mode (given that $f_{2-} \ll f_{2+}$). Both the numerator and the denominator of the fraction appearing in the last expression of  eq. (\ref{r-tot}) have been written as $1+{\cal P}_{\rm sourced} / {\cal P}_{\rm vacuum}$, so that we can immediately compare the impact on $r$ of the sourced tensor modes vs. the sourced scalar modes. We see that, relatively to the sourced scalars, the sourced tensors are more relevant at smaller values of $\epsilon_\phi$. This is the regime of greatest interest for our study, since it corresponds to a small $r_{\rm vac}$.

\begin{figure}[tbp]
\centering 
\includegraphics[width=0.45\textwidth,angle=0]{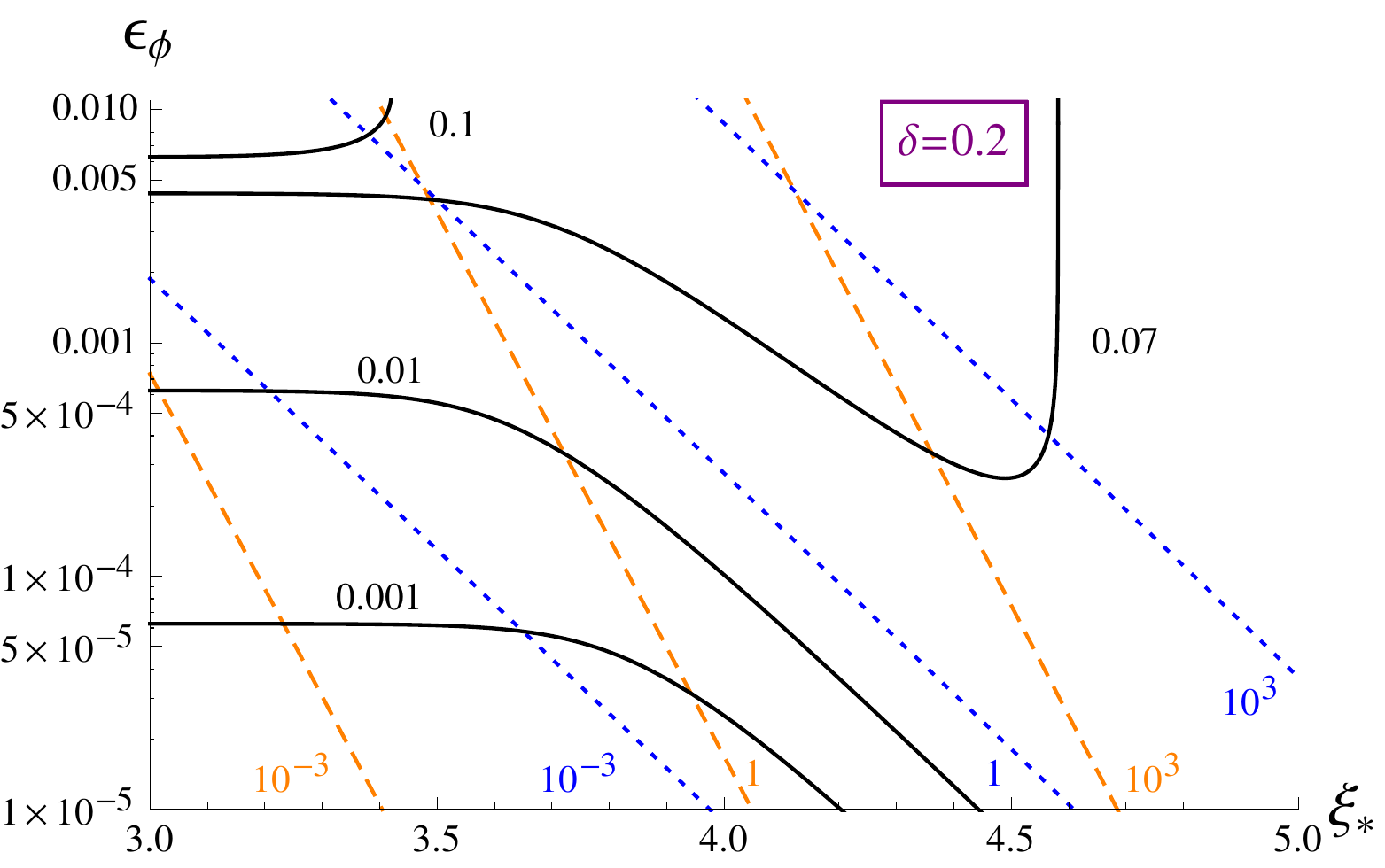}
\includegraphics[width=0.45\textwidth,angle=0]{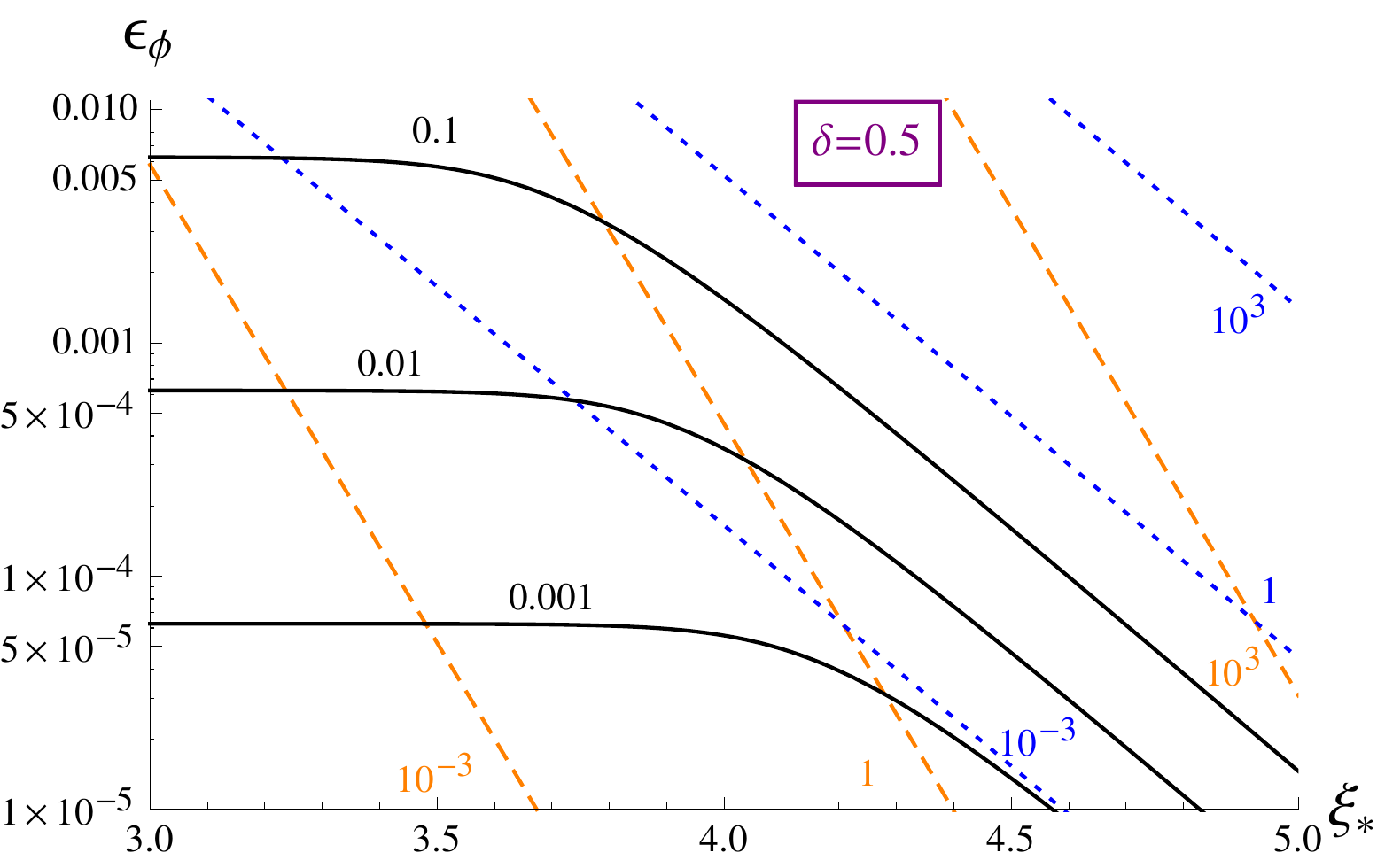}
\hfill
\caption{Black-solid lines: tensor-to-scalar ratio (\ref{r-tot}). Orange dashed (respectively, blue dotted) lines: ratio between the sourced and the vacuum tensor (respectively, scalar) power spectrum. All ratios are evaluated at the peak of the sourced GW power spectrum. 
}
\label{fig:ratios}
\end{figure}

This is confirmed by the contour lines shown in Figure \ref{fig:ratios}, where we show the tensor-to-scalar ratio (black solid lines), the ratio between the sourced and the vacuum tensor power spectrum (orange dashed lines), and the ratio between the sourced and the vacuum scalar power spectrum (blue dotted lines), evaluated at the peak of the sourced GW signal. The ratios are shown as functions of $\xi_*$ and $\epsilon_\phi$. The sourced modes are negligible at the smallest values of $\xi_*$ shown in the figure, so that $r \simeq r_{\rm vac} \simeq 16 \epsilon_\phi$ there (so that the contour lines of equal $r$ are $\xi_*$ independent - and, therefore, horizontal, in this region). At greater values of $\xi_*$, and, particularly, at the smallest $\epsilon_\phi$ shown, we see that $r \gg r_{\rm vac}$.

Figure \ref{fig:ratios}, and several of the following figures, are comprised of two panels. In the left (right) panel we show results for $\delta = 0.2$ ($\delta = 0.5$). The roll of $\sigma$ is substantial for a number of e-folds $\Delta N \simeq \frac{1}{\delta}$, cf. eq. (\ref{dot-sigma}). Increasing $\delta$ therefore decreases the amount of time during which $\sigma$ is rolling. This decreases the amplitude of the produced signal (encoded in the parameter $f_{i,j}^c$, cf. eq. (\ref{f23-fit})), as well as its width (encoded in $\sigma_{i,j}$). The latter effect is due to the fact that, the smaller $\Delta N$ is, the fewer are the modes that exited the horizon while $\dot{\sigma}$ was non-negligible. By comparing the results given in Tables \ref{tab:fij-d02} and \ref{tab:fij-d05},  we can also see that increasing $\delta$ decreases more the scalar production than the tensor one. This is also visible by comparing the left and the right panel of Figure \ref{fig:ratios}, as well as from the following figures, where we can notice that greater values of $\delta$ corresponds to a greater production of tensor vs. scalar modes. This occurs because the inflaton perturbations are sourced by the $\delta \sigma$ modes only as long as $\dot{\sigma} \neq 0$. So, a decrease of $\Delta N$ affects more the sourced scalar modes than the tensor modes because it decreases both (i) the number of modes that are sourced (namely, the width of the peak), and (ii)   the interval of time during which the  $\delta \sigma$ modes (produced by the gauge field) can be converted into inflaton perturbations. 

In the following analysis, we show that the model admits choices of parameters that result in a sourced signal visible in correlators involving the B-mode of the CMB polarization, and that are consistent with  the well-analyzed   TT and  TTT correlators.  To have visible effects from the sourced GW, we choose values of $k_*$ leading to a peak of the sourced signal  at  large  CMB scales. Specifically, we choose  $k_*$ ranging from  $7 \times 10^{-5} \, {\rm Mpc}^{-1}$ (affecting only the first few multipoles) to $5 \times 10^{-3} \, {\rm Mpc}^{-1}$ (affecting  multipoles up to the first  acoustic peak).

\subsubsection{CMB power spectra}

\begin{figure}
\centering{ 
\includegraphics[width=.45\textwidth,clip]{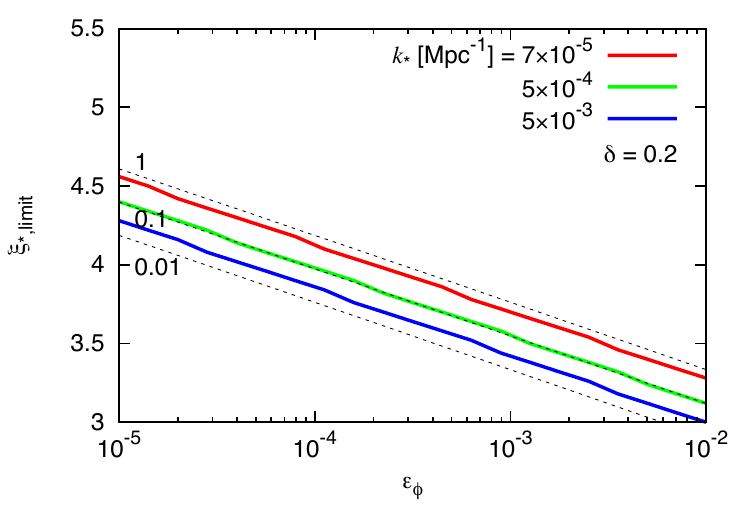}
\includegraphics[width=.45\textwidth,clip]{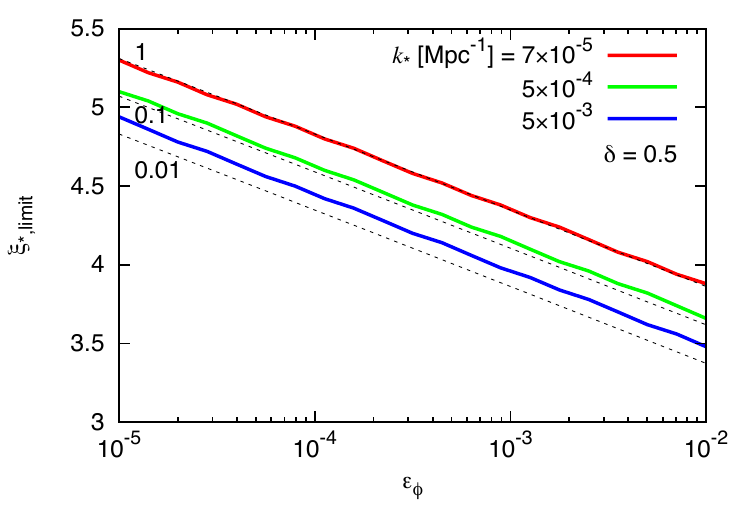}
}
\caption{Solid lines: Largest  value of  $\xi_*$ (controlling the amount of produced quanta) allowed by the WMAP TT data, obtained as explained in the text, as a function of $\epsilon_\phi$, and for different values of $k_*$ 
(controlling the scale of the bump of the sourced modes, see the next figure). Dashed lines: ratio of ${\cal P}_\zeta^{(1)} / {\cal P}_\zeta^{(0)} $ at the peak of the GW bump.  
 }
\label{fig:TT1}
\end{figure}

The main goal of this work is to provide a concrete example of a particle physics process that can enhance the GW signal from inflation, while being ``consistent with the TT  and TTT data''. With this, we mean that this additional signal should not significantly worsen  the fit to the TT and the TTT data of the standard cosmological model which does not include this signal. The signal that we are studying manifests itself at the largest cosmological scales; we compare it against the latest WMAP data \cite{Bennett:2012zja,Hinshaw:2012aka}, which are  cosmic variance limited at such  scales.  We fix all the cosmological parameters consistently with the best fit values reported in  \cite{Hinshaw:2012aka}, denoting this set of values as ${\cal C}_0$;~\footnote{Specifically, we fix the physical baryon density $\Omega_b h^2 = 0.02264$, the physical cold dark matter density $ \Omega_c \, h^2 = 0.1138$, the ionization optical depth $ \tau = 0.089$, the Hubble constant $H_0 = 70.0~{\rm km/sec/Mpc}$. The parameters relevant to the initial condition, i.e., the scalar amplitude $\Delta_\zeta^2 = 2.41 \times 10^{-9}$, the scalar spectral index $n_s = 0.972$ and the WMAP pivot scale $k_0 = 0.002 ~ {\rm Mpc^{-1}}$, fix our vacuum power spectrum ${\cal P}_\zeta^{(0)}(k) = \Delta_\zeta^2(k/k_0)^{n_s - 1}$. We have also assumed a flat universe, with $3.046$ relativistic species and no massive neutrino.} we denote by ${\cal L}_0$ the likelihood of the fit  to the WMAP TT data of the standard cosmological model with these values of the parameters:  
\begin{equation}
{\cal L}_0 = {\rm e}^{-\chi^2_{\xi_* = 0}  / 2} \;\;. 
\end{equation}
We then add the sourced scalar and tensor modes obtained from the mechanism presented in the previous sections. For illustrative purposes, we fix three values of $k_*$ (controlling the scale at which the sourced signals are peaked) and two values of $\delta$ (controlling the width of the sourced peak). For any of these six choices, we then increase $\xi_*$ (the parameter that controls the amplitude of the sourced signal) until the likelihood of the fit decreases by a factor ${\rm e}^{2}$ with respect to ${\cal L}_0$, namely, we find the value  $\xi_* = \xi_{*,{\rm limit}}$ for which 
 $\chi^2_{\xi_{*,{\rm limit}}} = \chi^2_{\xi_* = 0} + 4$.~\footnote{We use the WMAP power spectrum likelihood code: \url{http://lambda.gsfc.nasa.gov/}. } An experimentalist fitting the WMAP data with the cosmological parameters ${\cal C}_0$, would obtain a fit that is two sigmas  worse if he/she includes the sourced signal instead of the standard cosmological model.  In fact, the value $\xi_{*,{\rm limit}}$ obtained in this way is conservative, as we do not vary any of the cosmological parameters ${\cal C}_0$. Varying them, we can only improve the fit of the WMAP data for that given value of $\xi_*$, and so likely obtain larger values for  $\xi_{*,{\rm limit}}$. However, changing values in ${\cal C}_0$ may give a disagreement with data at smaller angular scales than the WMAP ones, and for this reason we do not vary such parameters.  As we show below, in most cases the values of  $\xi_{*,{\rm limit}}$ obtained with this procedure already provide a visible GW signal, which is the goal of this present analysis (in summary, we are not interested in providing precise Bayesian limits on $\xi_*$ within this model - for which we should provide priors, marginalize over all the other parameters and include smaller scales data - but only in the goal specified at the beginning of this subsection). 
 
In Figure \ref{fig:TT1} we present the value of $\xi_{*,{\rm limit}}$ obtained with this procedure as a function of the slow roll parameter $\epsilon_\phi$. In the left (right) panel of the figure we set $\delta = 0.2$ ($0.5$), corresponding to a significant $\dot{\sigma}$ for about $5$ ($2$) e-folds. In each panel we fix  $k_* = 7 \times 10^{-5} \, {\rm Mpc}^{-1} \;, k_* = 5 \times 10^{-4} \, {\rm Mpc}^{-1} \;,$ and $k_* = 5 \times 10^{-3} \, {\rm Mpc}^{-1}$, producing a bump, respectively at the very largest angular scales  ($\ell \la 5$), on the rise of the first peak, and in the region around the first two peaks.~\footnote{We verified that, apart from the $\left\{ \delta = 0.5,\, k_* = 7 \times 10^{-5} \, {\rm Mpc}^{-1} \right\}$ case,  the sourced GW give a negligible contribution to the TT signal, and in all the other cases the limits shown in the figure are due to $\zeta_{\rm sourced}$. } In the same figure we also show with black dashed lines the ratio $ {\cal P}_\zeta^{(1)} / {\cal P}_\zeta^{(0)} $ at the peak of the GW bump. We see that the allowed amount of scalar signal strongly depends on the scale. In the examples with the bump at the largest scales, the sourced signal can be as large as the vacuum one at the peak, due to the large cosmic variance present at those scales. A significantly smaller fraction,  ${\rm O } \left( 1 \% - 10 \% \right)$, is allowed in the examples in which the signal affects the acoustic peaks.  
 
By comparing the left and right panel of Figure \ref{fig:TT1} we again see that, for any fixed value of $\xi_*$, the sourced signal is stronger at small values of $\delta$. 

\begin{figure}[tbp]
\centering 
\includegraphics[width=0.45\textwidth,angle=0]{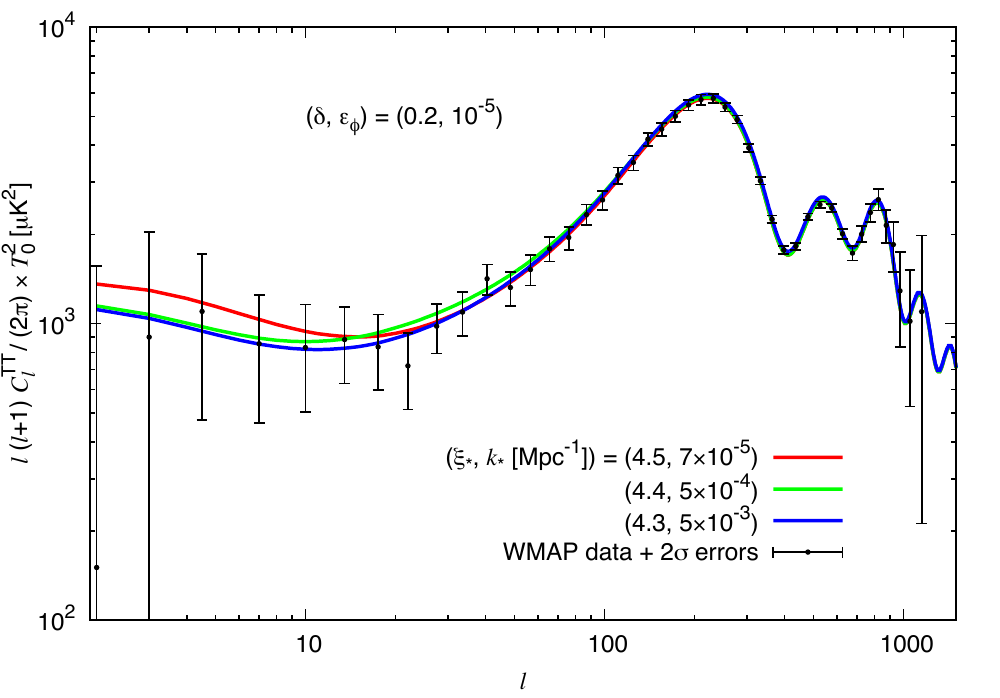}
\includegraphics[width=0.45\textwidth,angle=0]{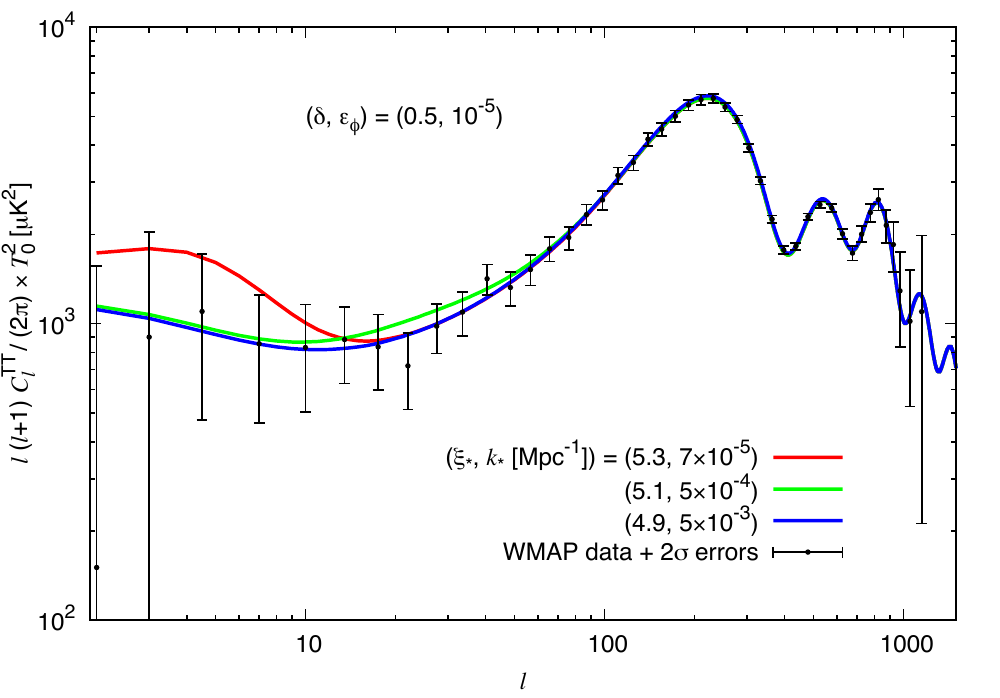}
\hfill
\caption{Temperature-temperature CMB power spectrum. The final WMAP data  \cite{Bennett:2012zja, Hinshaw:2012aka} are compared against the theoretical curves, evaluated for $\epsilon_\phi = 10^{-5}$ and for $\delta = 0.2$ (left panel) and $\delta = 0.5$ (right panel). In each panel we show the theoretical curves for three different values of $k_*$ (corresponding to three different locations of the peak of the sourced signal) and for the limiting value $\xi_* = \xi_{*,{\rm limit}}$, obtained as explained in the text. 
}
\label{fig:TT2}
\end{figure}

In Figure \ref{fig:TT2} we show the TT power spectrum obtained for the same choices of $\delta $ and $k_*$ as in Figure  \ref{fig:TT1}, for $\epsilon_\phi = 10^{-5}$, and for the corresponding value of $\xi_* = \xi_{*,{\rm limit}}$. The theoretical curves present a bump due to the sourced scalar modes. As we already mentioned, the bump ranges from the lowest $\ell$ multipoles (for the smallest $k_*$ chosen) to $\simeq$ the first two acoustic peaks (for the largest $k_*$ chosen). 

\begin{figure}
  \centering{ 
\includegraphics[width=.45\textwidth,clip]{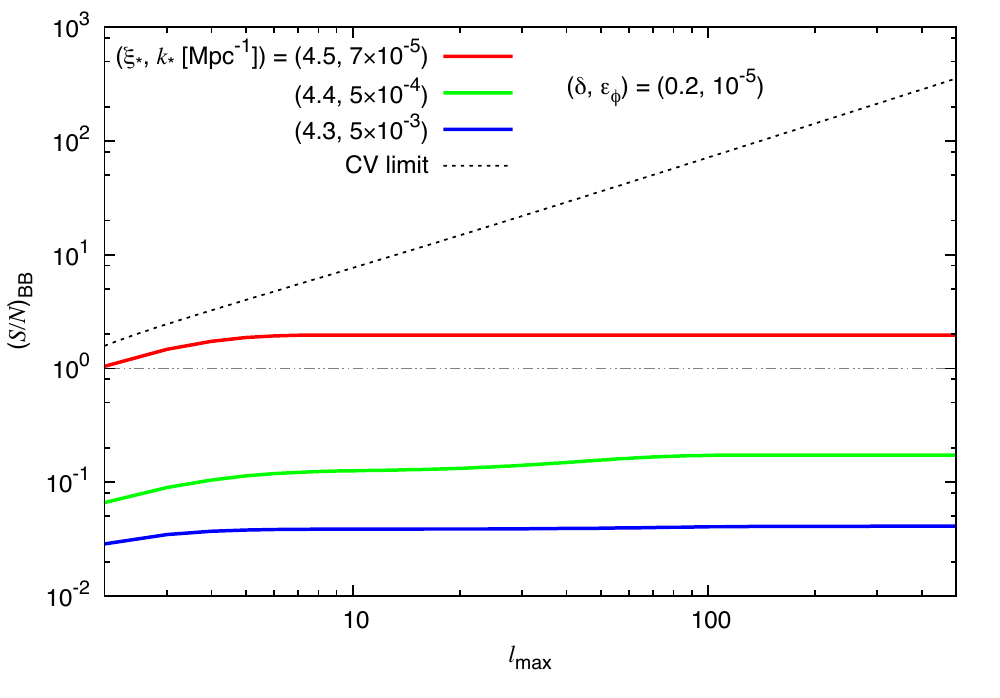}
\includegraphics[width=.45\textwidth,clip]{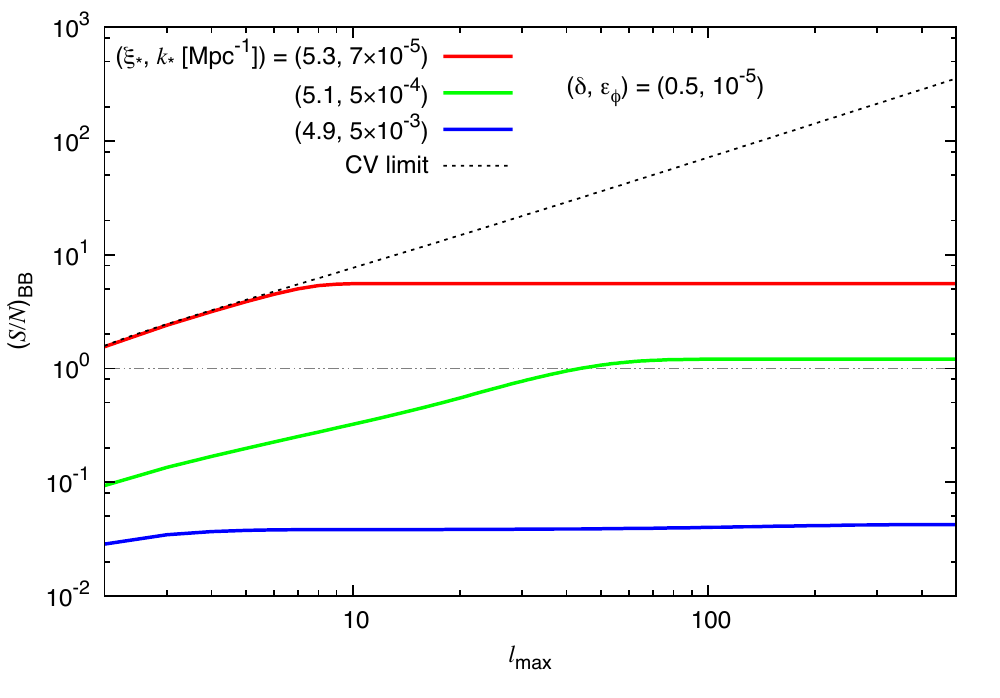}
}
\centering{ 
\includegraphics[width=.45\textwidth,clip]{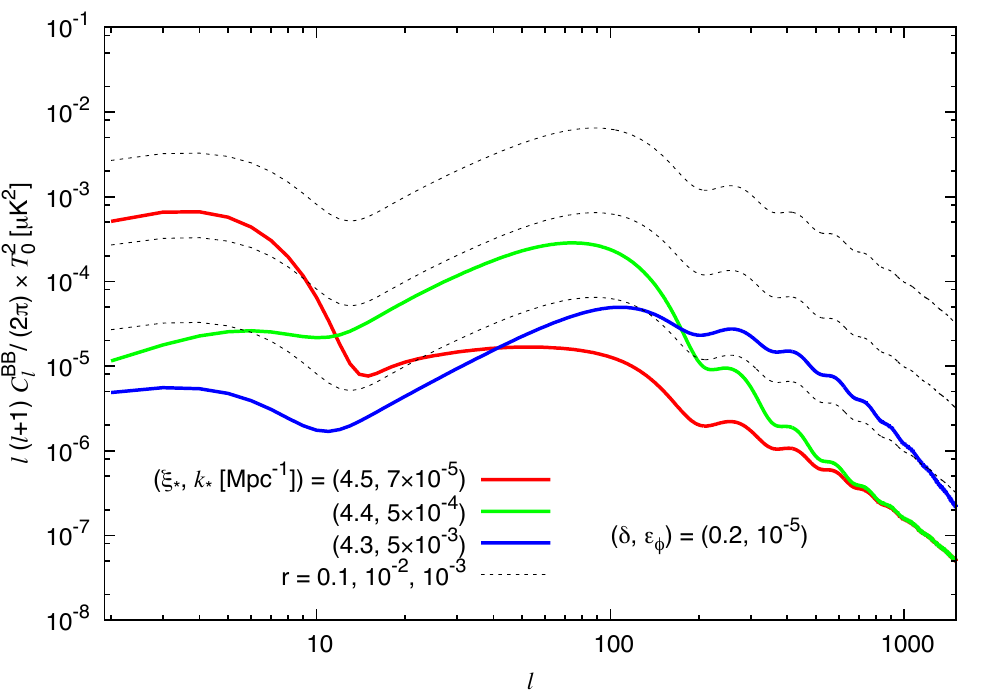}
\includegraphics[width=.45\textwidth,clip]{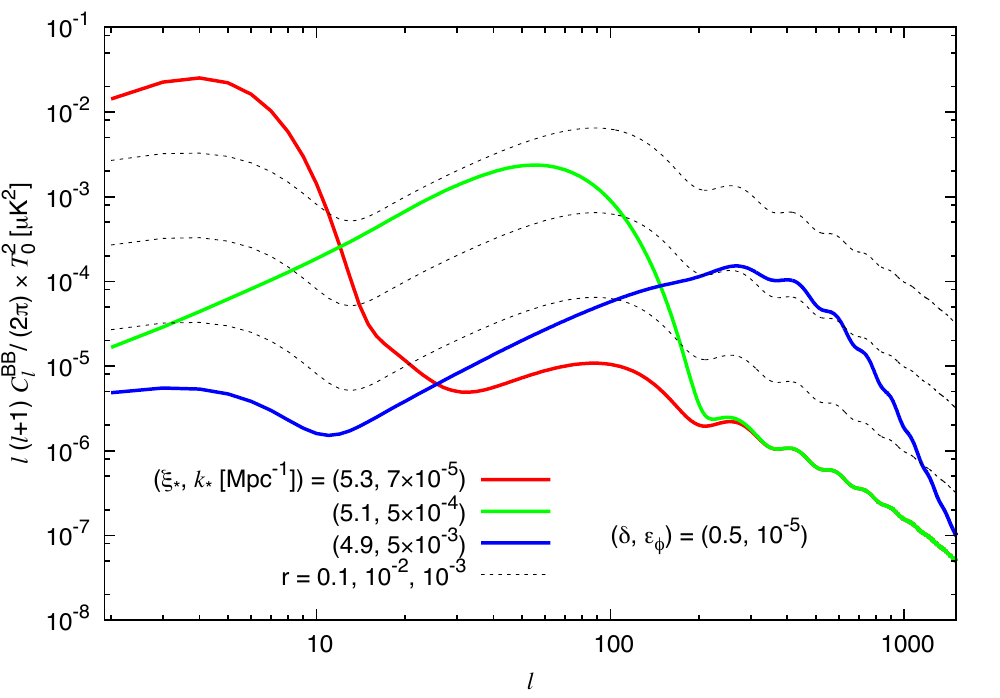}
}
\caption{First row: Forecasted signal-to-noise ratio for the detection of the B-mode auto power spectrum in a realistic CMB experiment with  {\it Planck}-like sensitivity (Colored solid lines), and in a cosmic variance limited (namely ideal noise-free) CMB experiment (Black dotted line), as a function of the maximum multipole $\ell$ included in the analysis. Second row: $C_l^{BB}$ coefficients. The model parameters are chosen as in the previous figure. 
}
\label{fig:BB}
\end{figure}

In Figure \ref{fig:BB} we show the self-correlations between the B modes of the CMB polarizations (BB) sourced by the tensor modes when $\xi_* = \xi_{*,{\rm limit}}$. The parameters $\epsilon_\phi ,\, \delta ,\,$ and $k_*$ are chosen as in the two previous figures. For comparison we also show with black dashed lines the BB correlation obtained for a scale invariant $r$ of $0.1 ,\, 10^{-2} ,\,$ and $10^{-3}$ (from top to bottom, respectively). As we already mentioned in the Introduction,  the proposed stage $4$ CMB experiments claim an expected statistical uncertainty  $\sigma \left( r \right) = 10^{-3}$ or below \cite{Abazajian:2013vfg} in the scale invariant case. If this is achieved, the theoretical curves  chosen in Figure \ref{fig:BB} appear to be within observational reach. We recall that  $\epsilon_\phi= 10^{-5}$ corresponds to a vacuum tensor-to-scalar ratio $r_{\rm vac} \simeq 1.6 \cdot 10^{-4}$. Therefore, the enhancement of the BB signal visible in the figure is entirely due to the sourced tensor modes. The enhancement is present at progressively larger $\ell$ for increasing values of $k_*$ shown (namely for bumps of gauge field production at progressively smaller scales). By comparing the left and the right panel of Figure \ref{fig:BB} we observe that BB can reach greater values at increasing $\delta$. This is consistent with what we have already mentioned: at fixed $\xi_*$, both the sourced scalar and tensor modes decrease with increasing $\delta$. However, the scalar mode decreases more. Therefore, at larger values of $\delta$, larger values of $\xi_*$ can be compatible with the WMAP TT bounds (cf. Figure \ref{fig:TT1}). Such values lead to a larger amount of sourced tensor modes. 

The signal-to-noise (S/N) ratio shown in the figure is evaluated through 
\begin{eqnarray}
  \left( \frac{S}{N} \right)_{BB}^2
  = \sum_{\ell = 2}^{\ell_{\rm max}} \frac{2\ell + 1}{2} \left( \frac{C_\ell^{BB}}{C_{\ell, \rm dat}^{BB}} \right)^2 ~.
\end{eqnarray}
Here, $(C_\ell^{BB})^2$ corresponds to the signal given by our theory, while the other terms in this relation account for the uncertainty of the BB power spectrum in a given experiment. For simplicity, we here (and also in the other S/N estimations) assume a full-sky isotropic CMB measurement, thus, the summations in terms of $m$ disappear in the S/N formula. The data spectrum in a given experiment is regarded as the sum of the signal and instrumental noise, reading $C_{\ell, \rm dat}^{BB} = C_\ell^{BB} + N_\ell^{BB}$. In the paper, we analyze  two different types of  measurements: a realistic measurement including a {\it Planck}-level noise spectrum \cite{Planck:2006aa} (as described in Appendix A of \cite{Shiraishi:2013vha}) and an ideal noise-free cosmic variance dominated measurement (i.e., $N_\ell^{BB} = 0$).

 The results are shown in the upper panels of Figure~\ref{fig:BB}. In the cosmic variance limited case, because of $C_{\ell}^{BB} / C_{\ell, \rm dat}^{BB} = 1$, S/N becomes a simple increasing function: $(S/N)_{BB} = \sqrt{(\ell_{\rm max} + 3)(\ell_{\rm max} - 1)/2}$ (black dotted lines), independently of the shape of $C_{\ell}^{BB}$ and the values of input model parameters. The S/N is lower in 
 the {\it Planck}-like realistic experiment; however it can exceed one in the examples with the smallest values of $k_*$ shown. A greater BB signal can be obtained at larger values of $\delta$.

\begin{figure}
\centering{ 
\includegraphics[width=.45\textwidth,clip]{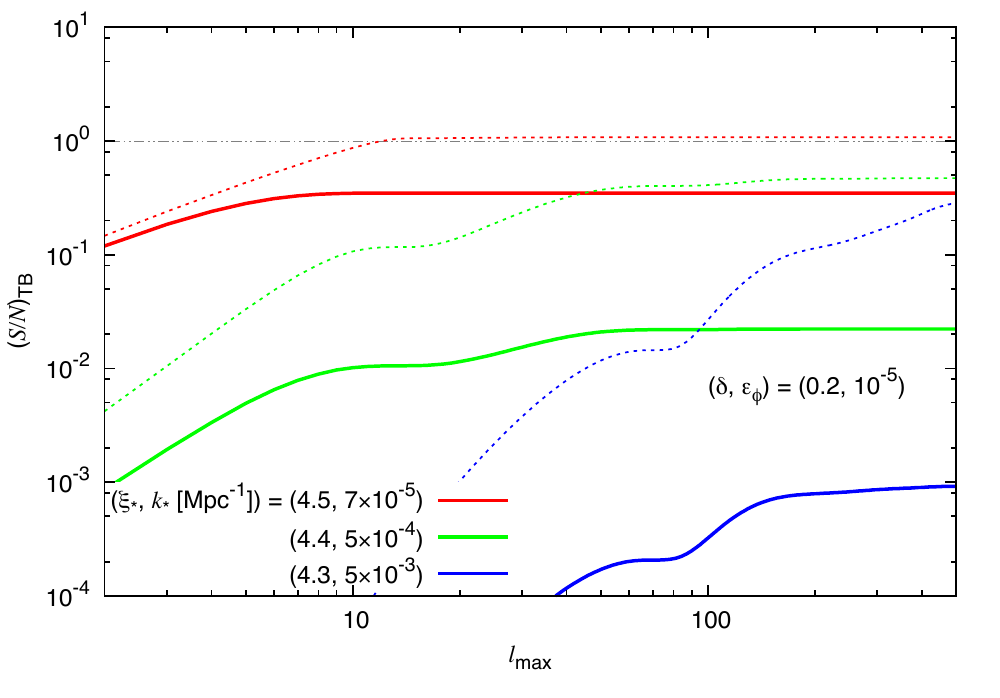}
\includegraphics[width=.45\textwidth,clip]{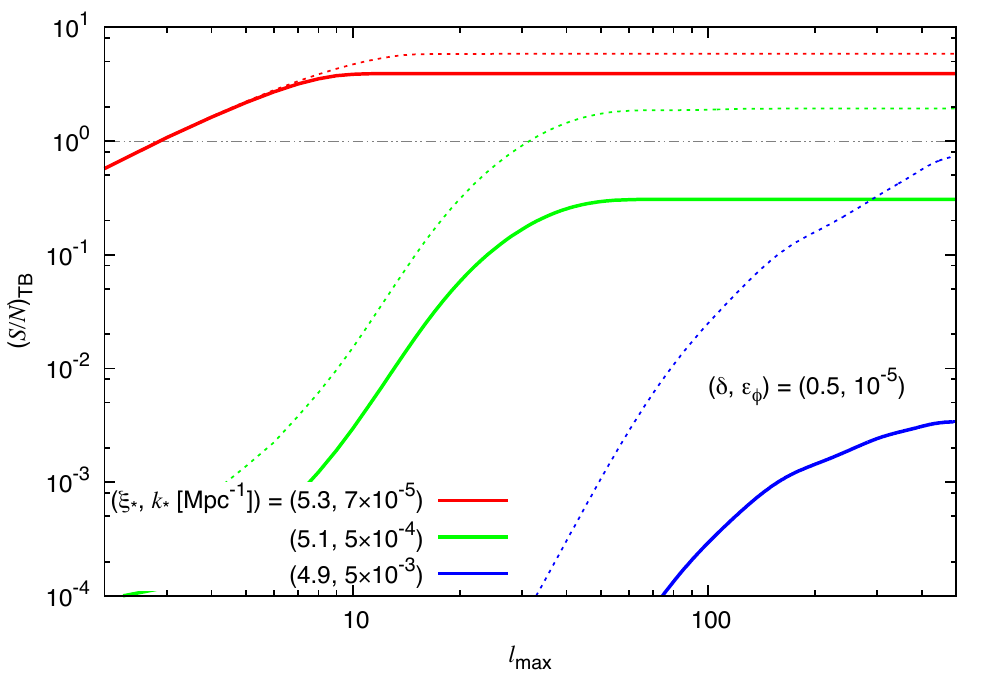}
}
\centering{ 
\includegraphics[width=.45\textwidth,clip]{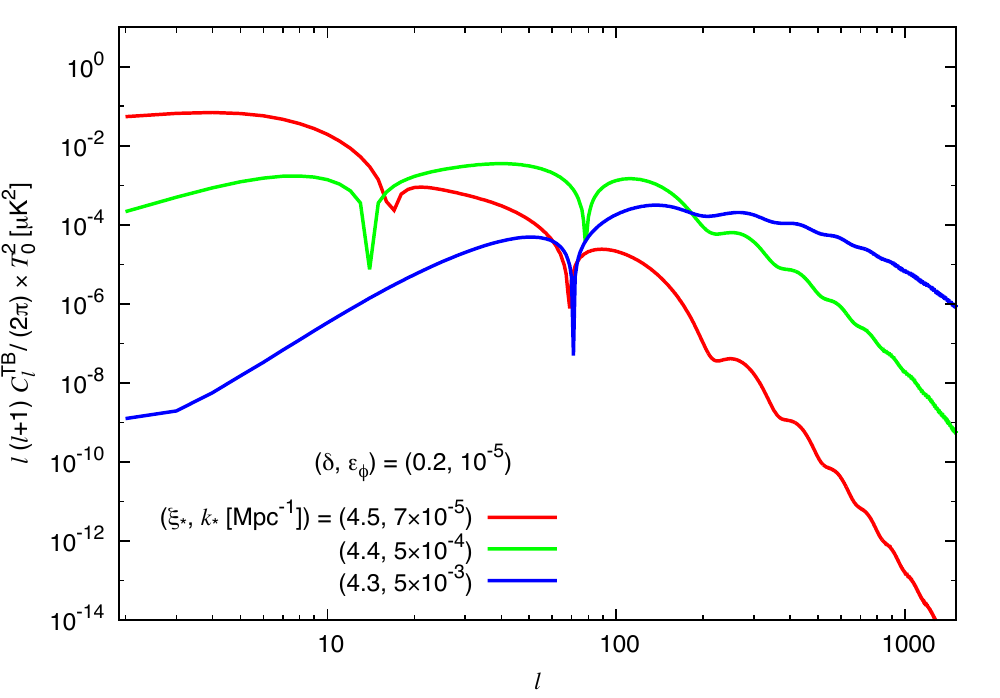}
\includegraphics[width=.45\textwidth,clip]{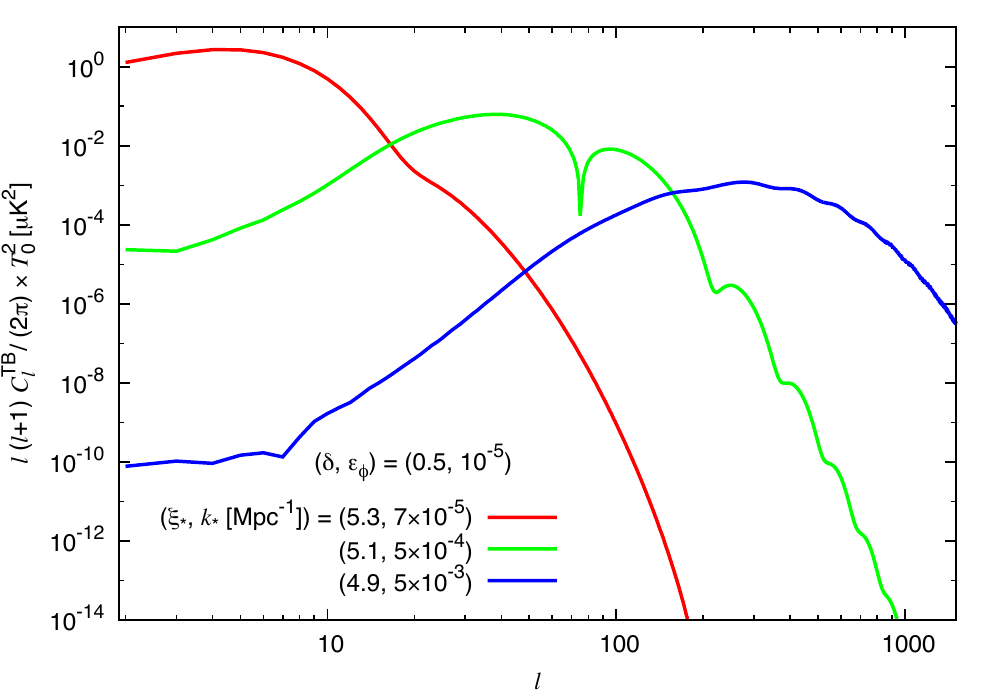}
}
\caption{
  First row: Forecasted signal-to-noise ratio for the detection of the TB correlation in a realistic CMB experiment with  {\it Planck}-like sensitivity (Colored solid lines), and in a cosmic variance limited, CMB experiment (Colored dotted lines), as a function of the maximum multipole $\ell$ included in the analysis. Second row: TB spectra. The parameters are chosen as in the  previous  figures. 
}
\label{fig:TB-SN}
\end{figure}

The sourced GW signal breaks parity, generating a nonvanishing correlation between the CMB temperature anisotropy and B-mode polarization (TB) \cite{Lue:1998mq,Saito:2007kt,Gluscevic:2010vv,Shiraishi:2013kxa,Ferte:2014gja}. In  Figure  \ref{fig:TB-SN} we compute the forecasted signal-to-noise ratio for the detection of such signal:
\begin{eqnarray}
  \left( \frac{S}{N} \right)_{TB}^2
  = \sum_{\ell = 2}^{\ell_{\rm max}} (2\ell + 1)
  \frac{\left(C_\ell^{TB}\right)^2}{C_{\ell}^{TT} C_{\ell, \rm dat}^{BB}} ~, \label{eq:SN_TB}
\end{eqnarray}
in the {\it Planck}-like realistic experiment and the ideal cosmic variance-limited experiment. In the {\it Planck}-like measurement, we can neglect the noise spectrum of temperature mode $N_\ell^{TT}$, since this is negligibly small compared with the signal $C_\ell^{TT}$ on our interesting scales ($\ell \leq 500$). We see that, among the examples shown, such a signature can be marginally detected only for the examples shown at $\delta = 0.5$ in the cosmic-variance-limited experiment. Not surprisingly, these are the cases that also lead to a stronger BB signal, cf. Figure \ref{fig:BB}. Among the examples we considered, these are the only cases for which $r_{\rm peak}$ (namely the tensor-to-scalar ratio evaluated at the top of the GW signal) is substantially greater than $10^{-2}$. Based on the difference between the $\delta = 0.2$ and $\delta =0.5$ cases, it is very likely that a large TB correlation can be produced at greater values of $\delta$, possibly at a detectable level even in a  {\it Planck}-like experiment. As a comparison, we note that  \cite{Gluscevic:2010vv} forecasted that a one $\sigma$ detection is possible for  $r \ga 0.002$ in the scale invariant case, and a few $\sigma$ detection for the largest values of $r\sim .05$ that can be achieved in our scenario. The results of~\cite{Gluscevic:2010vv} are slightly more optimistic than ours because of the scale invariance of the correlator considered in that paper, that allows to use information from a larger number of multipoles.

\subsubsection{CMB bispectra}

\begin{figure}
\centering{ 
\includegraphics[width=.45\textwidth,clip]{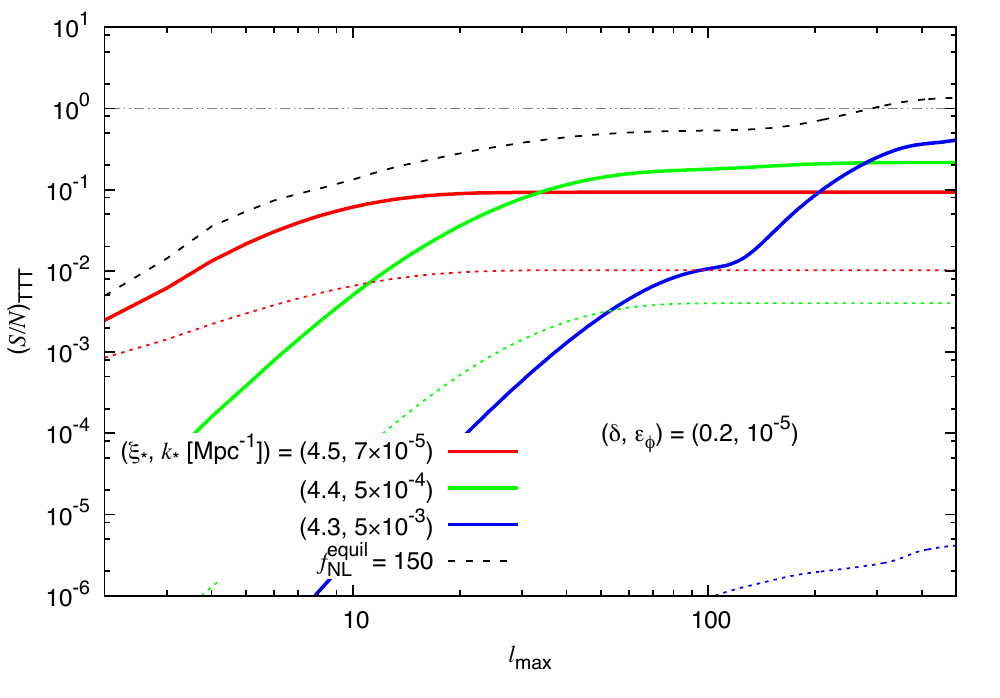}
\includegraphics[width=.45\textwidth,clip]{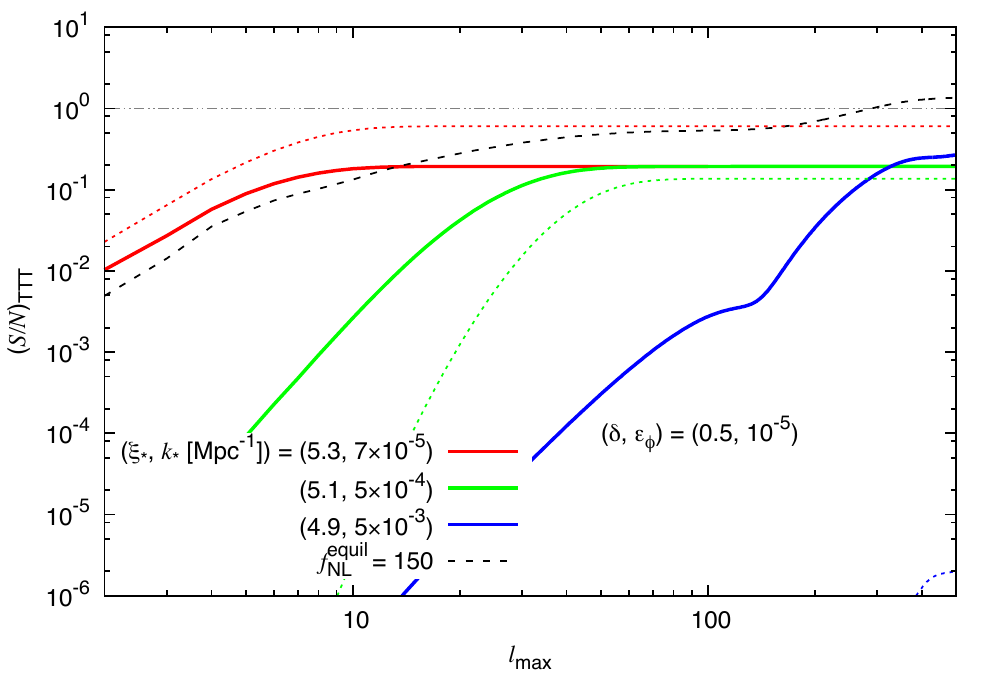}
}
\centering{ 
\includegraphics[width=.45\textwidth,clip]{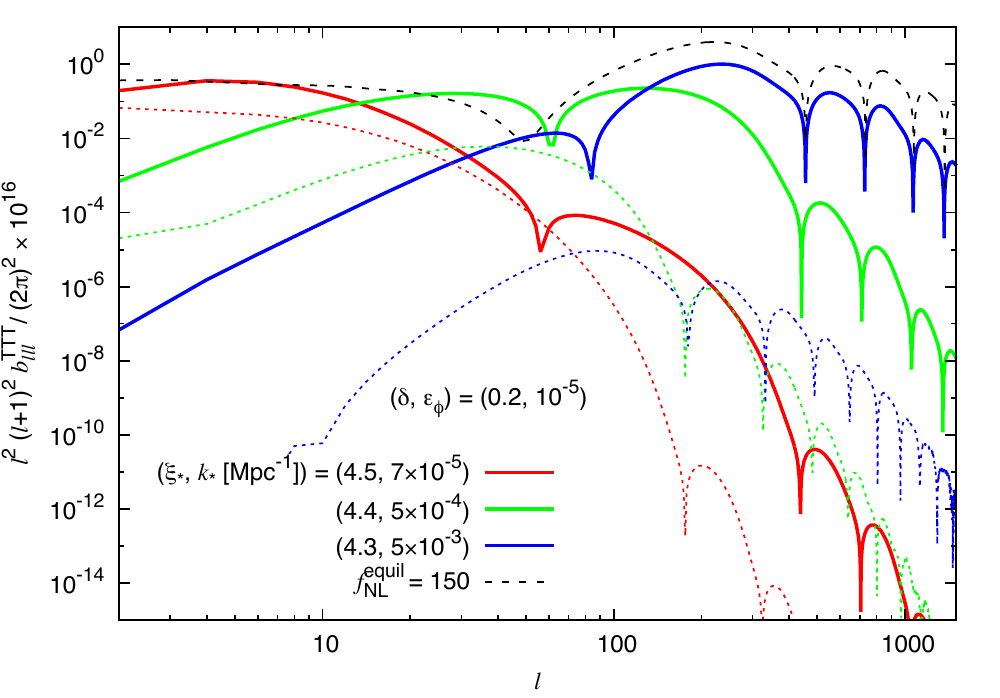}
\includegraphics[width=.45\textwidth,clip]{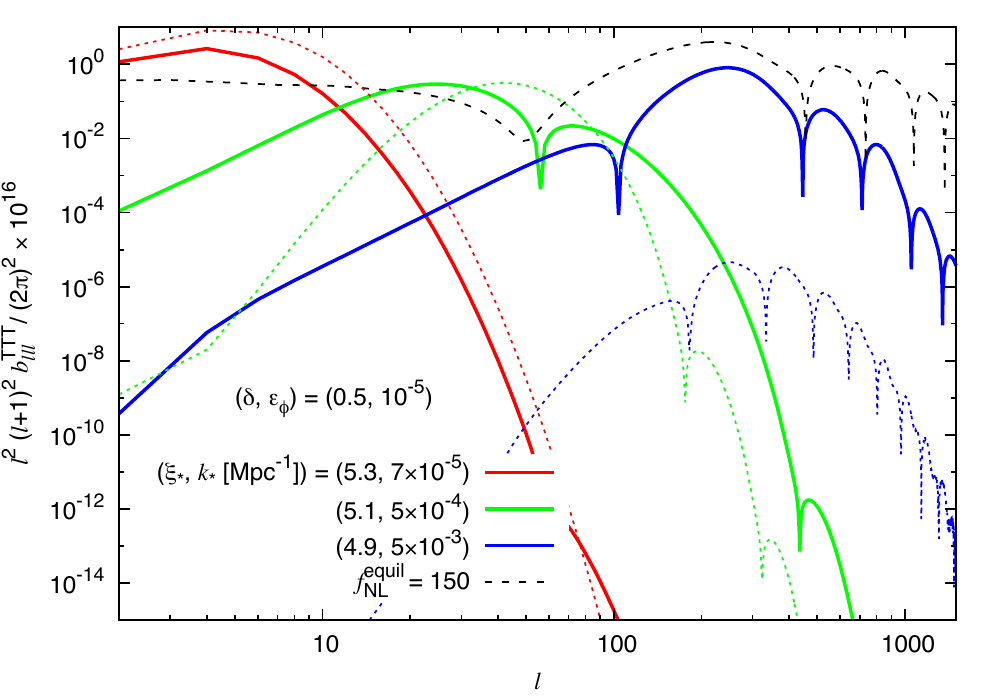}
}
\caption{First row: Forecasted signal-to-noise ratio for the detection of the sourced TTT bispectrum in a cosmic variance-limited, CMB experiment, as a function of the maximum multipole $\ell$ included in the analysis. Second row: TTT reduced bispectrum for equilateral triangles of multipole $\ell$. The colored lines are for the same parameters as the previous  figures. Solid (dotted) lines refer to the contribution to TTT from the sourced scalar (tensor) modes. For comparison, the black dashed line indicates the forecasted S/N ratio for scale-invariant equilateral non-gaussianity of magnitude $f_{\rm NL}^{\rm equil} = 150$ (a rough indication of the level that can be reached by {\it Planck} in the scale-invariant case \cite{Ade:2015ava}). 
}
\label{fig:TTT-SN}
\end{figure}

  Let us now study the possibility of detecting the non-gaussian statistics of the sourced modes. All the CMB temperature and polarization bispectra due to the scalar and tensor non-gaussianities, discussed in the following analysis, are computed by means of the techniques outlined in \cite{Shiraishi:2010sm,Shiraishi:2010kd}.

In Figure \ref{fig:TTT-SN} we show the forecasted signal-to-noise ratio for the detection of the sourced TTT bispectrum as a function of the maximum multipole $\ell$ included in the analysis, reading
\begin{eqnarray}
  \left( \frac{S}{N} \right)_{TTT}^2
  = \sum_{\ell_1, \ell_2, \ell_3 = 2}^{\ell_{\rm max}}
  \frac{\left| B_{\ell_1 \ell_2 \ell_3}^{TTT} \right|^2}{6 C_{\ell_1}^{TT} C_{\ell_2}^{TT} C_{\ell_3}^{TT}} ~,
\label{S/N-TTT}
\end{eqnarray}
with
\begin{eqnarray}
  B_{\ell_1 \ell_2 \ell_3} \equiv \sum_{m_1 m_2 m_3}
  \left(
\begin{array}{ccc}
  \ell_1 & \ell_2 & \ell_3 \\
  m_1 & m_2 & m_3 
  \end{array}
\right)
\langle a_{\ell_1 m_1} a_{\ell_2 m_2} a_{\ell_3 m_3} \rangle
\end{eqnarray}
denoting the angle-averaged bispectrum. Here we show only the results in a cosmic variance-limited CMB experiment. The same results will be obtained also in the {\it Planck}-like experiment, since the instrumental noise is perfectly negligible for $\ell \leq 500$. All the lines shown are for the maximum amount of particle production allowed by the WMAP TT data (namely, for $\xi_* = \xi_{*,\, {\rm limit}}$ shown in Figure \ref{fig:TT1}). 

Given that the sourced signal has a much greater deviation from gaussianity that the vacuum one   \cite{Barnaby:2010vf}, and given the stringent limit on non-gaussianity, one may have expected to find detectable non-gaussianity in the examples shown. Figure  \ref{fig:TTT-SN} shows that this is not the case. The main reason is that, for the values of  $k_*$ that we have chosen, the sourced signal manifests itself only at the largest CMB scales, so that only those scales are relevant for the phenomenological study of the bispectrum. This is confirmed by the colored lines shown in the Figure, where the signal-to-noise ratio saturates at values of $\ell_{\rm max}$ well below those included in the {\it Planck} studies (in particular, the smaller $k_*$ is, the smaller is the value of $\ell_{\rm max}$ at which the signal-to-noise ratio saturates). This contrasts with the scale invariant case shown by the dashed line, where we can see that S/N continues to grow with increasing $\ell_{\rm max}$.  The strong weakening of the non-gaussianity limits with decreasing values of $\ell_{\rm max}$ can be also seen for example in Figure 11 of \cite{Ade:2015ava}. 

In the analysis,   we distinguish between the scalar $\left\langle \zeta^{(1) 3} \right\rangle$ and  tensor $\left\langle h_+^{(1) 3} \right\rangle$ contribution to the TTT signal. In the tensor case, both even ($\ell_1 + \ell_2 + \ell_3 = {\rm even}$) and odd  ($\ell_1 + \ell_2 + \ell_3 = {\rm odd}$) are included in the analysis \cite{Cook:2013xea,Shiraishi:2013kxa}.  We can also see all this from the second row of Figure \ref{fig:TTT-SN}, which presents the TTT reduced bispectrum, defined as
\begin{eqnarray}
 b_{\ell_1 \ell_2 \ell_3} \equiv  \left[ \sqrt{\frac{(2\ell_1 + 1)(2\ell_2 + 1)(2\ell_3 + 1)}{4\pi}}  \left(
\begin{array}{ccc}
  \ell_1 & \ell_2 & \ell_3 \\
  0 & 0 & 0
  \end{array}
\right) \right]^{-1} B_{\ell_1 \ell_2 \ell_3} ~,
\end{eqnarray}
on equilateral triangles as a function of the multipole $\ell$. We observe that, contrary to the scale invariant case, the bispectrum significantly decreases at the higher multipoles shown in the figure (also in this case, the smaller $k_*$ is, the sooner in $\ell$ space the bispectrum starts decreasing). Also in the second row we distinguish between the contribution of the scalar and the tensor mode.  While in the examples with $\delta=0.2$ the scalar contribution dominates over the tensor one, the tensor mode plays a non-negligible or even dominant role in two of the examples shown   at $\delta = 0.5$. 

\begin{figure}
\centering{ 
\includegraphics[width=.45\textwidth,clip]{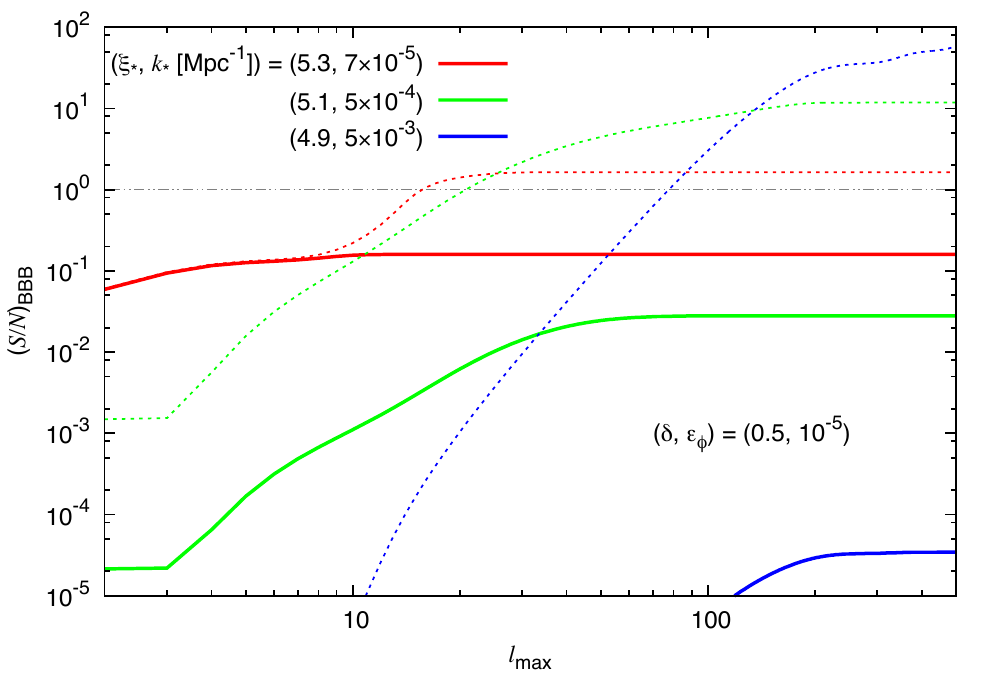}
\includegraphics[width=.45\textwidth,clip]{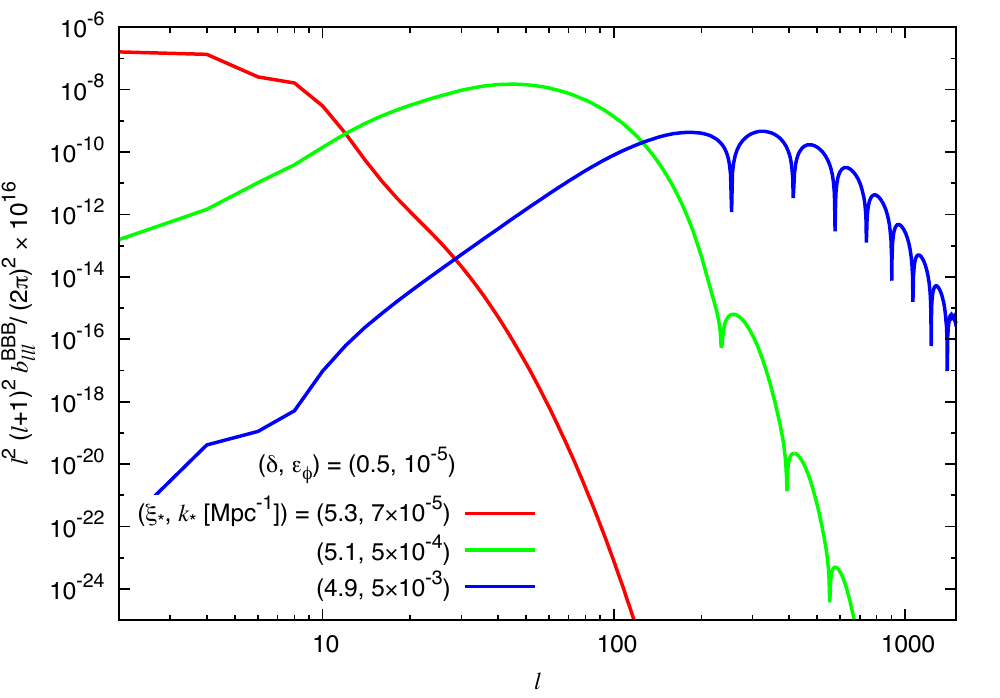}
}
\caption{Left panel: Forecasted signal-to-noise ratio for the detection of the sourced BBB bispectrum in a realistic CMB experiment with  {\it Planck}-like sensitivity (Colored solid lines), and in a cosmic variance limited, CMB experiment (Colored dotted lines), as a function of the maximum multipole $\ell$ included in the analysis. Right panel: BBB reduced bispectrum for equilateral triangles of multipole $\ell$. The examples shown are for $\delta = 0.5$; the other  parameters are chosen as in  the previous  figures.  
}
\label{fig:BBB-SN}
\end{figure}

Another interesting non-gaussian signature of the sourced tensor mode is the BBB correlation. The signal-to-noise ratio is computed as
\begin{eqnarray}
  \left( \frac{S}{N} \right)_{BBB}^2
  = \sum_{\ell_1, \ell_2, \ell_3 = 2}^{\ell_{\rm max}}
  \frac{\left| B_{\ell_1 \ell_2 \ell_3}^{BBB} \right|^2}{6 C_{\ell_1, \rm dat}^{BB} C_{\ell_2, \rm dat}^{BB} C_{\ell_3, \rm dat}^{BB}} ~.
\label{S/N-BBB}
\end{eqnarray}
Like the BB case, we here analyze the detectability both in a {\it Planck}-like experiment and in a cosmic-variance-limited experiment. 
In Figure \ref{fig:BBB-SN} we show that this signal has a much higher prospect of being detected than the TB one in the cosmic-variance-limited measurement. Such large S/N seen in the figure is due to the fact that no instrumental error is assumed, so that all the noise is due to the small BB signal (instead, the largest TT signal due to the vacuum scalar modes constitutes a stronger noise for the TB measurement in the cosmic variance-dominated case). 

We also computed the B-mode bispectrum for $\delta = 0.2$, obtaining a larger S/N ratio than in the $\delta =0.5$ case shown in  Figure \ref{fig:BBB-SN}. We do not show this result, as we are concerned about its reliability:  the expression for the noise in 
(\ref{S/N-TTT}) and (\ref{S/N-BBB}) assumes that the signal is approximately gaussian. We investigate whether this is the case in 
Appendix \ref{app:NG}, where we estimate the departure from gaussianity of the full $\zeta$ and $h_+$ signals. Our estimate suggests that the departure is very small in the scalar case, and marginally small for  $h_+$ at  $\delta = 0.5$. However, this is not the case  for $\delta = 0.2$. The result performed there suggests that the noise in this case may increase by a ${\rm O } \left( 1 \right)$ factor with respect to what we computed using  (\ref{S/N-BBB}). This still likely results in a potentially visible signal in a  cosmic-variance-limited experiment, but, due to this uncertainty, we do not show this result here. 

As in the TTT case, the S/N shown in the Figure \ref{fig:BBB-SN} is computed by summing all the signals in $\ell_1 + \ell_2 + \ell_3 = {\rm even}$ and $\ell_1 + \ell_2 + \ell_3 = {\rm odd}$. The sum enhances the S/N. On the other hand, the analysis in the limited domain, $\ell_1 + \ell_2 + \ell_3 = {\rm odd}$ in TTT or $\ell_1 + \ell_2 + \ell_3 = {\rm even}$ in BBB, is also very informative, since such signals originate from nonvanishing parity-odd non-gaussianity \cite{Kamionkowski:2010rb,Shiraishi:2011st}, namely a distinctive signature of this model.~\footnote{See \cite{Maldacena:2011nz,Soda:2011am,Shiraishi:2011st,Shiraishi:2012sn,Cook:2013xea,Shiraishi:2013kxa} for the other possibilities to generate parity-violating CMB bispectra. See also \cite{Shiraishi:2014roa,Shiraishi:2014ila,Ade:2015ava} for the WMAP and {\it Planck} constraints.} More comprehensive analysis with the other auto- and cross-bispectra including the E-mode polarization, not discussed above, may also  help to improve the detectability \cite{Babich:2004yc,Yadav:2007ny,Yadav:2007rk,Shiraishi:2013vha,Shiraishi:2013kxa,Ade:2015ava}.

\section{Conclusions}
\label{sec:conclusions}

Given the expected improvements in sensitivity of the measurements aiming at the detection of primordial tensor modes, it is crucial to determine whether there are viable alternatives to the standard mechanism of creation of GW through the amplification of vacuum fluctuations.

In the present paper we have shown that, if an axion-like field that is only gravitationally coupled to the inflaton experiences a transient epoch of relatively fast roll, then tensor perturbations can receive an additional contribution whose amplitude is not proportional to the energy scale of inflation. This implies that, for a broad range of parameters, tensors of inflationary origin might be detectable even if inflation happens at a low scale, and even for subplanckian inflaton displacements, without contradicting the existing constraints. Remarkably, the simple potential~(\ref{V-sigma}) naturally satisfies our requirements. To leave an observable GW signal, the transient epoch must occur while the large scale CMB modes leave the horizon. 

The system we have considered comes with a rich set of signatures. 
First, the amplitude of the induced tensor modes can be strongly scale dependent. As a consequence significant $B$-mode CMB fluctuations might be detected only at scales related  to the reionization bump, or only at the B-peak at $\ell\sim 80$, but not at both angular scales. In particular, detection at $\ell\sim 80$ without a corresponding detection at $\ell\sim 10$ would be interpreted as a locally blue tilt (similarly to the situation considered in~\cite{Mukohyama:2014gba}) for the primordial tensor spectrum, that is usually considered as a signature that would falsify the inflationary scenario. 

Moreover, the parity-violating nature of the process generates a characteristic $\langle TB\rangle$ correlator, that, in an ideal experiment, might be marginally detectable if the amplitude of the induced tensors is large enough.

Finally, the statistics of the produced gravitational waves has a departure from gaussianity that is significantly more marked than for the vacuum modes. We have shown some choices of parameters for which a cosmic variance limited experiment could detect a nonvanishing $\langle BBB\rangle$ correlator at a very high level of significance. Moreover, we have computed the phenomenology of the model only for a very few choices of parameters. As we mentioned in the previous section, a larger tensor mode can be produced for increasing values of the axion mass (greater $\delta$), and it is likely that this can lead to larger TB and BBB correlators, which may be observable (particularly in the BBB case) already in a  {\it Planck}-like experiment.  A more comprehensive analysis with the other possible auto- and cross-bispectra including the E-mode polarization may further enhance a signal-to-noise ratio. We believe that our result motivates a dedicated study of such localized signal in the {\it Planck} data. 

For the examples we have studied the effect on the temperature  correlators is statistically small. This is the case  because {\em (i)} due to helicity conservation, the amplitude of the sourced scalar perturbations is smaller than that of the tensors, and {\em (ii)} all relevant effects take place at multipoles with $\ell\lesssim {\cal O}(10^2)$, where cosmic variance is large. We have thus explicitly verified that the model studied here is one of the very few existing examples for which a sourced GW signal is produced, while respecting the strong limits imposed from the observed T  modes of the CMB. 

On the other hand, different choice of parameters can be made in the same model that can result in a stronger scalar mode production than a tensor mode. This can easily be the case if the field $\sigma$ rolls for a considerable amount to e-folds during inflation. This would generate a bump in the scalar modes with significant departure from gaussianity 
and with a greater width than in the examples considered here. By tuning the moment at which the roll of $\sigma$ takes place, the bump can manifest itself either at CMB or LSS scales.

Remarkably, all the above features originate from the single operator appearing as the last term of the lagrangian~(\ref{lagr}), that is naturally expected to occur in theories with pseudo-scalar fields. It would be interesting to perform a more comprehensive study of the scenarios involving just Einstein's Relativity and a local quantum field theory that can lead to comparable phenomenology.

The next few years will witness major improvements in the observations aiming at the detection of primordial B-modes. It is important that all potential signatures are considered when the new data will arrive. 

\vskip.25cm
\noindent{\bf Acknowledgements:} 

We thank N. Bartolo, C. Caprini, M. Kleban, and M. Liguori for useful discussions. L.S. thanks the Institut the Physique Th\'eorique of the CEA, Saclay, for hospitality during the completion of this work. The work of M.P. is partially supported from the DOE grant DE-SC0011842  at the University of Minnesota. The work of L.S. is partially supported by the NSF grants PHY-1205986 and PHY-1520292. The work of M.S. is supported in part by a Grant-in-Aid for JSPS Research under Grant No. 27-10917. R.N. and M.S. are supported in part by World Premier International Research Center Initiative (WPI Initiative), MEXT, Japan.

\appendix 

\section{The expression of $A_+(\tau,k)$ in WKB approximation}
\label{app:WKB}

In this appendix we derive the expression~(\ref{approxaplus}) for the mode function of the photon. We start from the equation of motion~(\ref{eom-Apm-app}) for the positive helicity photon, that, after  defining $x=-k\tau$, $x_*=-k\tau_*$, reads
\begin{equation}
\frac{d^2A_+}{dx^2}+\left(1-\,\frac{2}{x}\,\frac{2\,\xi_*}{(x/x_*)^\delta+(x/x_*)^{-\delta}}\right)\,A_+=0\,,
\end{equation}
and that we solve using the WKB approximation. To do so, we write it in the general form
\begin{equation}\label{geneschro}
A_+''(x)+Q(x)\,A_+(x)=0\,,
\end{equation}
where $Q(x)$ vanishes for $x=\bar{x}$ and is positive for $x>\bar{x}$, so that we can define
\begin{align}
Q(x)\equiv\left\{
\begin{array}{cl}
p(x)^2 & ,\, x>\bar{x}\\
-\kappa(x)^2 & ,\, x<\bar{x}\,,
\end{array}
\right.
\end{align}
Then, for $x>\bar{x}$, in the adiabatic regime $|p'(x)|\ll p(x)^2$, the general solution reads
\begin{equation}
A_+({x>\bar{x}}) \cong \frac{\alpha}{\sqrt{p(x)}} \,
               \cos \left( \int_{\bar{x}}^{x} p(x) dx - \frac{\pi}{4} \right) -
               \frac{\beta}{\sqrt{p(x)}} \,
               \sin \left( \int_{\bar{x}}^{x} p(x) dx - \frac{\pi}{4} \right) \;, 
\end{equation}
with $\alpha$ and $\beta$ constants, and where the symbol $\cong$ denotes approximate equality in the WKB approximation. Similarly, for $x<\bar{x}$, the solution in the WKB regime  $|\kappa'(x)|\ll \kappa(x)^2$ will be a linear combination of $\exp \left( -\int_{x}^{\bar{x}} \kappa(x) dx \right)/\sqrt{\kappa (x)}$ and $\exp \left( \int_{x}^{\bar{x}} \kappa(x) dx \right)/\sqrt{\kappa (x)}$. To determine the coefficients of such a combination in terms of $\alpha$ and $\beta$ we expand eq.~(\ref{geneschro}) near $x=\bar{x}$ as 
\begin{equation}
A_+''(x)+Q'(\bar{x})\,(x-\bar{x})\,A_+(x)\simeq 0 \,, 
\end{equation}
that can be solved in terms of Airy functions, and that we join to the adiabatic solutions in the regimes $x-\bar{x}\to\pm\infty$, as discussed for instance in chapter 7 of~\cite{merzbacher}. The final result is
\begin{equation}
A_+ (x<\bar{x}) \cong \frac{\alpha/2}{\sqrt{\kappa (x)}} \,
               \exp \left( -\int_{x}^{\bar{x}} \kappa(x)\, dx \right) +
               \frac{\beta   }{\sqrt{\kappa (x)}} \,
               \exp \left( \int_{x}^{\bar{x}} \kappa(x)\, dx \right) \;.
\end{equation}

The requirement that the photons are in their adiabatic vacuum in the ultraviolet translates into the boundary condition $\alpha=\frac{1}{\sqrt{2\,k}}$, $\beta=-\frac{i}{\sqrt{2\,k}}$. In the regime $x\ll \bar{x}$ we keep only the term that corresponds to the growing mode, so that 
\begin{equation}\label{generalwkb}
A_+(x<\bar{x})\simeq -\frac{i}{\sqrt{2\,k\,\kappa(x)}}\,\exp\left\{\int_{x}^{\bar{x}} \kappa(x) dx \right\}\,.
\end{equation}

As a check of the validity of our procedure we can consider the limit $\delta\to 0$, so that $\frac{2\,\xi_*}{x^\delta+x^{-\delta}}\to \xi$, which brings us back to the case of a constant velocity for the rolling axion first studied in~\cite{Anber:2006xt}. Indeed, in this regime
\begin{equation}
\left.\int_{x}^{\bar{x}} \kappa(x)\, dx\right|_{\delta\to 0}=\int_x^{2\,\xi}\sqrt{\frac{2\,\xi}{x}-1}\,dx=2\,\xi\,\arccos(\sqrt{\frac{x}{2\xi}})-\sqrt{x\,(2\,\xi-x)}\simeq \pi\xi-2\sqrt{2\xi\,x}\,,
\end{equation}
(where in the last step we have expanded to first order in $\sqrt{x}$), that, once inserted into eq.~(\ref{generalwkb}), gives a result that coincides with the approximate expression given in~\cite{Anber:2006xt}.

In the general case $\delta\neq 0$ it is impossible to compute the integral of $\kappa(x)$ in closed form. Nevertheless, we can write
\begin{equation}
\int_x^{\bar{x}}\kappa(x)\,dx=\int_0^{\bar{x}}\kappa(x)\,dx-\int_0^x\kappa(x)\,dx\,,
\end{equation}
so that only the second term depends on $x$. Since we care only about the regime of moderately small $x$, we consider only the leading part of the integrand as $x\to 0$:
\begin{equation}\label{xdependence}
\int_0^x\kappa(x)\,dx\simeq \int_0^x\sqrt{\frac{4\,\xi_*}{x\,(x/x_*)^{-\delta}}}\,dx=\frac{4\,\sqrt{\xi_*}}{1+\delta}\,x^{\frac{1+\delta}{2}}\,x_*^{-\frac{\delta}{2}}\,.
\end{equation}
This equation, once inserted into eq.~(\ref{generalwkb}), gives the full dependence of the function $A_+(\tau,\,k)$ on the conformal time $\tau$ as it appears in eq.~(\ref{defatilde}) in  the main text. The overall normalization of $A_+(\tau,\,k)$, on the other hand, has to be computed numerically, and we determine it by matching the WKB approximate solution with the numerical solution of eq.~(\ref{eom-Apm-app}), yielding the  constant $N(\xi_*,x_*,\delta)$ defined in eq.~(\ref{defatilde}). We have checked that, in the part of parameter space we are interested in, the  expression found this way approximates the exact solution at the $30\%$ level or better. 

\section{Sourced scalar modes at constant $\xi$.}
\label{app:constant}

In this Appendix  we  show how, for constant values of $\xi$, the isocurvature perturbation associated to $\sigma$ is partially converted into curvature perturbation during superhorizon evolution~\cite{Ferreira:2014zia,Mirbabayi:2014jqa}. As a consequence, large fluctuations in the gauge mode can introduce a significant sourced component in the curvature perturbations. We start from the equations~(\ref{eom_Qi_exact}) already given in the main text, and we redefine 
\begin{equation}
M_{ij}^2 \equiv - \tau^2 {\tilde M}_{ij}^2 
\simeq  \left(
\begin{array}{cc}
2 + 9 \, \epsilon_\phi + 3 \, \epsilon_\sigma - 3 \, \eta_\phi &
6 \sqrt{\epsilon_\phi \epsilon_\sigma} \\
6 \sqrt{\epsilon_\phi \epsilon_\sigma} &
2 + 9 \, \epsilon_\sigma + 3 \, \epsilon_\phi - 3 \, \eta_\sigma
\end{array}
\right) \; ,
\label{Mij_rescaled}
\end{equation}
If $M^2_{ij}$ is approximately constant we can diagonalize the mass matrix as
\begin{align}\label{diago_mat}
M=C^T\,\Lambda\,C\,,\qquad C=\left(
\begin{array}{cc}
\cos\theta & \sin\theta \\
-\sin\theta & \cos\theta
\end{array}\right)\,,\qquad
\Lambda=\left(
\begin{array}{cc}
\lambda_\phi & 0 \\
0 & \lambda_\sigma
\end{array}\right)\,,
\end{align}
so that we can solve the system~(\ref{eom_Qi_exact}), obtaining
\begin{equation}
Q_\phi=\sin\theta\,\cos\theta\int d\tau' \left[G_k^{\lambda_\phi}(\tau,\,\tau')-G_k^{\lambda_\sigma}(\tau,\,\tau') \right] \, \hat{\cal S}_\sigma \left(\tau' , \vec k \right) \,,
\end{equation}
where $G_k^{\lambda_{\phi,\sigma}}(\tau,\,\tau')$ is the retarded Green function associated to the operator $\frac{\partial^2}{\partial\tau^2}+\left(k^2-\frac{\lambda_{\phi,\sigma}}{\tau^2}\right)$:
\begin{equation}
G_k^\lambda(\tau,\,\tau')= \Theta(\tau-\tau') \, \frac{\pi}{2}\sqrt{\tau\,\tau'}\left(J_\mu(-k\tau)\,Y_\mu(-k\tau')-J_\mu(-k\tau')\,Y_\mu(-k\tau)\,\right) \,,\qquad \mu=\sqrt{\lambda+\frac{1}{4}} \,. 
\end{equation}

Inspection of the equations given in subsection~\ref{gaugemodes} above shows that, for $\xi\gtrsim O(1)$, the source term ${\cal S}_\sigma \left(\tau' , \vec k \right)$  is exponentially suppressed for $|k\,\tau'|\gtrsim 1$. As a consequence, we can set $|k\,\tau'|\ll 1$ in the propagator. Also, since we are looking at modes far outside of the horizon, we have $|k\,\tau|\ll 1$, and we can use the approximate expression for the propagator
\begin{equation}\label{apprprop}
G^\lambda_k(\tau,\,\tau')\simeq \frac{\sqrt{\tau\,\tau'}}{2\,\mu}\,\left(\frac{\tau'}{\tau}\right)^\mu\,.
\end{equation}

To first order in the slow roll parameters, denoting $\lambda_\sigma=2+\delta\lambda_\sigma$ and $\lambda_\phi=2+\delta\lambda_\phi$ we can expand
\begin{equation}
G^\lambda_k(\tau,\,\tau')\simeq \frac{\sqrt{\tau\,\tau'}}{3}\,\left(\frac{\tau'}{\tau}\right)^{3/2}\,\left(1-\frac{\delta\lambda}{3}+\frac{\delta\lambda}{3}\,\log\frac{\tau'}{\tau}\right)
\end{equation}
so that, neglecting term $-\delta\lambda/3$ which is subdominant with respect to $\frac{\delta\lambda}{3}\,\log\frac{\tau'}{\tau}$,
\begin{equation}
Q_\phi\simeq\sin\theta\,\cos\theta\,\frac{\delta\lambda_\phi-\delta\lambda_\sigma}{3}\int d\tau'\,\left(\log\frac{\tau'}{\tau}\right)\, G_k^{2}(\tau,\,\tau') \, {\cal S}_\sigma \left(\tau' , \vec k \right) \,.
\end{equation}

Since most of the integral in $d\tau'$ gets contribution from $\tau'\simeq -1/k$, we can approximate $\log\frac{\tau'}{\tau}\simeq -\log(-k\,\tau)\simeq N_k$, the number of efoldings of inflation from the time the mode $k$ has left the horizon.

As a consequence the perturbation in $\delta\phi$ is the same one that would have been obtained in the case in which $\sigma\equiv\phi$ (first studied in~\cite{Barnaby:2010vf}) times the factor $\sin\theta\,\cos\theta\,\frac{\delta\lambda_\phi-\delta\lambda_\sigma}{3}\,N_k=2\sqrt{\epsilon_\sigma\,\epsilon_\phi}\,N_k$, that accounts for the fact that in this model $\sigma \neq \phi$, and that quanta of $\phi$ are obtained from the conversion of quanta of $\sigma$. Our computation reproduces the  result of~\cite{Ferreira:2014zia}, up to a factor of $2$. The origin of such a difference can be traced to the second equality in eq.~(5.17) of~\cite{Ferreira:2014zia}, where $\dot\sigma$ is treated as a constant. By properly accounting for the time dependence of $\dot\sigma=\sqrt{2\,\epsilon_\sigma}\,H\,M_P$, the quantity $2\epsilon_\phi-\lambda_2$ in the last term of that equation should read $3\,\epsilon_\phi-\lambda_2$, leading to a result equivalent to ours.

\section{Scalar Modes}
\label{app:scalar}

In this appendix  we show the explicit derivation of the scalar power spectrum and bispectrum in the model \eqref{lagr}, whose fitting functions used for the phenomenological studies are given in \eqref{f23-fit}.

Starting from \eqref{zeta-Qphi}, we split $\hat \zeta$ into the vacuum mode and sourced contribution in the same way as $\hat{Q}_\phi$ in \eqref{split_Qphi}. Using the solution \eqref{sol_Qphi1} for $\hat Q_\phi^{(1)}$, we find the sourced mode of $\hat{\zeta}$ as
\begin{equation}
\hat\zeta^{(1)} \left(\tau , \vec k \right) \simeq 
\frac{3\sqrt{2} \, H \tau}{M_p} \int d\tau' \, G_k(\tau, \tau') \, \frac{\sqrt{\epsilon_\sigma(\tau')}}{\tau'^2} \int d\tau'' \, G_k(\tau' , \tau'') \, \hat{\cal S}_\sigma \big(\tau'' , \vec k \big) \; .
\label{zeta1_1}
\end{equation}
where $\hat{\cal S}_\sigma$ and $G_k(\tau,\tau')$ are defined in \eqref{eom_Qi_app_sigma} and \eqref{Green_JY}, respectively. The explicit expression for $\hat{\cal S}_\sigma$ is obtained by using \eqref{EB_approx} as
\begin{eqnarray}
\hat{\cal S}_\sigma \big(\tau ,\vec k \big) & = &
\alpha\frac{a^3}{f} \int \frac{d^3x}{(2\pi)^{3/2}} \, {\rm e}^{-i \vec k \cdot \vec x} \, \vec{\hat E} \cdot \vec{\hat B}
\nonumber\\
& \simeq & \frac{\alpha H\tau}{4f}
\int \frac{d^3p}{(2\pi)^{3/2}} \,
\epsilon_i^{(+)} \big( \vec p \big) \, \epsilon_i^{(+)} \big( \vec k - \vec p \big) \, 
p^{1/4}
\vert \vec k - \vec p \vert^{1/4}
\left( p^{1/2} + \vert \vec k - \vec p \vert^{1/2} \right)
\nonumber\\ && \qquad\quad \times
\tilde A \left(\tau , p \right)
\tilde A \left( \tau , \vert \vec k - \vec p \vert \right)
\left[ {\hat a}_+ \left( \vec{p} \right) + {\hat a}_+^\dagger \left( - \vec{p} \right) \right]
\left[ {\hat a}_+ \left( \vec{k} - \vec{p} \right) + {\hat a}_+^\dagger \left( - \vec{k} + \vec{p} \right) \right] \; ,
\nonumber\\
\label{Ssig_explicit}
\end{eqnarray}
after symmetrizing $p$ and $\vert \vec k - \vec p \vert$, where
\begin{eqnarray}
\tilde A \left(\tau , p \right)
\tilde A \left( \tau , \vert \vec k - \vec p \vert \right)
&=& 
N\left[ \xi_* , -p\tau_* , \delta \right]
N\left[ \xi_* , -\vert \vec k - \vec p \vert \tau_* , \delta \right]
\nonumber\\ && \times
\exp \left[ - \frac{4 \, \xi_*^{1/2}}{1+\delta} \left( \frac{\tau}{\tau_*} \right)^{\delta/2} \left( \sqrt{- p \tau} + \sqrt{ - \vert \vec k - \vec p \vert \tau} \right) \right] .
\end{eqnarray}
Since we are interested in large-scale modes, we can safely assume $-k\tau \ll 1$, leading to
\begin{equation}
G_k\left(\tau,\tau'\right) \simeq \Theta\left(\tau-\tau'\right) \, \sqrt{\frac{\pi}{2}} \, \frac{\sqrt{-\tau'}}{-k^{3/2}\tau} \, J_{3/2} \left( -k \tau' \right) \; , \quad -k\tau \ll 1 \; ,
\label{Green_app_Bessel}
\end{equation}
while this approximation is not valid for $G_k\left(\tau',\tau''\right)$. plugging this and \eqref{Green_JY} into \eqref{zeta1_1}, we have
\begin{eqnarray}
\hat\zeta^{(1)} \left(\tau , \vec k \right) & \simeq &
\frac{3 \pi^{3/2} \, H}{2 M_p \, k^{3/2}}
\int_{-\infty}^\tau \frac{d\tau'}{\tau'} \, J_{3/2} \left( -k \tau' \right) \, \sqrt{\epsilon_\sigma(\tau')}
\int_{-\infty}^{\tau'} d\tau'' \sqrt{-\tau''} \,
\hat{\cal S}_\sigma \big(\tau'' , \vec k \big)
\nonumber\\ && \qquad \times
\Big[ J_{3/2} \left( -k\tau' \right) \, Y_{3/2} \left( -k\tau'' \right) 
- Y_{3/2} \left( -k\tau' \right) \, J_{3/2} \left( -k\tau'' \right) \Big]
\; .
\end{eqnarray}
We then plug \eqref{Ssig_explicit} into the above expression and rescale 
\begin{equation}
\vec{\tilde p} \equiv \frac{\vec p}{k} \; , \quad x' \equiv - k \tau' \; , \quad x'' \equiv -k \tau'' \; ,
\end{equation}
to obtain
\begin{eqnarray}
\hat\zeta^{(1)} \left(\tau , \vec k \right)  & \simeq &
\frac{3 \pi^{3/2} H^2 \alpha \sqrt{\epsilon_{\sigma,*}}}{8 M_p f}
\int \frac{d^3 \tilde p}{(2\pi)^{3/2}} \,
{\tilde p}^{1/4} \vert {\hat k} - \vec{\tilde p} \vert^{1/4}
\left( {\tilde p}^{1/2} + \vert \hat k - \vec{\tilde p} \vert^{1/2} \right)
\nonumber\\ && \times
N\left[ \xi_* ,\, \tilde p \, x_* ,\, \delta \right]
N\left[ \xi_* ,\, \vert \hat k - \vec{\tilde p} \vert \, x_* ,\, \delta \right]
\hat{\cal P}  \left[ \vec{p} ,\, \vec{k}  \right] 
{\cal T}_\zeta \left[ \xi_* ,\, x_* ,\, \delta ,\, \sqrt{\tilde p} + \sqrt{\vert \hat k - \vec{\tilde p} \vert} \right] \,, 
\nonumber\\
\label{zeta1_full}
\end{eqnarray}
where we have defined
\begin{eqnarray}
&& \!\!\!\!\! \hat{\cal P}  \left[ \vec{p} , \vec{k}  \right] \equiv 
\epsilon_i^{(+)} \left( \vec{\tilde p} \right)  \, \epsilon_i^{(+)} \left( \hat{k} - \vec{\tilde p} \right)
\left[ {\hat a}_+ \left( \vec{p} \right) + {\hat a}_+^\dagger \left( - \vec{p} \right) \right] 
\left[ {\hat a}_+ \left( \vec{k} - \vec{p} \right) + {\hat a}_+^\dagger \left( - \vec{k} +\vec{p} \right) \right]
\nonumber\\
&& \!\!\!\!\! {\cal T}_\zeta \left[ \xi_* ,\, x_* ,\, \delta ,\, Q \right] \equiv
\int_0^\infty \frac{dx'}{x'} 
J_{3/2} \left( x' \right)  
\sqrt{\frac{\epsilon_\sigma(x')}{\epsilon_{\sigma,*}}} 
\int_{x'}^\infty dx'' \, x''^{3/2} 
\exp \left[ - \frac{4 \, \xi_*^{1/2}}{1+\delta}  \frac{x''^{(1+\delta)/2}}{x_*^{\delta/2}} \, Q \right]
\nonumber\\
&& \qquad\qquad\qquad\qquad \times
\left[ J_{3/2} \left(x'\right) \, Y_{3/2} \left( x'' \right)
- Y_{3/2} \left( x' \right) \, J_{3/2} \left( x'' \right) \right]\,,
\end{eqnarray}
where we have sent $-k\tau \rightarrow 0$ for the lower bound of the integral.
The functional forms of $\xi$ and $\epsilon_\sigma$ is shown in \eqref{xi-def}. Using this, and the fact that 
$\sqrt{\epsilon_\sigma} = \frac{\dot{\sigma}}{\sqrt{2} M_p H} = {\rm const.} \times \xi$, we have 
\begin{equation}
\xi(x) = \frac{2\xi_*}{\left( x / x_* \right)^\delta + \left( x_* / x \right)^\delta} \; , \quad
\sqrt{\frac{\epsilon_\sigma(x)}{\epsilon_{\sigma,*}}} =
\frac{2}{\left( x / x_* \right)^\delta + \left( x_* / x \right)^\delta} \; ,
\end{equation}
where $x_* = - k \tau_* = \left[ k / a(\tau_*) \right] / H$ is the ratio between the physical momentum of the mode and the horizon at the moment in which $\xi$ is at its maximum. 
We emphasize that apart from the creation and annihilation operators, the dependence on $k$ arises only through $x_* = k / k_*$ in the expression \eqref{zeta1_full}.
Using \eqref{zeta1_full}, we proceed to the calculations of the scalar power spectrum and bispectrum in the following subsections.

\subsection{Scalar Power Spectrum}
\label{subapp:scalarPS}

The power spectrum of curvature perturbations $\hat\zeta$ can be defined as
\begin{equation} 
{\cal P}_\zeta \left( k \right) \, \delta^{(3)} \left( \vec{k} + \vec{k}' \right) \equiv 
\frac{k^3}{2\pi^2} 
\left\langle {\hat \zeta} \left( \vec{k}  \right)  {\hat \zeta} \left( \vec{k}'  \right) \right\rangle \,, 
\label{def_scalarPS} 
\end{equation} 
and it consists of vacuum mode and sourced contribution, namely,
\begin{equation}
{\cal P}_\zeta(k) = {\cal P}_\zeta^{(0)} (k) + {\cal P}_\zeta^{(1)} (k) \; .
\label{PSsca_dec}
\end{equation}
Notice that these two modes are uncorrelated, and thus there is no cross term.
Here ${\cal P}_\zeta^{(0)} (k)$ is the standard vacuum mode, for which, using eq.   \eqref{Q0-mode}, we find
\begin{equation}
{\cal P}_\zeta^{(0)} (k) = \frac{H^2}{8\pi^2\epsilon_\phi M_p^2} \; ,
\label{Pzeta0}
\end{equation}
neglecting the slow-roll corrections. For the sourced mode, we compute, using \eqref{zeta1_full},
\begin{eqnarray}
&& \!\!\!\!\!
\left\langle {\hat \zeta}^{(1)} \left( \vec{k}  \right)  {\hat \zeta}^{(1)} \left( \vec{k}'  \right) \right\rangle \simeq
\frac{9 \pi^3 H^4 \alpha^2 \epsilon_{\sigma,*}}{64 M_p^2 f^2}
\int \frac{d^3 \tilde p \, d^3 \tilde p'}{(2\pi)^3}
\left\langle \hat{\cal P} \left[ \vec{p} , \vec{k}  \right] \hat{\cal P} \left[ \vec{p} \, {}' , \vec{k}'  \right] \right\rangle
\nonumber\\ && \qquad\qquad\quad \times
\left( \tilde p \, \tilde p' \vert {\hat k} - \vec{\tilde p} \vert \vert {\hat k}' - \vec{\tilde p}' \vert \right)^{1/4}
\left( {\tilde p}^{1/2} + \vert \hat k - \vec{\tilde p} \vert^{1/2} \right)
\left( {\tilde p}'^{1/2} + \vert \hat k' - \vec{\tilde p}' \vert^{1/2} \right)
\nonumber\\ && \qquad\qquad\quad \times
N\left[ \xi_* ,\, \tilde p \, x_* ,\, \delta \right]
N\left[ \xi_* ,\, \vert \hat k - \vec{\tilde p} \vert \, x_* ,\, \delta \right]
N\left[ \xi_* ,\, \tilde p' \, x_* ,\, \delta \right]
N\left[ \xi_* ,\, \vert \hat k' - \vec{\tilde p}' \vert \, x_* ,\, \delta \right]
\nonumber\\ && \qquad\qquad\quad \times
{\cal T}_\zeta \left[ \xi_* ,\, \frac{k}{k_*} ,\, \delta , \sqrt{\tilde p} + \sqrt{\vert \hat k - \vec{\tilde p} \vert} \right]
{\cal T}_\zeta \left[ \xi_* ,\, \frac{k'}{k_*} ,\, \delta ,\, \xi_* , \sqrt{\tilde p'} + \sqrt{\vert \hat k' - \vec{\tilde p}' \vert} \right] \; ,
\end{eqnarray}
where $k_*$ is the mode that exits the horizon at the moment when $\xi$ takes its maximum value $\xi_*$, namely $k_* = a(\tau_*) H$. Evaluating
\begin{eqnarray}
\left\langle \hat{\cal P} \left[ \vec{p} , \vec{k}  \right] \hat{\cal P} \left[ \vec{p} \, {}' , \vec{k}'  \right] \right\rangle & = &
\delta^{(3)} \left( \vec k + \vec k' \right)
\left[ \delta^{(3)} \left( \vec p + \vec p \,{}' \right) + \delta^{(3)} \left( \vec k - \vec p + \vec p \, {}' \right) \right]
\nonumber\\ && \times
\epsilon_i^{(+)} \left( \vec{\tilde p} \right)  \, \epsilon_i^{(+)} \left( \hat{k} - \vec{\tilde p} \right) 
\epsilon_j^{(+)} \left( \vec{\tilde p} \,{}' \right)  \, \epsilon_j^{(+)} \left( \hat{k}' - \vec{\tilde p} \, {}' \right) \,
 \; ,
\end{eqnarray}
and using the definition \eqref{def_scalarPS}, we obtain
\begin{eqnarray}
{\cal P}_\zeta^{(1)}(k) & \simeq & \frac{9 \pi \, H^4 \alpha^2 \epsilon_{\sigma,*}}{64 M_p^2 f^2}
\int \frac{d^3 \tilde p}{(2\pi)^3}
\left\vert \epsilon_i^{(+)} \left( \vec{\tilde p} \right)  \, \epsilon_i^{(+)} \left( \hat{k} - \vec{\tilde p} \right) \right\vert^2
\tilde p^{1/2} \vert \hat k - \vec{\tilde p} \vert^{1/2} 
\left( \tilde p^{1/2} + \vert \hat k - \vec{\tilde p} \vert^{1/2} \right)^2
\nonumber\\ && \times
N^2\left[ \xi_* ,\, \tilde p \, x_* ,\, \delta \right]
N^2\left[ \xi_* ,\, \vert \hat k - \vec{\tilde p} \vert \, x_* ,\, \delta \right]
{\cal T}_\zeta^2 \left[ \xi_* ,\, x_* ,\, \delta , \sqrt{\tilde p} + \sqrt{\vert \hat k - \vec{\tilde p} \vert} \right] \; .
\end{eqnarray}
Expressing $H^2/M_p^2$ and $\alpha^2\epsilon_{\sigma,*} /f^2$ in terms of ${\cal P}_\zeta^{(0)}$ and $\xi_*$, respectively, by \eqref{Pzeta0} and $\alpha^2\epsilon_{\sigma,*} / f^2 = \dot{\sigma}^2(\tau_*) / (2M_p^2 H^2 f^2) = 2 \xi_*^2 / M_p^2$, we find
\begin{equation} 
{\cal P}_\zeta^{(1)}(k) \simeq \left[ \epsilon_\phi {\cal P}_\zeta^{(0)} \right]^2 \, f_{2,\zeta} \left(\xi_* ,\, x_* ,\, \delta \right) \,, 
\label{Pzeta1_full} 
\end{equation} 
where
\begin{eqnarray}
f_{2, \zeta} \left(\xi_* ,\, x_* ,\, \delta \right) & \equiv & \frac{9 \pi^5}{2} \, \xi_*^2
\int \frac{d^3 \tilde p}{(2\pi)^3}
\left( 1 - \frac{\vec{\tilde p}}{\tilde p} \cdot \frac{\hat{k} - \vec{\tilde p}}{\vert \hat{k} - \vec{\tilde p} \vert} \right)^2
\tilde p^{1/2} \vert \hat k - \vec{\tilde p} \vert^{1/2} 
\left( \tilde p^{1/2} + \vert \hat k - \vec{\tilde p} \vert^{1/2} \right)^2
\nonumber\\ && \times
N^2\left[ \xi_* ,\, \tilde p \, x_* ,\, \delta \right]
N^2\left[ \xi_* ,\, \vert \hat k - \vec{\tilde p} \vert \, x_* ,\, \delta \right]
{\cal T}_\zeta^2 \left[ \xi_* ,\, x_* ,\, \delta , \sqrt{\tilde p} + \sqrt{\vert \hat k - \vec{\tilde p} \vert} \right] \; ,
\nonumber\\
\end{eqnarray}
using the property of the polarization vector, $\big\vert \epsilon_i^{(\lambda)} \left( \vec{p} \right)  \, \epsilon_i^{(\lambda')} \left( \vec{q} \right) \big\vert^2 = \left( 1 - \lambda\lambda'\hat{p} \cdot \hat{q} \right)^2 / 4$. For the concrete evaluation of the $\tilde p$ integral, we denote the cosine of the angle between $\vec{\tilde p}$ and $\hat{k}$ by $\eta$. After taking the trivial angular integral, we have
\begin{eqnarray}
f_{2,\zeta} \left(\xi_* ,\, x_* ,\, \delta \right) && =
\frac{9 \pi^3}{8} \, \xi_*^2
\int_0^\infty d\tilde{p} \int_{-1}^1 d\eta \,
\tilde{p}^{5/2} \left( 1 - 2 \tilde{p} \eta + \tilde{p}^2 \right)^{1/4}
\left[ \tilde{p}^{1/2} + \left( 1 - 2 \tilde{p} \eta + \tilde{p}^2 \right)^{1/4} \right]^2
\nonumber\\ && \times
\left[ 1 + \frac{\tilde p - \eta}{\left( 1 - 2 \tilde{p} \eta + \tilde{p}^2 \right)^{1/2}} \right]^2
N^2\left[ \xi_* ,\, \tilde p \, x_* ,\, \delta \right]
N^2\left[ \xi_* ,\,  \left( 1 - 2 \tilde{p} \eta + \tilde{p}^2 \right)^{1/2} x_* ,\, \delta \right]
\nonumber\\ && \times
{\cal T}_\zeta^2 \left[ \xi_* ,\, x_* ,\, \delta ,\, \tilde{p}^{1/2} + \left( 1 - 2 \tilde{p} \eta + \tilde{p}^2 \right)^{1/4} \right] \; .
\label{f2_peta}
\end{eqnarray}
Alternatively, we can change the variables of integration from $\tilde p$ and $\eta$ to $x$ and $y$ such that $x = \tilde p + \vert \hat{k} - \vec{\tilde p} \vert$ and  $y = \tilde p - \vert \hat{k} - \vec{\tilde p} \vert$. Then the integral reduces to
\begin{eqnarray}
f_{2,\zeta}\left(\xi_* ,\, x_* ,\, \delta \right) & = &
\frac{9 \pi^3}{32} \, \xi_*^2
\int_1^\infty dx \int_0^1 dy \,
\frac{\left( 1-x^2 \right)^2 \left( \sqrt{x+y} + \sqrt{x-y} \right)^2}{\sqrt{x+y} \sqrt{x-y}}
\nonumber\\ & \times &
N^2\left[ \xi_* ,\, \frac{x+y}{2} \, x_* ,\, \delta \right]
N^2\left[ \xi_* ,\,  \frac{x-y}{2} \, x_* ,\, \delta \right]
{\cal T}_\zeta^2 \left[ \xi_* ,\, x_* ,\, \delta ,\, \frac{\sqrt{x+y} + \sqrt{x-y}}{\sqrt{2}}\right] \; .
\nonumber\\
\label{f2_xy}
\end{eqnarray}
We evaluate $f_2$ numerically using either \eqref{f2_peta} or \eqref{f2_xy}.
The total power spectrum of the curvature perturbations is, from \eqref{PSsca_dec} and \eqref{Pzeta1_full},
\begin{equation}
{\cal P}_\zeta = {\cal P}_\zeta^{(0)} \left[ 1 + \epsilon_\phi^2 {\cal P}_\zeta^{(0)} \, f_{2,\zeta} \left( \xi_* ,\, x_* ,\, \delta \right) \right] \,, 
\label{scapower_total}
\end{equation}
where ${\cal P}_\zeta^{(0)}$ is given in \eqref{Pzeta0}.

\subsection{Scalar Bispectrum}
\label{subapp:scalarBS}

We define the bispectrum of the curvature perturbations as
\begin{equation}
{\cal B}_\zeta \left(k_1 , k_2 , k_3\right) \, \delta^{(3)} \left( \vec{k}_1 + \vec{k}_2 + \vec{k}_3 \right) \equiv
\left\langle \hat{\zeta} \left( \vec{k}_1 \right) \hat{\zeta} \left( \vec{k}_2 \right) \hat{\zeta} \left( \vec{k}_3 \right) \right\rangle \; ,
\end{equation}
where ${\cal B}_\zeta$ depends only on the magnitudes of the three momenta, under the restriction that they form a triangle. The bispectrum in principle consists of the vacuum and sourced modes, similarly to the power spectrum. However the vacuum mode bispectrum ${\cal B}_\zeta^{(0)}$ is small, and we focus on the contribution from the source, i.e.
\begin{equation}
{\cal B}_\zeta \cong {\cal B}_\zeta^{(1)} \; .
\end{equation}
Taking the $3$-point function of $\hat\zeta^{(1)}$ in \eqref{zeta1_full}, we have
\begin{eqnarray}
&& \!\!\!\!\! \!\!\!\!\!
\left\langle \hat{\zeta}^{(1)} \left( \vec{k}_1 \right) \hat{\zeta}^{(1)} \left( \vec{k}_2 \right) \hat{\zeta}^{(1)} \left( \vec{k}_3 \right) \right\rangle 
\simeq
\frac{3^3 \pi^{9/2} H^6 \alpha^3\epsilon_{\sigma,*}^{3/2}}{2^9 M_p^3 f^3}
\int \frac{d^3\tilde p_1 \, d^3\tilde p_2 \, d^3\tilde p_3}{(2\pi)^{9/2}}
\left\langle 
\hat{\cal P} \left[ \vec{p}_1 , \vec{k}_1 \right]
\hat{\cal P} \left[ \vec{p}_2 , \vec{k}_2 \right]
\hat{\cal P} \left[ \vec{p}_3 , \vec{k}_3 \right]
\right\rangle
\nonumber\\ && \qquad\qquad\quad \times
\prod_{i=1}^{3}
\left( \tilde p_i \vert {\hat k}_i - \vec{\tilde p}_i \vert \right)^{1/4}
\left( {\tilde p}_i^{1/2} + \vert \hat k_i - \vec{\tilde p}_i \vert^{1/2} \right)
N\left[ \xi_* ,\, \tilde p_i \, x_* ,\, \delta \right]
N\left[ \xi_* ,\, \vert \hat k_i - \vec{\tilde p}_i \vert \, x_* ,\, \delta \right]
\nonumber\\ && \qquad\qquad\qquad \times
{\cal T}_\zeta \left[ \xi_* ,\, \frac{k_i}{k_*} ,\, \delta ,\, \sqrt{\tilde p}_i + \sqrt{\vert \hat k_i - \vec{\tilde p}_i \vert} \right] \; .
\end{eqnarray}
Evaluating the vacuum expectation value, we find
\begin{eqnarray}
{\cal B}_\zeta^{(1)} \left( \vec{k}_1 ,\,  \vec{k}_2 ,\,  \vec{k}_3 \right) & \simeq &
\frac{3^3 \pi^{9/2} H^6 \alpha^3 \epsilon_{\sigma,*}^{3/2}}{2^6 M_p^3 f^3} \,
\frac{1}{k_1^4 k_2^4 k_3^4}
\int \frac{d^3p}{(2\pi)^{9/2}} \,
\tilde{P}\left[ \vec{p} , \vec{p} + \vec{k}_1 , \vec{p} - \vec{k}_3 \right]
\sqrt{p \big\vert \vec{p} + \vec{k}_1 \big\vert \big\vert \vec{p} - \vec{k}_3 \big\vert}
\nonumber\\ & & \times
\left( \sqrt{p} + \sqrt{\big\vert \vec{p} + \vec{k}_1 \big\vert} \right)
\left( \sqrt{\big\vert \vec{p} + \vec{k}_1 \big\vert} + \sqrt{\big\vert \vec{p} - \vec{k}_3 \big\vert} \right)
\left( \sqrt{\big\vert \vec{p} - \vec{k}_3 \big\vert} + \sqrt{p} \right)
\nonumber\\ & & \times
N^2\left[ \xi_* ,\, \frac{p}{k_*} ,\, \delta \right]
N^2\left[ \xi_* ,\, \frac{\vert \vec p + \vec k_1 \vert}{k_*} ,\, \delta \right]
N^2\left[ \xi_* ,\, \frac{\vert \vec p - \vec k_3 \vert}{k_*} ,\, \delta \right]
\nonumber\\ && \times
{\cal T}_\zeta \left[ \xi_* ,\, \frac{k_1}{k_*} ,\, \delta ,\, \frac{\sqrt{p} + \sqrt{\vert \vec{p} + \vec k_1 \vert}}{\sqrt{k_1}} \right]
{\cal T}_\zeta \left[ \xi_* ,\, \frac{k_2}{k_*} ,\, \delta ,\, \frac{\sqrt{\big\vert \vec{p} + \vec{k}_1 \big\vert} + \sqrt{\big\vert \vec{p} - \vec{k}_3 \big\vert}}{\sqrt{k_2}} \right]
\nonumber\\ & & \times
{\cal T}_\zeta \left[ \xi_* ,\, \frac{k_3}{k_*} ,\, \delta ,\, \frac{\sqrt{\big\vert \vec{p} - \vec{k}_3 \big\vert} + \sqrt{p}}{\sqrt{k_3}} \right] \; ,
\label{B1-par}
\end{eqnarray}
where
\begin{eqnarray}
&& \!\!\!\!\! \!\!\!\!\!
\tilde{P}\left[ \vec{v}_1 , \vec{v}_2 , \vec{v}_3 \right] \equiv
\epsilon_i^{(+)*} \left( \vec{v}_1 \right) \epsilon_i^{(+)} \left( \vec{v}_2 \right) 
\epsilon_j^{(+)*} \left( \vec{v}_2 \right) \epsilon_j^{(+)} \left( \vec{v}_3 \right) 
\epsilon_k^{(+)*} \left( \vec{v}_3 \right) \epsilon_k^{(+)} \left( \vec{v}_1 \right) 
\nonumber\\ && \quad =
\frac{1}{8} \Big[ 
\hat v_1 \cdot \hat v_2 + \hat v_2 \cdot \hat v_3 + \hat v_3 \cdot \hat v_1
+ \left( \hat v_1 \cdot \hat v_2 \right)^2 + \left( \hat v_2 \cdot \hat v_3 \right)^2 + \left( \hat v_3 \cdot \hat v_1 \right)^2
\nonumber\\ && \qquad
+ \left( \hat v_1 \cdot \hat v_2 \right) \left( \hat v_2 \cdot \hat v_3 \right)
+ \left( \hat v_2 \cdot \hat v_3 \right) \left( \hat v_3 \cdot \hat v_1 \right)
+ \left( \hat v_3 \cdot \hat v_1 \right) \left( \hat v_1 \cdot \hat v_2 \right)
- \left( \hat v_1 \cdot \hat v_2 \right) \left( \hat v_2 \cdot \hat v_3 \right) \left( \hat v_3 \cdot \hat v_1 \right)
\Big]
\nonumber\\ && \qquad
+ \frac{i}{8} \, \hat{v}_1 \cdot \left( \hat{v}_2 \times \hat{v}_3 \right)
\left( 1 + \hat v_1 \cdot \hat v_2 + \hat v_2 \cdot \hat v_3 + \hat v_3 \cdot \hat v_1 \right) \; .
\end{eqnarray}
We can disregard the imaginary part as the bispectrum is real (see Appendix \ref{app:BS}). Rescaling
\begin{equation}
k \equiv k_1 \; , \quad x_2 \equiv \frac{k_2}{k} \; , \quad x_3 \equiv \frac{k_3}{k} \; , \quad x_* \equiv \frac{k}{k_*} \; , \quad \vec{\tilde p} \equiv \frac{\vec{p}}{k} \; ,
\end{equation}
and using the relations
\begin{equation}
\frac{H^2}{M_p^2} = 8 \pi^2 \epsilon_\phi {\cal P}_\zeta^{(0)} \; , \quad \frac{\alpha \sqrt{\epsilon_{\sigma,*}}}{f} = \frac{\sqrt{2} \, \xi_*}{M_p} \; ,
\end{equation}
we obtain
\begin{equation}
{\cal B}_\zeta^{(1)} \simeq \frac{\left[ \epsilon_\phi {\cal P}_\zeta^{(0)} \right]^3}{k_1^2 k_2^2 k_3^2} \, 
f_{3,\zeta} \left( \xi_* , x_* , \delta , x_2 , x_3 \right) \; ,
\end{equation}
where
\begin{eqnarray}
&& \!\!\!\!\!
f_{3,\zeta} \left( \xi_* , x_* , \delta , x_2 , x_3 \right) = 
2^{9/2} 3^3 \pi^{21/2} \, \frac{\xi_*^3}{x_2^2 x_3^2}
\int \frac{d^3 \tilde{p}}{(2\pi)^{9/2}} \,
\Re\left( \tilde{P}\left[ \vec{\tilde p} , \vec{\tilde p} + \hat{k}_1 , \vec{\tilde p} - x_3 \hat{k}_3 \right] \right)
\nonumber\\ && \times
\sqrt{\tilde p \big\vert \vec{\tilde p} + \hat{k}_1 \big\vert \big\vert \vec{\tilde p} - x_3 \hat{k}_3 \big\vert}
\left( \sqrt{\tilde p} + \sqrt{\big\vert \vec{\tilde p} + \hat{k}_1 \big\vert} \right)
\left( \sqrt{\big\vert \vec{\tilde p} + \hat{k}_1 \big\vert} + \sqrt{\big\vert \vec{\tilde p} - x_3 \hat{k}_3 \big\vert} \right)
\left( \sqrt{\big\vert \vec{\tilde p} - x_3 \hat{k}_3 \big\vert} + \sqrt{\tilde p} \right)
\nonumber\\ && \times
N^2\left[ \xi_* ,\, \tilde p \, x_* ,\, \delta \right]
N^2\left[ \xi_* ,\, \vert \vec{\tilde p} + \hat k_1 \vert \, x_* ,\, \delta \right]
N^2\left[ \xi_* ,\, \vert \vec{\tilde p} - x_3 \hat k_3 \vert \, x_* ,\, \delta \right]
{\cal T}_\zeta \left[ \xi_* ,\, x_* ,\, \delta , \sqrt{\tilde p} + \sqrt{\vert \vec{\tilde p} + \hat k_1 \vert} \right]
\nonumber\\ & & \times
{\cal T}_\zeta \left[ \xi_* ,\, x_2 x_* ,\, \delta , \frac{\sqrt{\big\vert \vec{\tilde p} + \hat{k}_1 \big\vert} + \sqrt{\big\vert \vec{\tilde p} - x_3 \hat{k}_3 \big\vert}}{\sqrt{x_2}} \right]
{\cal T}_\zeta \left[ \xi_* ,\, x_3 x_* ,\, \delta ,\, \frac{\sqrt{\big\vert \vec{\tilde p} - x_3 \hat{k}_3 \big\vert} + \sqrt{\tilde p}}{\sqrt{x_3}} \right] \; ,
\nonumber\\
\end{eqnarray}
where $\Re$ denotes the real part. 

In order to evaluate the integrals, we can orient $\vec{k}_1$ along the $x$ axis and express $\vec{k}_2$ and $\vec{k}_3$ in terms of $x_2$ and $x_3$, namely,
\begin{eqnarray}
\vec{k}_1 & =&  k \left( 1 ,\, 0 ,\, 0 \right) \; , \nonumber\\
\vec{k}_2 &=& \frac{k}{2} \left( - 1 - x_2^2 + x_3^2 ,\, \sqrt{ - \left( 1 - x_2 + x_3 \right)  \left( 1 + x_2 - x_3 \right)  \left( 1 - x_2 - x_3 \right)  \left( 1 + x_2 + x_3 \right) } ,\, 0 \right) \; , \nonumber\\ 
\vec{k}_3 &=& \frac{k }{ 2 } \left( - 1 + x_2^2 - x_3^2 ,\, - \sqrt{ - \left( 1 - x_2 + x_3 \right)  \left( 1 + x_2 - x_3 \right)  \left( 1 - x_2 - x_3 \right)  \left( 1 + x_2 + x_3 \right) } ,\, 0 \right)  \; ,
\nonumber\\
\end{eqnarray}
and then perform numerical integration. The phenomenology of this result is extensively studied in Section \ref{sec:pheno}.

\section{Tensor Modes}
\label{app:tensor}

In this appendix, we present the detailed derivation of the tensor power spectrum and bispectrum from the source, parallel to the scalar counterparts derived in Appendix \ref{app:scalar}. The phenomenological study in Section \ref{sec:pheno} is done with the use of fitting functions \eqref{f23-fit} for the results derived in this appendix.

We define the tensor mode operators in the basis of circular polarization,
\begin{equation}
\hat h_\lambda \left(\tau , \vec{k} \right) \equiv \Pi_{ij, \lambda} \left( \hat{h} \right) \, \hat h_{ij} \left(\tau , \vec{k} \right) = \frac{2}{M_p \, a(\tau)} \, \hat Q_\lambda \left(\tau , \vec{k} \right) \; ,
\label{def_hlambda}
\end{equation}
where the tensor canonical mode $\hat Q_\lambda$ and the polarization tensor $\Pi_{ij,\lambda}$ are defined in \eqref{deco-hij} and \eqref{pol_tensor}, respectively. The equation of motion for the canonical operator $\hat Q_\lambda$ is shown in \eqref{eom_Qlambda}. We can decompose the solution into the homogeneous and particular ones, corresponding to the vacuum and sourced modes, respectively, in the same way as for the case of the scalar perturbations \eqref{split_Qphi}. The solution for the vacuum mode is given in \eqref{sol_tensorvac}, and the sourced contribution $\hat Q_\lambda^{(1)}$ can be solved formally by \eqref{Q1lambda_formal}. This, together with \eqref{def_hlambda}, provides the solution for the sourced mode $\hat h_\lambda^{(1)}$ as
\begin{equation}
\hat h_\lambda^{(1)} \left( \tau , \vec{k} \right) \simeq 
- \frac{2H\tau}{M_p} \int_{-\infty}^\infty d\tau' \, G_k(\tau,\tau') \, \hat{\cal S}_\lambda \left( \tau' , \vec{k} \right) \; ,
\label{hhat_formal}
\end{equation}
where we have used $a(\tau) \simeq - 1 / (H\tau)$, the Green function is given in \eqref{Green_JY}, and the source $\hat{\cal S}_\lambda$ is defined in \eqref{eom_Qlambda}. 
The explicit expression for $\hat{\cal S}_\lambda$ can be found by using \eqref{def_EB} as
\begin{eqnarray}
\hat{\cal S}_\lambda \left(\tau , \vec{k} \right) & = &
- \frac{a^3}{M_p} \, \Pi_{ij,\lambda} \left( {\hat k} \right)\int \frac{d^3x}{(2\pi)^{3/2}} \, {\rm e}^{-i \vec k \cdot \vec x}  \left[\hat{E}_i \, \hat{E}_j +  \hat{B}_i \, \hat{B}_j  \right]
\nonumber\\
& \simeq &
- \frac{H \sqrt{-\tau \, \xi(\tau)}}{\sqrt{2} M_p}
\int \frac{d^3p}{(2\pi)^{3/2}} \, 
{\cal P}_\lambda \left[ \vec{k} , \vec{p} , \vec{k}-\vec{p} \right] 
p^{1/4} \big\vert \vec{k} - \vec{p} \big\vert^{1/4}
\left( 1 + \frac{-\tau}{2 \, \xi(\tau)} \sqrt{p \big\vert \vec{k} - \vec{p} \big\vert} \right)
\nonumber\\ & & \times
\tilde A \left( \tau , p \right) \, \tilde A \left( \tau , \vert \vec k - \vec p \vert \right)
\left[ \hat{a}_\lambda\left(\vec{p}\right) + \hat{a}_\lambda^\dagger\left( -\vec{p}\right) \right]
\left[ \hat{a}_\lambda\left( \vec{k} - \vec{p} \right) + \hat{a}_\lambda^\dagger\left( - \vec{k} + \vec{p} \right) \right] \; ,
\nonumber\\
\label{Slambda_explicit}
\end{eqnarray}
where $\hat{a}_\lambda^\dagger$ and $\hat{a}_\lambda$ are the creation and annihilation operators, respectively, of the helicity $2\lambda$ graviton modes. Recalling the relation \eqref{pol_tensor}, we have defined
\begin{equation}
{\cal P}_\lambda \left[ \vec{k} , \vec{p} , \vec{k}-\vec{p} \right] \equiv 
\epsilon_i^{(\lambda)*} \left( \vec{k} \right) \,
\epsilon_i^{(+)} \left( \vec{p} \right) \,
\epsilon_j^{(\lambda)*} \left( \vec{k} \right) \,
\epsilon_j^{(+)} \left( \vec{k} - \vec{p} \right) \; .
\label{def_Plambda}
\end{equation}
Since we are interested in large-scale modes $-k\tau \ll 1$, we can approximate
\begin{equation}
G_k\left(\tau , \tau'\right) \simeq 
\frac{\Theta\left(\tau - \tau'\right)}{k^3 \tau \tau'} \left[ k \tau' \cos \left( k \tau' \right) - \sin \left( k\tau' \right) \right] \; , \quad
-k\tau \ll 1 \; ,
\label{Green_app2}
\end{equation}
which is identical to \eqref{Green_app_Bessel}. Combining \eqref{hhat_formal} with \eqref{Slambda_explicit} and \eqref{Green_app2}, we obtain
\begin{eqnarray}
\hat h_\lambda^{(1)} \left( \tau , \vec{k} \right) & \simeq &
\frac{\sqrt{2} \, H^2}{M_p^2 \, k^{7/2}} 
\int \frac{d^3p}{(2\pi)^{3/2}} \, 
{\cal P}_\lambda \left[ \vec{k} , \vec{p} , \vec{k}-\vec{p} \right] 
p^{1/4} \big\vert \vec{k} - \vec{p} \big\vert^{1/4} \,
\nonumber\\ && \qquad \times
N\left[ \xi_* ,\, \frac{p}{k_*} ,\, \delta \right]
N\left[ \xi_* ,\, \frac{\vert \vec k - \vec{p} \vert}{k_*} ,\, \delta \right]
{\cal T}_h \left[ \xi_* ,\, x_* ,\, \delta ,\, \frac{p}{k} ,\, \frac{\vert \vec{k} - \vec{p} \vert}{k} \right]
\nonumber\\ && \qquad \times
\left[ \hat{a}_\lambda\left(\vec{p}\right) + \hat{a}_\lambda^\dagger\left( -\vec{p}\right) \right]
\left[ \hat{a}_\lambda\left( \vec{k} - \vec{p} \right) + \hat{a}_\lambda^\dagger\left( - \vec{k} + \vec{p} \right) \right] \; ,
\label{h1_full}
\end{eqnarray}
where
\begin{eqnarray}
&&{\cal T}_h \left[ \xi_* ,\, x_* ,\, \delta ,\, , \tilde p ,\, \tilde q \right] \equiv 
{\cal T}_{h1} \left[ \xi_* ,\, x_* ,\, \delta ,\, \sqrt{\tilde p} + \sqrt{\tilde q} \right] 
+ \frac{\sqrt{\tilde{p} \, \tilde{q}}}{2} \, {\cal T}_{h2} \left[ \xi_* ,\, x_* ,\, \delta ,\, \sqrt{\tilde p} + \sqrt{\tilde q} \right] \; , 
\nonumber\\
&&{\cal T}_{h1} \left[ \xi_* ,\, x_* ,\, \delta ,\, Q \right] \equiv
\int_0^{\infty} dx' \left( x' \cos x' - \sin x' \right) \sqrt{\frac{\xi(x')}{x'}} \,
\exp \left[ - \frac{4 \, \xi_*^{1/2}}{1+\delta}  \frac{x'^{(1+\delta)/2}}{x_*^{\delta/2}} \, Q \right],
\nonumber\\
&&{\cal T}_{h2} \left[ \xi_* ,\, x_* ,\, \delta ,\, Q \right] \equiv
\int_0^{\infty} dx' \left( x' \cos x' - \sin x' \right) \sqrt{\frac{x'}{\xi(x')}} \,
\exp \left[ - \frac{4 \, \xi_*^{1/2}}{1+\delta}  \frac{x'^{(1+\delta)/2}}{x_*^{\delta/2}} \, Q \right].
\end{eqnarray}
Here we have rescaled the variables $x' \equiv - k \tau'$, and $x_* = - k \tau_* = \left[ k / a(\tau_*) \right] / H$ is the ratio between the physical momentum of the mode and the horizon at the moment in which $\xi$ is at its maximum. For the evaluation of the integral, recall
\begin{equation}
\xi(x) = \frac{2 \xi_*}{\left(x/x_* \right)^\delta + \left(x_*/x \right)^\delta} \; .
\end{equation}
Using the expression \eqref{h1_full}, we compute the tensor power spectrum and bispectrum in the following subsections.

\subsection{Tensor Power Spectrum}
\label{subapp:tensorPS}

We define the power spectrum of the tensor mode as
\begin{equation}
{\cal P}_\lambda (k) \, \delta_{\lambda\lambda'} \delta^{(3)} \left( \vec{k} + \vec{k}' \right) \equiv 
\frac{k^3}{2\pi^2} 
\left\langle \hat{h}_\lambda \left( \vec{k} \right) \hat{h}_{\lambda'} \left( \vec{k}' \right) \right\rangle \; ,
\end{equation}
and separate it into the vacuum mode and the component sourced by the gauge field, as in \eqref{dec_zetahpm}. The vacuum mode is the standard one, and we focus on the computation of the sourced contribution in this appendix. Using \eqref{h1_full} and taking the two-point correlator of $\hat{h}^{(1)}_\lambda$, we find
\begin{eqnarray}
\left\langle \hat{h}_\lambda^{(1)} \left( \vec{k} \right) \hat{h}_{\lambda'}^{(1)} \left( \vec{k}' \right) \right\rangle & \simeq &
\delta^{(3)} \left( \vec{k} + \vec{k}' \right)
\frac{4 \, H^4}{M_p^4 \, k^7} 
\int \frac{d^3p}{(2\pi)^3} \, 
{\cal P}_\lambda \left[ \vec{k} , \vec{p} , \vec{k}-\vec{p} \right]
{\cal P}_{\lambda'}^* \left[ \vec{k} , \vec{p} , \vec{k}-\vec{p} \right]
\sqrt{p \big\vert \vec{k} - \vec{p} \big\vert}
\nonumber\\ && \times
N^2\left[ \xi_* ,\, \frac{p}{k_*} ,\, \delta \right]
N^2\left[ \xi_* ,\, \frac{\vert \vec k - \vec{p} \vert}{k_*} ,\, \delta \right]
{\cal T}_h^2 \left[ \xi_* ,\, x_* ,\, \delta ,\, \frac{p}{k} ,\, \frac{\vert \vec{k} - \vec{p} \vert}{k} \right] \; .
\end{eqnarray}
One can show that 
\begin{equation}
\int d\phi \,
{\cal P}_\lambda \left[ \vec{k} , \vec{p} , \vec{q} \right]
{\cal P}_{\lambda'}^* \left[ \vec{k} , \vec{p} , \vec{q} \right] =
\frac{\delta_{\lambda\lambda'}}{16} 
\int d\phi \left( 1 + \lambda \, \hat{k} \cdot \hat{p} \right)^2 \left( 1 + \lambda \, \hat{k} \cdot \hat{q} \right)^2 \; ,
\end{equation}
and therefore we find the sourced part of the tensor power spectrum
\begin{eqnarray}
{\cal P}_\lambda^{(1)} \left( k \right) & \simeq & 
\frac{H^4}{8 \pi^2 M_p^4 \, k^4}
\int \frac{d^3p}{(2\pi)^3} \, 
\left( 1 + \lambda \, \hat{k} \cdot \hat{p} \right)^2 \left( 1 + \lambda \, \hat{k} \cdot \frac{\vec{k}-\vec{p}}{\big\vert \vec{k} - \vec{p} \big\vert} \right)^2
\sqrt{p \big\vert \vec{k} - \vec{p} \big\vert}
\nonumber\\ && \times
N^2\left[ \xi_* ,\, \frac{p}{k_*} ,\, \delta \right]
N^2\left[ \xi_* ,\, \frac{\vert \vec k - \vec{p} \vert}{k_*} ,\, \delta \right]
{\cal T}_h^2 \left[ \xi_* ,\, x_* ,\, \delta ,\, \frac{p}{k} ,\, \frac{\vert \vec{k} - \vec{p} \vert}{k} \right] .
\end{eqnarray}
After rescaling $\vec{\tilde p} \equiv \vec{p} / k$, we arrive at the expression
\begin{equation}
{\cal P}_\lambda^{(1)} \left( k \right) \simeq 
\left[ \epsilon_\phi {\cal P}_\zeta^{(0)} \right]^2 \, 
f_{2,\lambda} \left( \xi_* ,\, x_* ,\, \delta \right) \; ,
\end{equation}
where ${\cal P}_\zeta^{(0)}$ is given in \eqref{Pzeta0} and
\begin{eqnarray}
f_{2,\lambda} \left( \xi_* ,\, x_* ,\, \delta \right) & = &
8 \pi^2
\int \frac{d^3 \tilde p}{(2\pi)^3} \, 
\left( 1 + \lambda \, \hat{k} \cdot \hat{\tilde p} \right)^2 \left( 1 + \lambda \, \hat{k} \cdot \frac{\hat{k} - \vec{\tilde p}}{\big\vert \hat{k} - \vec{\tilde p} \big\vert} \right)^2
\sqrt{\tilde p \big\vert \hat{k} - \vec{\tilde p} \big\vert} 
\nonumber\\ && \times
N^2\left[ \xi_* ,\, \tilde p \, x_* ,\, \delta \right]
N^2\left[ \xi_* ,\, \vert \hat k - \vec{\tilde p} \vert \, x_* ,\, \delta \right]
{\cal T}_h^2 \left[ \xi_* ,\, x_* ,\, \delta ,\, \tilde p ,\, \vert \hat{k} - \vec{\tilde p} \vert \right]
\; .
\end{eqnarray}
For the concrete evaluation of the integrals, we denote by $\eta$ the cosine of the angle between $\vec{\tilde p}$ and $\hat{k}$. After taking the trivial angular integral, we have
\begin{eqnarray}
f_{2,\lambda} \left( \xi_* ,\, x_* ,\, \delta \right) & = &
2
\int_0^\infty d \tilde p \int_{-1}^{1} d\eta \,
\frac{\tilde p^{5/2} \left( 1 + \lambda \eta \right)^2 \left( 1 - \tilde p \eta + \lambda \sqrt{1 - 2 \tilde p \eta + \tilde p^2} \right)^2}{\left( 1- 2 \tilde p \eta + \tilde p^2 \right)^{3/4}}
\nonumber\\ && \!\!\!\!\! \!\!\!\!\! \!\!\!\!\! \!\!\!\!\! \!\!\!\!\! \times
N^2\left[ \xi_* ,\, \tilde p \, x_* ,\, \delta \right]
N^2\left[ \xi_* ,\, \sqrt{1- 2 \tilde p \eta + \tilde p^2} \, x_* ,\, \delta \right]
{\cal T}_h^2 \left[ \xi_* ,\, x_* ,\, \delta ,\, \tilde p ,\, \sqrt{1- 2 \tilde p \eta + \tilde p^2} \right] \; .
\nonumber\\
\end{eqnarray}
Alternatively, as is done for \eqref{f2_xy}, we can change the variables of integration from $\tilde p$ and $\eta$ to $x$ and $y$ such that $x = \tilde p + \vert \hat{k} - \vec{\tilde p} \vert$ and  $y = \tilde p - \vert \hat{k} - \vec{\tilde p} \vert$. Then $f_{2,\lambda}$ takes the form
\begin{eqnarray}
f_{2,\lambda} \left( \xi_* ,\, x_* ,\, \delta \right) & = &
\frac{1}{4} \int_1^\infty dx \int_0^1 dy \,
\frac{\left(1-y^2\right)^2 \left( 1 + \lambda x \right)^4}{\sqrt{x+y} \sqrt{x-y}} \,
N^2\left[ \xi_* ,\, \frac{x+y}{2} \, x_* ,\, \delta \right]
\nonumber\\ && \times
N^2\left[ \xi_* ,\, \frac{x-y}{2} \, x_* ,\, \delta \right] \,
{\cal T}_h^2 \left[ \xi_* ,\, x_* ,\, \delta ,\, \frac{x+y}{2} ,\, \frac{x-y}{2} \right] \; .
\end{eqnarray}
We can now readily evaluate $f_{2,\lambda}$ numerically. All the phenomenological features of the tensor power spectrum in this model are captured in $f_{2,\lambda}$. Since we are choosing $\xi>0$ (i.e. $\dot{\sigma}>0$) so that the $A_+$ modes are produced through the tachyonic instability (see discussion around \eqref{approxaplus}), only the $+$ helicity state of the tensor modes is efficiently sourced by the gauge field, leading to the hierarchy $f_{2,+} \gg f_{2,-}$.

The total tensor power spectrum for each helicity state is
\begin{equation}
{\cal P}_\lambda = \frac{H^2}{\pi^2 M_p^2} \left[ 1 + \frac{H^2}{64 \pi^2 M_p^2} \, f_{2,\lambda} \left( \xi_* ,\, x_* ,\, \delta \right) \right] 
= 8 \epsilon_\phi {\cal P}_\zeta^{(0)} \left[ 1 + \frac{\epsilon_\phi}{8} \, {\cal P}_\zeta^{(0)} \,  f_{2,\lambda} \left( \xi_* ,\, x_* ,\, \delta \right) \right] \; ,
\end{equation}
and the tensor-to-scalar ratio is, using \eqref{scapower_total},
\begin{equation}
r \equiv \frac{\sum_{\lambda} {\cal P}_\lambda}{{\cal P}_\zeta} =
16 \epsilon_\phi \,
\frac{1 +  \frac{\epsilon_\phi}{16} {\cal P}_\zeta^{(0)} \left( f_{2,+} + f_{2,-} \right)}{1 + \epsilon_\phi^2 {\cal P}_\zeta^{(0)} f_{2,\zeta}} \; .
\end{equation}
Interesting phenomenological features arise when the sourced contribution dominates the tensor spectrum while it is subdominant for the scalar spectrum.

\subsection{Tensor Bispectrum}
\label{subapp:tensorBS}

We define the tensor bispectrum as
\begin{equation}
{\cal B}_{\lambda_1 \lambda_2 \lambda_3} \left(k_1 , k_2 , k_3\right) \, \delta^{(3)} \left( \vec{k}_1 + \vec{k}_2 + \vec{k}_3 \right) \equiv
\left\langle \hat{h}_{\lambda_1} \left( \vec{k}_1 \right) \hat{h}_{\lambda_2} \left( \vec{k}_2 \right) \hat{h}_{\lambda_3} \left( \vec{k}_3 \right) \right\rangle \; ,
\end{equation}
where in general ${\cal B}_{\lambda_1 \lambda_2 \lambda_3}$ is real and depends only on the magnitude of the three momenta, forming a triangle. While it consists of the two contributions from the vacuum fluctuations and the source effects, the former is unobservable, and we focus on the latter. Moreover, since only one polarization state of the gauge field is enhanced, the produced gauge quanta efficiently source only one of the tensor helicity states. We thus define and have
\begin{equation}
{\cal B}_\lambda^{(1)} \equiv {\cal B}_{\lambda\lambda\lambda}^{(1)} \cong {\cal B}_{\lambda\lambda\lambda} \,, 
\end{equation}
where the superscript $(1)$ denotes the sourced mode. We compute the $3$-point correlation function of $\hat{h}_\lambda^{(1)}$ with the same $\lambda$ using \eqref{h1_full},
\begin{eqnarray}
&& \!\!\!\!\! 
\left\langle \hat{h}_{\lambda}^{(1)} \left( \vec{k}_1 \right) \hat{h}_{\lambda}^{(1)} \left( \vec{k}_2 \right) \hat{h}_{\lambda}^{(1)} \left( \vec{k}_3 \right) \right\rangle \simeq
\delta^{(3)} \left( \vec{k}_1 + \vec{k}_2 + \vec{k}_3 \right) \,
\frac{2^{9/2} \, H^6}{M_p^6 \left( k_1 k_2 k_3 \right)^{7/2}} 
\int \frac{d^3p}{(2\pi)^{9/2}} \,
\nonumber\\ && \times
{\cal P}_\lambda \left[ \vec{k}_1 ,\, -\vec{p} ,\, \vec{p} + \vec{k}_1 \right] 
{\cal P}_\lambda \left[ \vec{k}_2 ,\, -\vec{p} - \vec{k}_1 ,\, \vec{p}-\vec{k}_3 \right] 
{\cal P}_\lambda \left[ \vec{k}_3 ,\, -\vec{p} + \vec{k}_3 ,\, \vec{p} \right] 
\sqrt{p \big\vert \vec{p} + \vec{k}_1 \big\vert \big\vert \vec{p} - \vec{k}_3 \big\vert}
\nonumber\\ && \times
N^2\left[ \xi_* ,\, \frac{p}{k_*} ,\, \delta \right]
N^2\left[ \xi_* ,\, \frac{\vert \vec p + \vec k_1 \vert}{k_*} ,\, \delta \right]
N^2\left[ \xi_* ,\, \frac{\vert \vec p - \vec k_3 \vert}{k_*} ,\, \delta \right]
\nonumber\\ && \times
{\cal T}_h \left[ \xi_* ,\, \frac{k_1}{k_*} ,\, \delta ,\, \frac{p}{k_1} , \frac{\big\vert \vec{p} + \vec{k}_1 \big\vert}{k_1} \right]
{\cal T}_h \left[ \xi_* ,\, \frac{k_2}{k_*} ,\, \delta ,\, \frac{\big\vert \vec{p} + \vec{k}_1 \big\vert}{k_2} , \frac{\big\vert \vec{p} - \vec{k}_3 \big\vert}{k_2} \right]
{\cal T}_h \left[ \xi_* ,\, \frac{k_3}{k_*} ,\, \delta ,\, \frac{\big\vert \vec{p} - \vec{k}_3 \big\vert}{k_3} , \frac{p}{k_3} \right] .
\nonumber\\
\end{eqnarray}
After rescaling
\begin{equation}
k \equiv k_1 \; , \quad x_2 \equiv \frac{k_2}{k} \; , \quad x_3 \equiv \frac{k_3}{k} \; , \quad x_* \equiv \frac{k}{k_*} \; , \quad \vec{\tilde p} \equiv \frac{\vec{p}}{k} \; ,
\end{equation}
and recalling \eqref{Pzeta0}, we obtain
\begin{equation}
{\cal B}_\lambda^{(1)} \simeq \frac{\left[ \epsilon_\phi {\cal P}_\zeta^{(0)} \right]^3}{k_1^2 k_2^2 k_3^2} \, f_{3 , \lambda} \left( \xi_* ,\, x_* ,\, \delta ,\, x_2 ,\, x_3 \right) \; ,
\end{equation}
where
\begin{eqnarray}
&& f_{3 , \lambda} \left( \xi_* ,\, x_* ,\, \delta ,\, x_2 ,\, x_3 \right) =
\frac{2^{27/2} \pi^6}{x_2^{3/2} x_3^{3/2}} 
\int \frac{d^3 \tilde p}{(2\pi)^{9/2}} \,
\sqrt{\tilde p \big\vert \vec{\tilde p} + \hat{k}_1 \big\vert \big\vert \vec{\tilde p} - x_3 \hat{k}_3 \big\vert} \,
{\cal P}_{\lambda\lambda\lambda} \left[ \vec{k}_i ,\, \vec{\tilde p} \right]
\nonumber\\ && \qquad\qquad\qquad \times
N^2\left[ \xi_* ,\, \tilde p \, x_* ,\, \delta \right]
N^2\left[ \xi_* ,\, \vert \vec{\tilde p} + \hat k_1 \vert \, x_* ,\, \delta \right]
N^2\left[ \xi_* ,\, \vert \vec{\tilde p} - x_3 \hat k_3 \vert \, x_* ,\, \delta \right]
\nonumber\\ && \qquad\qquad\qquad \times
{\cal T}_h \left[ \xi_* ,\, x_1 x_* ,\, \delta ,\, \tilde p , \big\vert \vec{\tilde p} + \hat{k}_1 \big\vert \right]
{\cal T}_h \left[ \xi_* ,\, x_2 x_* ,\, \delta ,\, \frac{\big\vert \vec{\tilde p} + \hat{k}_1 \big\vert}{x_2} , \frac{\big\vert \vec{\tilde p} - x_3 \hat{k}_3 \big\vert}{x_2} \right]
\nonumber\\ && \qquad\qquad\qquad \times
{\cal T}_h \left[ \xi_* ,\, x_3 x_* ,\, \delta ,\, \frac{\big\vert \vec{\tilde p} - x_3 \hat{k}_3 \big\vert}{x_3} , \frac{\tilde p}{x_3} \right] \,, 
\nonumber\\
\label{f3lambda}
\end{eqnarray}
and
\begin{eqnarray}
{\cal P}_{\lambda\lambda\lambda} \left[ \vec{k}_i ,\, \vec{\tilde p} \right] & \equiv &
{\cal P}_\lambda \left[ \vec{k}_1 ,\, -\vec{p} ,\, \vec{p} + \vec{k}_1 \right] 
{\cal P}_\lambda \left[ \vec{k}_2 ,\, -\vec{p} - \vec{k}_1 ,\, \vec{p}-\vec{k}_3 \right] 
{\cal P}_\lambda \left[ \vec{k}_3 ,\, -\vec{p} + \vec{k}_3 ,\, \vec{p} \right] 
\nonumber\\
& = &
\epsilon_i^{(\lambda)*} \left( \hat{k}_1 \right) \,
\epsilon_j^{(\lambda)*} \left( \hat{k}_1 \right) \,
\epsilon_j^{(+)} \left( \vec{\tilde p} + \hat{k}_1 \right) \,
\epsilon_k^{(+)*} \left( \vec{\tilde p} + \hat{k}_1 \right) \,
\epsilon_k^{(\lambda)*} \left( \hat{k}_2 \right) \,
\epsilon_l^{(\lambda)*} \left( \hat{k}_2 \right) \,
\nonumber\\ && \times
\epsilon_l^{(+)} \left( \vec{\tilde p} - x_3 \hat{k}_3 \right) \,
\epsilon_m^{(+)*} \left( \vec{\tilde p} - x_3 \hat{k}_3 \right) \,
\epsilon_m^{(\lambda)*} \left( \hat{k}_3 \right) \,
\epsilon_n^{(\lambda)*} \left( \hat{k}_3 \right) \,
\epsilon_n^{(+)} \left( \vec{\tilde p} \right) \,
\epsilon_i^{(+)*} \left( \vec{\tilde p} \right) \; .
\nonumber\\
\end{eqnarray}
In order to perform explicit evaluations of the integrals in \eqref{f3lambda}, we fix $\vec{k}_1$ along the $x$ axis and write $\vec{k}_2$ and $\vec{k}_3$ in terms of $x_2$ and $x_3$, namely,
\begin{eqnarray}
\vec{k}_1 & =&  k \left( 1 ,\, 0 ,\, 0 \right) \; , \nonumber\\
\vec{k}_2 &=& \frac{k}{2} \left( - 1 - x_2^2 + x_3^2 ,\, \sqrt{ - \left( 1 - x_2 + x_3 \right)  \left( 1 + x_2 - x_3 \right)  \left( 1 - x_2 - x_3 \right)  \left( 1 + x_2 + x_3 \right) } ,\, 0 \right) \; , \nonumber\\ 
\vec{k}_3 &=& \frac{k }{ 2 } \left( - 1 + x_2^2 - x_3^2 ,\, - \sqrt{ - \left( 1 - x_2 + x_3 \right)  \left( 1 + x_2 - x_3 \right)  \left( 1 - x_2 - x_3 \right)  \left( 1 + x_2 + x_3 \right) } ,\, 0 \right)  \; ,
\nonumber\\
\end{eqnarray}
and the polarization vector for a given momentum $\vec{q}$ can be written in terms of its components as
\begin{equation}
\epsilon^{(\pm)} \left( \vec{q} \right) = 
\frac{1}{\sqrt{2}} \left(
\frac{q_x q_z \mp i \, q_y \sqrt{q_x^2+q_y^2+q_z^2}}{\sqrt{q_x^2+q_y^2} \sqrt{q_x^2+q_y^2+q_z^2}} ,\,
\frac{q_y q_z \pm i \, q_x \sqrt{q_x^2+q_y^2+q_z^2}}{\sqrt{q_x^2+q_y^2} \sqrt{q_x^2+q_y^2+q_z^2}} ,\,
- \frac{\sqrt{q_x^2+q_y^2}}{\sqrt{q_x^2+q_y^2+q_z^2}} 
\right) \; .
\end{equation}
Using these explicit forms, we can evaluate \eqref{f3lambda} numerically for any given set of parameters. In the present case with $\xi>0$ (i.e. $\dot{\sigma}>0$), the positive helicity state of the gauge field is produced, and consequently only the positive tensor mode $\hat{h}_+^{(1)}$ is efficiently sourced, resulting in $f_{3,+} \gg f_{3,-}$. The phenomenology of the tensor bispectrum is featured by $f_{3,\lambda}$ and is discussed in detail in Section \ref{sec:pheno}.

\section{Shape and properties of the bispectrum}
\label{app:BS}

Let us first verify that our expressions for the scalar and tensor bispectrum given above are symmetric under the exchange of the external momenta $\vec{k}_i$, and real. In this discussion, we refer to the scalar bispectrum for definiteness. However, all our statements  also  apply to the tensor bispectrum. To show that the bispectrum is invariant under the exchange of any two momenta it is enough to show that ${\cal B}_\zeta^{(1)} \left( k_3 ,\, k_1 ,\, k_2 \right) = {\cal B}_\zeta^{(1)} \left( k_1 ,\, k_2 ,\, k_3 \right)$
and that ${\cal B}_\zeta^{(1)} \left( k_1 ,\, k_3 ,\, k_2 \right) = {\cal B}_\zeta^{(1)} \left( k_1 ,\, k_2 ,\, k_3 \right)$. To verify the first equality, we express ${\cal B}_\zeta^{(1)} \left( \vec{k}_3 ,\, \vec{k}_1 ,\, \vec{k}_2 \right)$ through expression (\ref{B1-par}) (namely, we replace $\vec{k}_1 \rightarrow \vec{k}_3 ,\, \vec{k}_2 \rightarrow \vec{k}_1$ and $\vec{k}_3 \rightarrow \vec{k}_2$ both at the left and right hand sides of eq.  (\ref{B1-par})), and we relabel the integration variable as $\vec{p} \rightarrow \vec{p} - \vec{k}_3$. Using the fact that the three external momenta add up to zero, it is immediate to verify that the resulting expression coincides with the expression (\ref{B1-par}) for ${\cal B}_\zeta^{(1)} \left( k_1 ,\, k_2 ,\, k_3 \right)$. To verify the second equality, we again start from the expression  (\ref{B1-par}) to express  ${\cal B}_\zeta^{(1)} \left( \vec{k}_1 ,\, \vec{k}_3 ,\, \vec{k}_2 \right)$ and we relabel $\vec{p} \rightarrow - \vec{p} - \vec{k}_1$. Also in this case, we recover the expression  (\ref{B1-par}) for  ${\cal B}_\zeta^{(1)} \left( k_1 ,\, k_2 ,\, k_3 \right)$. 

To verify that ${\cal B}_\zeta$ is real we instead use the reality of ${\hat \zeta} \left( \vec{x} \right)$, and the consequent identity ${\hat \zeta} \left( \vec{k} \right) = {\hat \zeta}^* \left( -\vec{k} \right)$, to write ${\cal B}_\zeta \left( \vec{k}_1 ,\, \vec{k}_2 ,\, \vec{k}_3 \right) =  {\cal B}_\zeta^* \left( -\vec{k}_1 ,\, -\vec{k}_2 ,\, -\vec{k}_3 \right)$. We then perform a $180^0$ rotation around the axis perpendicular to the plane defined by the three vectors $\vec{k}_i$ (recall that they define a plane since they add up to zero). Under this rotation, the three vectors change sign. Due to statistical isotropy, the rotation does not change the bispectrum, and so ${\cal B}_\zeta^* \left( - \vec{k}_1 ,\, - \vec{k}_2 ,\, - \vec{k}_3 \right) = {\cal B}_\zeta^* \left( \vec{k}_1 ,\, \vec{k}_2 ,\, \vec{k}_3 \right)$. From the two identities that we have just written we see that the bispectrum is real. 

As mentioned, identical properties apply also for ${\cal B}_\lambda$, due to statistical isotropy, and to the fact that $h_\lambda \left( \vec{x} \right)$ is scalar and real. 

We just proved exact identities, that apply to our explicit final results for $f_{3,\zeta}$ and  $f_{3,\lambda}$. We now provide an approximate expression for the bispectra, written as a sum of factorized terms. Namely, each of these terms appears as a product of three functions, each of which depends on one and only one of the external momenta. As well known, a factorized approximate expression for the bispectrum considerably speeds up its use in the data analysis. We conjecture this expression by noting that we expect the bispectrum to have a bump at the value of momenta $k_1 \simeq k_2 \simeq k_3 \simeq k_*$ at which the particle production is maximum, namely at which also the two point function is maximum. This leads us to expect a shape dependence of the type 
\begin{equation}
f_{3,j} \left( k_1 ,\, k_2 ,\, k_3 \right) \simeq {\rm Normalization} \times \left[ f_{2,j} \left( k_1 \right) \,  f_{2,j} \left( k_2 \right) \,  f_{2,j} \left( k_3 \right) \right]^{1/2} \,, 
\end{equation}
both in the scalar ($j=\zeta$) and tensor ($j=\lambda$) case. The normalization factor can be obtained by evaluating the bispectrum in the exact equilateral limit. A form that is factorized, and symmetric in the three momenta is 
\begin{eqnarray}
f_{3,j} \left( k_1 ,\, k_2 ,\, k_3 \right) &\simeq& \left[ \frac{f_{3,j} \left( k_1 ,\, k_1 ,\, k_1 \right)}{3 f_{2,j}^{3/2} \left( k_1 \right) } 
+  \frac{f_{3,j} \left( k_2 ,\, k_2 ,\, k_2 \right)}{3 f_{2,j}^{3/2} \left( k_2 \right)  } +   \frac{f_{3,j} \left( k_3 ,\, k_3 ,\, k_3 \right)}{3 f_{2,j}^{3/2} \left( k_3 \right)  } \right] \nonumber\\ 
&& \quad\quad \times \left[ f_{2,j} \left( k_1 \right) \,  f_{2,j} \left( k_2 \right) \,  f_{2,j} \left( k_3 \right) \right]^{1/2} \,, 
\label{shape}
\end{eqnarray}
The approximate expression (\ref{shape}) has the additional computational advantage that we simply need to evaluate $f_3$ in the equilateral limit. 

\begin{figure}
\centering{ 
\includegraphics[width=.45\textwidth,clip]{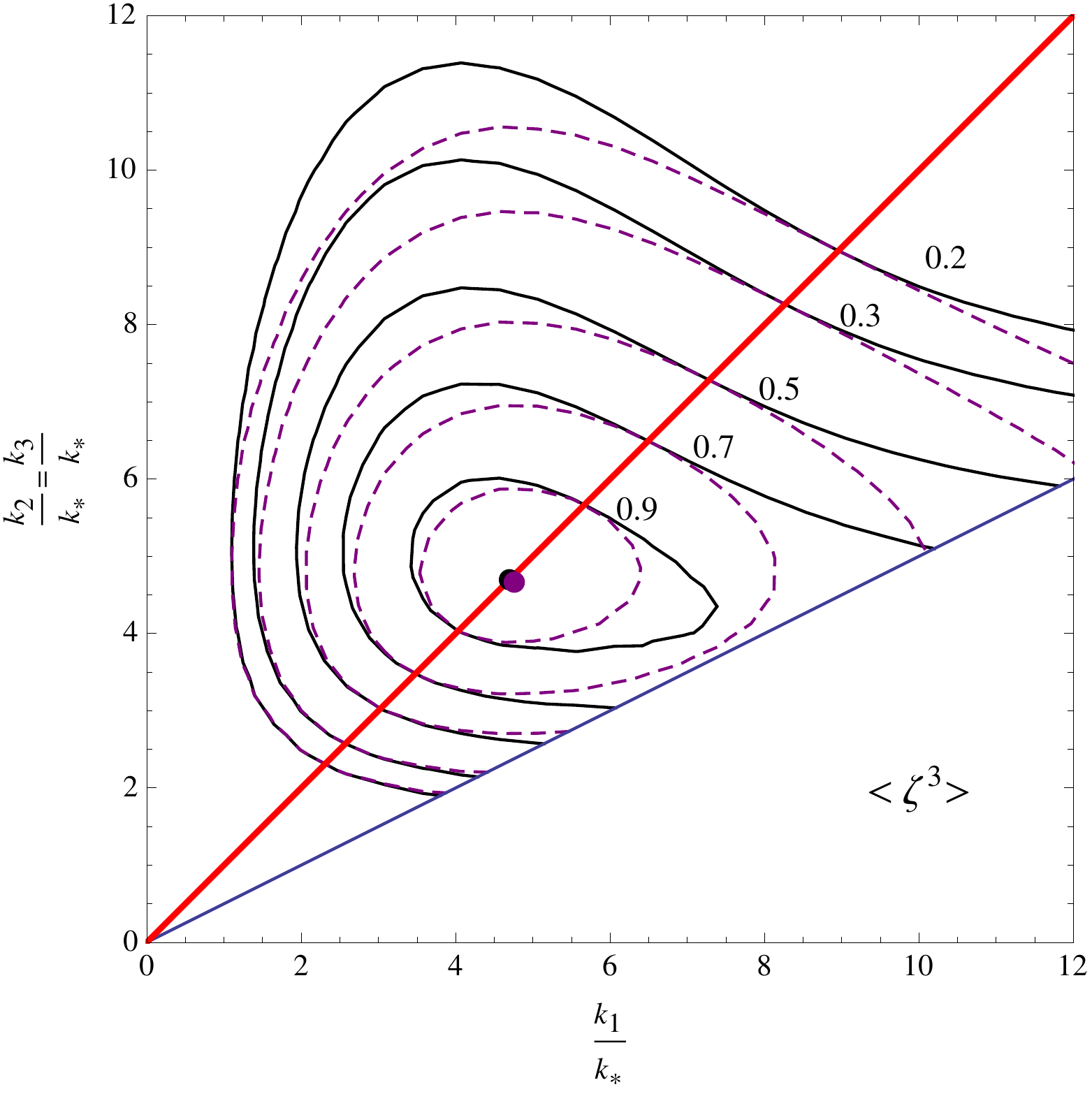}
\includegraphics[width=.45\textwidth,clip]{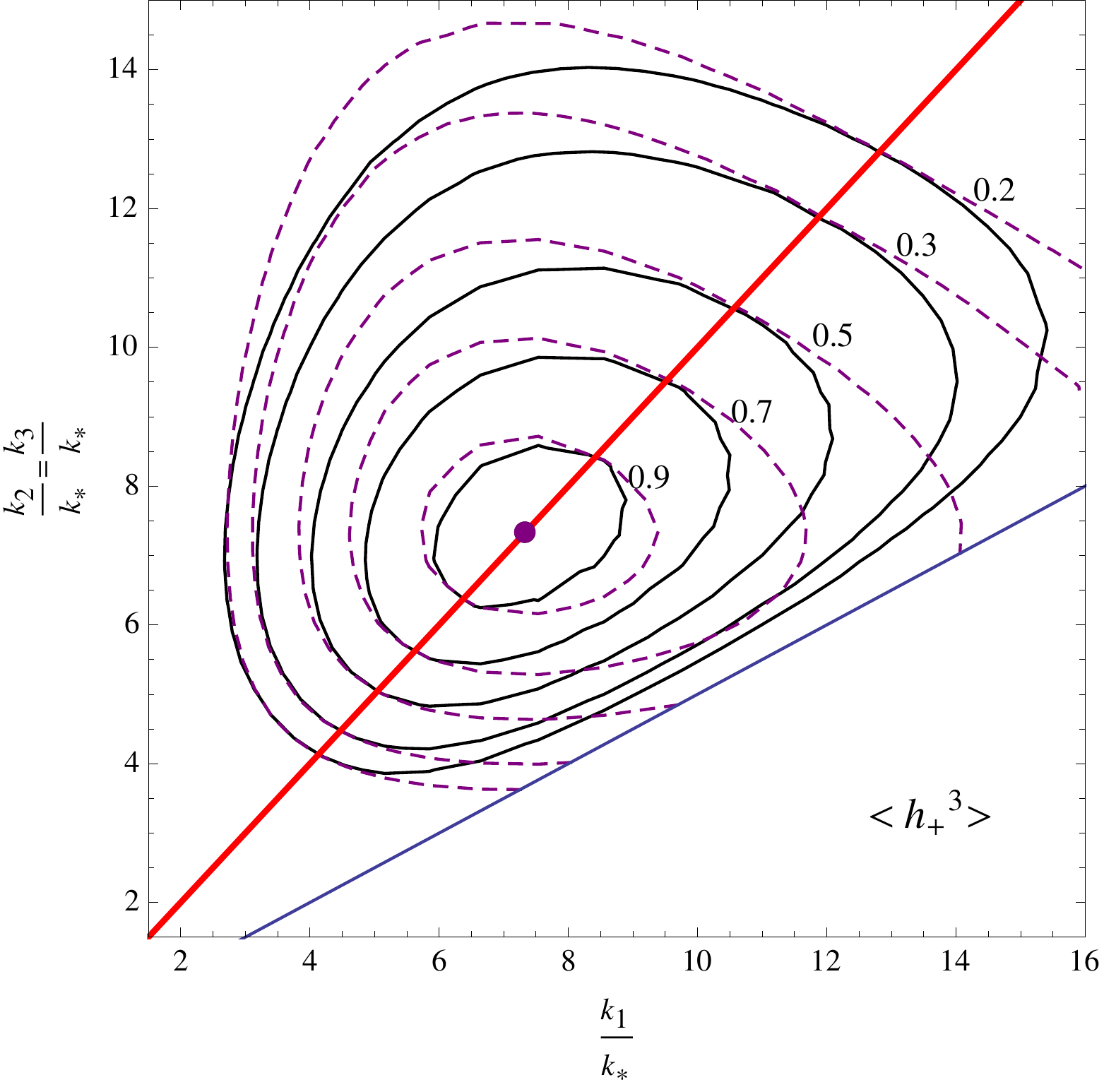}
}
\caption{Exact (black solid lines) vs. approximate (purple dashed lines; eq. (\ref{shape})) expression for the scalar (left panel) and tensor (right panel) bispectrum, for isosceles triangles, $k_2 = k_3$, and for $\xi_* = 5$ and $\delta= 0.5$. The black (purple) dot indicates the triangle for which the exact (approximate) bispectrum is maximum.  The contour lines indicate the triangles for which the bispectrum evaluates to (from inner to outer, respectively) a fraction of $0.9, 0.7, 0.5, 0.3,0.2$  than the maximum value of the exact bispectrum. The red line at $x_2 / x_ 1 = 1$, indicates the equilateral triangles, while the lower slope $x_2 = x_1 / 2$ line indicates folded triangles (smaller ratios are not possible for isosceles triangles). 
}
\label{fig:isosceles}
\end{figure}

To verify the accuracy of (\ref{shape}), in Figure  \ref{fig:isosceles} we show the comparison between the exact and the approximate bispectrum on isosceles triangles, $k_2=k_3$. The black/solid contours refer to the left hand side of (\ref{shape}), while the purple/dashed contours refer to the left hand side. The contour lines shown in the the figures are obtained by evaluating the bispectra on a grid of values  in the region shown, and by interpolation  (for the functions $f_{2,\zeta}$ and $f_{2,\lambda}$ used to produce the dashed lines we instead employ the fitting functions given in Subsection \ref{subsec:phenoresults}). In each panel, a black (purple) dot indicates the  triangle for which the exact (approximate) bispectrum is maximum. In both cases, the two dots actually appear as nearly superimposed or as superimposed to each other, as the locations of the two maxima are nearly coincident. We see that indeed the bispectrum is maximum on an equilateral triangle of scale $k_1 = k_2 = k_3 \simeq {\rm few} \times \, k_*$  approximately equal to the scale at which the sourced power spectrum is also maximum. Most importantly,  as shown by the figure, the approximate bispectrum provides a very accurate description of the exact one (particularly, close to the maximum). 

\section{Deviation from gaussianity}
\label{app:NG}

In this Appendix we estimate the departure from gaussianity of the statistics of the sourced  ${\hat \zeta}^{(1)}$ and $h_+^{(1)}$ modes. For brevity we denote either of the two fields as  ${\hat f}$, and we distinguish between the two fields only at the end of the appendix, when we make explicit evaluations. We assume that ${\hat f}$ is approximately gaussian  (to be verified by this computation) and, to estimate the departure from gaussianity we start from the conventional definition of the local $f_{\rm NL}^{\rm local}$ parameter 
\begin{equation}
{\hat f} \left( \vec{x} \right) = \hat f_g \left( \vec{x} \right) + f_{\rm NL}^{\rm local} \left[ {\hat f}_g^2 \left( \vec{x} \right) - \left\langle {\hat f}_g^2 \left( \vec{x} \right) \right\rangle \right] \,, 
\label{f-fg-fnl}
\end{equation} 
where ${\hat f}_g$ is gaussian. The statistics of ${\hat f}$ is nearly gaussian if the first term in this decomposition dominates, namely if 
\begin{equation}
f_{\rm NL}^{\rm local} \,  {\hat f}_g \left( \vec{x} \right) < 1 \;\; \leftrightarrow \;\;  
{\cal R} \equiv \left( f_{\rm NL}^{\rm local} \right)^2 \, \left\langle \hat f \left( \vec{x} \right) \,  \hat f \left( \vec{x} \right) \right\rangle < 1 \;. 
\label{cond}
\end{equation} 
When this condition is satisfied, we are also estimating that the presence of a nonvanishing three-point function does not change the two-point function significantly (namely, we are in this way estimating the loop diagram where the nonvanishing three-point function enters as a vertex). 

In Fourier space, the decomposition (\ref{f-fg-fnl}) reads 
\begin{equation}
{\hat f} \left( \vec{k} \right) = {\hat f}_g \left( \vec{k} \right) + f_{\rm NL}^{\rm local} \, \int \frac{d^3 p}{\left( 2 \pi \right)^{3/2}} \, {\hat f}_g \left( \vec{p} \right) \, 
 {\hat f}_g \left( \vec{k} - \vec{p} \right) \,.   
\label{local}
\end{equation} 
Under the assumption of small departure from gaussianity, namely that the first term dominates in (\ref{local}), we obtain the three point correlation function 
\begin{equation}
{\cal B} \left( k_1 ,\, k_2 ,\, k_3 \right) = 2 \sqrt{2} \pi^{5/2} f_{\rm NL}^{\rm local} \left[ 
\frac{ {\cal P }  \left( k_1 \right)}{k_1^3} \, \frac{  {\cal P }  \left( k_2 \right)}{k_2^3} + 
\frac{ {\cal P }  \left( k_1 \right)}{k_1^3} \, \frac{  {\cal P }  \left( k_3 \right)}{k_3^3} + 
\frac{ {\cal P }  \left( k_2 \right)}{k_2^3} \, \frac{  {\cal P }  \left( k_3 \right)}{k_3^3} \right] \,, 
\label{B-P-fNL}
\end{equation} 
where the two- and three-point correlation functions are expressed, respectively, in terms of ${\cal P}$ and ${\cal B}$ as in eq. (\ref{def-P-B}). 
The constant $f_{\rm NL}^{\rm local}$ corresponds to a specific shape of the bispectrum, denoted as local non-gaussianity (the standard local non-gaussianity parameter is obtained from using ${\hat \zeta} $ in (\ref{local}), and from rescaling $ f_{\rm NL}^{\rm local} \rightarrow \frac{3}{5} \,  f_{\rm NL}^{\rm local}$). In the present case, we promote $f_{\rm NL}^{\rm local}$ to an effective momentum-dependence nonlinear parameter $f_{\rm NL} \left( k_i \right)$, by inserting in (\ref{B-P-fNL}) the power spectra and bispectra of the sourced functions ${\hat \zeta}^{(1)}$ and  ${\hat h}_+^{(1)}$. We evaluate this expression in the equilateral limit, where the bispectrum of our sourced signals is maximum (this likely overestimates the departure of gaussianity of the sourced signals). In this way we obtain  
\begin{equation}
f_{{\rm NL},i} \left( k \right) \equiv 
\frac{k^6 \, {\cal B}_i \left( k,\, k,\, k \right)}{6 \sqrt{2} \pi^{5/2} {\cal P}_i^2 \left( k \right)}  \;\;,\;\; i = \zeta ,\, \lambda  \,. 
\label{fnl}
\end{equation} 

To estimate the departure from gaussianity we write the analogous of (\ref{cond}) in momentum space, accounting for the momentum dependence of $f_{{\rm NL},i} $. We thus obtain the condition   
\begin{equation} 
R_i  = \int_0^\infty \frac{dk}{k} f_{{\rm NL},i}^2 \left( k \right) \, {\cal P}_i \left( k \right) = 
 \int_0^\infty \frac{dk}{k} \, \frac{k^{12} \, {\cal B}_i^2 \left( k,\, k\, k \right)}{72 \, \pi^5 \, {\cal P}_i^3 \left( k \right) } 
 \equiv \int_0^\infty  \frac{d k}{k} \, r_i \left( k \right)  < 1 \;. 
\end{equation} 
Using the parametrization (\ref{f23-def}), we obtain 
\begin{equation}
r_i \left( k \right) \equiv \frac{k^{12} \, {\cal B}_i^2 \left( k,\, k\, k \right)}{72 \, \pi^5 \, {\cal P}_i^3 \left( k \right) } = 
\left\{ \begin{array}{l} 
 \frac{\epsilon_\phi^6 \, {\cal P}_\zeta^{(0) 3} \, f_{3,\zeta}^2}{72 \, \pi^5
\left( 1 + \epsilon_\phi^2  {\cal P}_\zeta^{(0) } \, f_{2,\zeta} \right)^3} \;\;,\;\; i = \zeta \;, \\ 
\frac{\epsilon_\phi^3 \, {\cal P}_\zeta^{(0) 3} \, f_{3,\lambda}^2}{
 36,864 \pi^5 \left( 1 +  \frac{1}{8} \, \epsilon_\phi {\cal P}_\zeta^{(0) } \, f_{2,\lambda} \right)^3} \;\;,\;\; i = \lambda \;. 
\label{R-zeta-lambda}
\end{array} \right. 
\end{equation} 

\begin{figure}
\centering{ 
\includegraphics[width=.4\textwidth,clip]{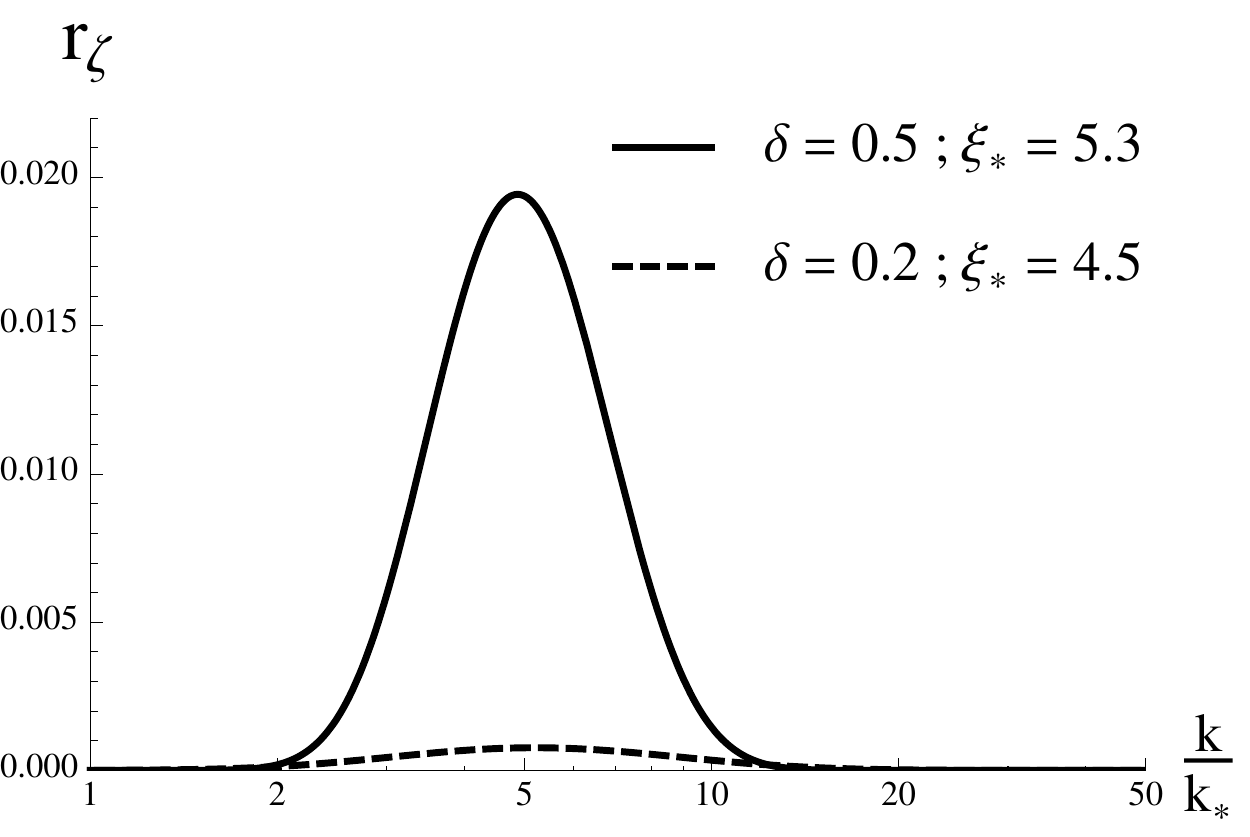}
\includegraphics[width=.4\textwidth,clip]{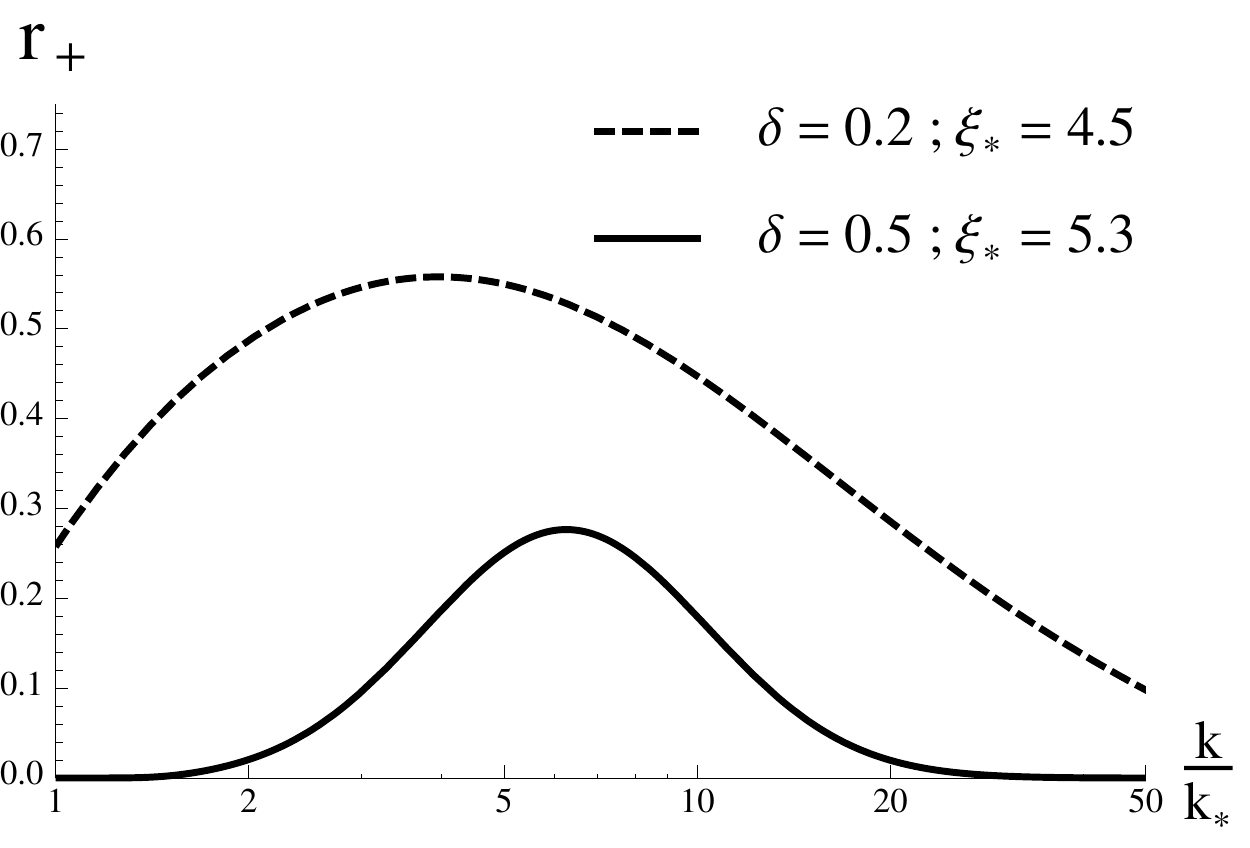}
}
\caption{Left (right) panel: parameter $r_\zeta \left( k \right)$ ($r_+ \left( k \right)$) for the greatest value of $\xi_*$ used in the previous figures, for $\delta = 0.2,\,0.5$ and for $\epsilon_\phi = 10^{-5}$. 
}
\label{fig:NG}
\end{figure}

In Figure \ref{fig:NG} we show the quantity $r_\zeta \left( k \right)$ (left panel) and  $r_+ \left( k \right)$ (right panel) for the greatest values of $\xi_*$ considered in the previous figures, and for which the departure from gaussianity is therefore greatest. The  quantity $r_i \left( k \right)$ is maximized at the bump of the sourced signal, $k \simeq {\rm few} \times \, k_*$. We note that $r_\zeta \left( k \right) \ll 1$, while $r_+ \left( k \right) < 1$.  The integrals of the curve shown evaluate to 
\begin{eqnarray}
&& R_\zeta \left( \delta = 0.2 ,\, \xi_* = 4.5 \right) \simeq 0.00098 \;\;,\;\; 
R_\zeta \left( \delta = 0.5 ,\, \xi_* = 5.3 \right) \simeq 0.015 \;\;,\;\; \nonumber\\ 
&& R_+ \left( \delta = 0.2 ,\, \xi_* = 4.5 \right) \simeq 1.6 \;\;,\;\; 
R_+ \left( \delta = 0.5 ,\, \xi_* = 5.3 \right) \simeq 0.35 \;. 
\end{eqnarray}

The $\zeta$ signal is approximately gaussian, as the two-point function is dominated by the vacuum mode (which thus dilutes the non-gaussianity of the sourced mode). The tensor signal is mildly gaussian (we recall that the expression (\ref{fnl}) likely overestimates the departure from gaussianity). This explains why in the examples we have studied in the main text the S/N ratio for detecting BBB is greater than that for TTT. It is possible that the large BBB signal can impact (at one loop) the BB correlator, thus inducing a greater two-point function (and that, in general, higher order correlators may increase the lower order ones). However, the fact that $R_+ \simeq 1$ suggests that this cannot change the order of magnitude of the result we have computed (nor the conclusion that the model can produce a visible BB signal, while being compatible with the limits from TT).


\begin{thebibliography}{99}

\bibitem{Ade:2015tva} 
  P.~A.~R.~Ade {\it et al.} [BICEP2 and Planck Collaborations],
  Phys.\ Rev.\ Lett.\  {\bf 114}, 101301 (2015)
  [arXiv:1502.00612 [astro-ph.CO]].

\bibitem{Abazajian:2013vfg} 
  K.~N.~Abazajian {\it et al.},
  Astropart.\ Phys.\  {\bf 63}, 55 (2015)
  [arXiv:1309.5381 [astro-ph.CO]].

\bibitem{Lyth:1996im} 
  D.~H.~Lyth,
  Phys.\ Rev.\ Lett.\  {\bf 78}, 1861 (1997)
  [hep-ph/9606387].

\bibitem{Biagetti:2013kwa} 
  M.~Biagetti, M.~Fasiello and A.~Riotto,
  Phys.\ Rev.\ D {\bf 88}, 103518 (2013)
  [arXiv:1305.7241 [astro-ph.CO]].

\bibitem{Biagetti:2014asa} 
  M.~Biagetti, E.~Dimastrogiovanni, M.~Fasiello and M.~Peloso,
  JCAP {\bf 1504}, 011 (2015)
  [arXiv:1411.3029 [astro-ph.CO]].

\bibitem{Fujita:2014oba} 
  T.~Fujita, J.~Yokoyama and S.~Yokoyama,
  PTEP {\bf 2015}, 043E01 (2015)
  [arXiv:1411.3658 [astro-ph.CO]].

\bibitem{Cannone:2014uqa} 
  D.~Cannone, G.~Tasinato and D.~Wands,
  JCAP {\bf 1501}, no. 01, 029 (2015)
  [arXiv:1409.6568 [astro-ph.CO]].

\bibitem{Cannone:2015rra} 
  D.~Cannone, J.~O.~Gong and G.~Tasinato,
  arXiv:1505.05773 [hep-th].

\bibitem{Cai:2015dta} 
  Y.~Cai, Y.~T.~Wang and Y.~S.~Piao,
  Phys.\ Rev.\ D {\bf 91}, 103001 (2015)
  [arXiv:1501.06345 [astro-ph.CO]].

\bibitem{Dimastrogiovanni:2012ew} 
  E.~Dimastrogiovanni and M.~Peloso,
  Phys.\ Rev.\ D {\bf 87}, no. 10, 103501 (2013)
  [arXiv:1212.5184 [astro-ph.CO]].

\bibitem{Adshead:2013nka} 
  P.~Adshead, E.~Martinec and M.~Wyman,
  JHEP {\bf 1309}, 087 (2013)
  [arXiv:1305.2930 [hep-th]].

\bibitem{Obata:2014loa} 
  I.~Obata, T.~Miura and J.~Soda,
  arXiv:1412.7620 [hep-ph].

\bibitem{Namba:2013kia} 
  R.~Namba, E.~Dimastrogiovanni and M.~Peloso,
  JCAP {\bf 1311}, 045 (2013)
  [arXiv:1308.1366 [astro-ph.CO]].

\bibitem{Khlebnikov:1997di} 
  S.~Y.~Khlebnikov and I.~I.~Tkachev,
  Phys.\ Rev.\ D {\bf 56}, 653 (1997)
  [hep-ph/9701423].

\bibitem{Easther:2006gt} 
  R.~Easther and E.~A.~Lim,
  JCAP {\bf 0604}, 010 (2006)
  [astro-ph/0601617].

\bibitem{GarciaBellido:2007dg} 
  J.~Garcia-Bellido and D.~G.~Figueroa,
  Phys.\ Rev.\ Lett.\  {\bf 98}, 061302 (2007)
  [astro-ph/0701014].

\bibitem{Dufaux:2007pt} 
  J.~F.~Dufaux, A.~Bergman, G.~N.~Felder, L.~Kofman and J.~P.~Uzan,
  Phys.\ Rev.\ D {\bf 76}, 123517 (2007)
  [arXiv:0707.0875 [astro-ph]].

\bibitem{Dufaux:2008dn} 
  J.~F.~Dufaux, G.~Felder, L.~Kofman and O.~Navros,
  JCAP {\bf 0903}, 001 (2009)
  [arXiv:0812.2917 [astro-ph]].

\bibitem{Dufaux:2009wn} 
  J.~F.~Dufaux,
  Phys.\ Rev.\ Lett.\  {\bf 103}, 041301 (2009)
  [arXiv:0902.2574 [astro-ph.CO]].

\bibitem{Dufaux:2010cf} 
  J.~F.~Dufaux, D.~G.~Figueroa and J.~Garcia-Bellido,
  Phys.\ Rev.\ D {\bf 82}, 083518 (2010)
  [arXiv:1006.0217 [astro-ph.CO]].

\bibitem{Figueroa:2013vif} 
  D.~G.~Figueroa and T.~Meriniemi,
  JHEP {\bf 1310}, 101 (2013)
  [arXiv:1306.6911 [astro-ph.CO]].

\bibitem{Antoniadis:2014xva} 
  I.~Antoniadis and S.~P.~Patil,
  Eur.\ Phys.\ J.\ C {\bf 75}, 182 (2015)
  [arXiv:1410.8845 [hep-th]].

\bibitem{Kleban:2015daa} 
  M.~Kleban, M.~Mirbabayi and M.~Porrati,
  arXiv:1508.01527 [hep-th].

\bibitem{Cook:2011hg} 
  J.~L.~Cook and L.~Sorbo,
  Phys.\ Rev.\ D {\bf 85}, 023534 (2012)
  [Phys.\ Rev.\ D {\bf 86}, 069901 (2012)]
  [arXiv:1109.0022 [astro-ph.CO]].

\bibitem{Senatore:2011sp} 
  L.~Senatore, E.~Silverstein and M.~Zaldarriaga,
  JCAP {\bf 1408}, 016 (2014)
  [arXiv:1109.0542 [hep-th]].

\bibitem{Barnaby:2012xt} 
  N.~Barnaby, J.~Moxon, R.~Namba, M.~Peloso, G.~Shiu and P.~Zhou,
  Phys.\ Rev.\ D {\bf 86}, 103508 (2012)
  [arXiv:1206.6117 [astro-ph.CO]].

\bibitem{Carney:2012pk} 
  D.~Carney, W.~Fischler, E.~D.~Kovetz, D.~Lorshbough and S.~Paban,
  JHEP {\bf 1211}, 042 (2012)
  [arXiv:1209.3848 [hep-th]].

\bibitem{Chung:1999ve} 
  D.~J.~H.~Chung, E.~W.~Kolb, A.~Riotto and I.~I.~Tkachev,
  Phys.\ Rev.\ D {\bf 62}, 043508 (2000)
  [hep-ph/9910437].

\bibitem{Berera:1995ie} 
  A.~Berera,
  Phys.\ Rev.\ Lett.\  {\bf 75}, 3218 (1995)
  [astro-ph/9509049].
  
\bibitem{BasteroGil:2009ec} 
  M.~Bastero-Gil and A.~Berera,
  Int.\ J.\ Mod.\ Phys.\ A {\bf 24}, 2207 (2009)
  [arXiv:0902.0521 [hep-ph]].
  
\bibitem{Sorbo:2011rz} 
  L.~Sorbo,
  JCAP {\bf 1106}, 003 (2011)
  [arXiv:1101.1525 [astro-ph.CO]].

\bibitem{Anber:2006xt} 
  M.~M.~Anber and L.~Sorbo,
  JCAP {\bf 0610}, 018 (2006)
  [astro-ph/0606534].

\bibitem{Lyth:2005jf} 
  D.~H.~Lyth, C.~Quimbay and Y.~Rodriguez,
  JHEP {\bf 0503}, 016 (2005)
  [hep-th/0501153].

\bibitem{Satoh:2010ep} 
  M.~Satoh,
  JCAP {\bf 1011}, 024 (2010)
  [arXiv:1008.2724 [astro-ph.CO]].

\bibitem{Lue:1998mq} 
  A.~Lue, L.~M.~Wang and M.~Kamionkowski,
  Phys.\ Rev.\ Lett.\  {\bf 83}, 1506 (1999)
  [astro-ph/9812088].

\bibitem{Saito:2007kt} 
  S.~Saito, K.~Ichiki and A.~Taruya,
  JCAP {\bf 0709}, 002 (2007)
  [arXiv:0705.3701 [astro-ph]].

\bibitem{Gluscevic:2010vv} 
  V.~Gluscevic and M.~Kamionkowski,
  Phys.\ Rev.\ D {\bf 81}, 123529 (2010)
  [arXiv:1002.1308 [astro-ph.CO]].

\bibitem{Ferte:2014gja} 
  A.~Fert\'e and J.~Grain,
  Phys.\ Rev.\ D {\bf 89}, no. 10, 103516 (2014)
  [arXiv:1404.6660 [astro-ph.CO]].

\bibitem{Caprini:2003vc} 
  C.~Caprini, R.~Durrer and T.~Kahniashvili,
  Phys.\ Rev.\ D {\bf 69}, 063006 (2004)
  [astro-ph/0304556].

\bibitem{Caprini:2014mja} 
  C.~Caprini and L.~Sorbo,
  JCAP {\bf 1410}, no. 10, 056 (2014)
  [arXiv:1407.2809 [astro-ph.CO]].

\bibitem{Barnaby:2010vf} 
  N.~Barnaby and M.~Peloso,
  Phys.\ Rev.\ Lett.\  {\bf 106}, 181301 (2011)
  [arXiv:1011.1500 [hep-ph]].

\bibitem{Barnaby:2011vw} 
  N.~Barnaby, R.~Namba and M.~Peloso,
  JCAP {\bf 1104}, 009 (2011)
  [arXiv:1102.4333 [astro-ph.CO]].

\bibitem{Choi:2015wva} 
  K.~Choi, K.~Y.~Choi, H.~Kim and C.~S.~Shin,
  arXiv:1507.04977 [astro-ph.CO].

\bibitem{Ratra:1991bn} 
  B.~Ratra,
  Astrophys.\ J.\  {\bf 391}, L1 (1992).

\bibitem{Mukohyama:2014gba} 
  S.~Mukohyama, R.~Namba, M.~Peloso and G.~Shiu,
  JCAP {\bf 1408}, 036 (2014)
  [arXiv:1405.0346 [astro-ph.CO]].

\bibitem{Ade:2014xna} 
  P.~A.~R.~Ade {\it et al.} [BICEP2 Collaboration],
  Phys.\ Rev.\ Lett.\  {\bf 112}, no. 24, 241101 (2014)
  [arXiv:1403.3985 [astro-ph.CO]].

\bibitem{Ferreira:2014zia} 
  R.~Z.~Ferreira and M.~S.~Sloth,
  JHEP {\bf 1412}, 139 (2014)
  [arXiv:1409.5799 [hep-ph]].

\bibitem{Mirbabayi:2014jqa} 
  M.~Mirbabayi, L.~Senatore, E.~Silverstein and M.~Zaldarriaga,
  Phys.\ Rev.\ D {\bf 91}, 063518 (2015)
  [arXiv:1412.0665 [hep-th]].

\bibitem{Eccles:2015ipa} 
  S.~Eccles, W.~Fischler, D.~Lorshbough and B.~A.~Stephens,
  arXiv:1505.04686 [astro-ph.CO].

\bibitem{Riotto:2002yw} 
  A.~Riotto,
  hep-ph/0210162.

\bibitem{Bennett:2012zja} 
  C.~L.~Bennett {\it et al.} [WMAP Collaboration],
  Astrophys.\ J.\ Suppl.\  {\bf 208}, 20 (2013)
  [arXiv:1212.5225 [astro-ph.CO]].
  
\bibitem{Hinshaw:2012aka} 
  G.~Hinshaw {\it et al.} [WMAP Collaboration],
  Astrophys.\ J.\ Suppl.\  {\bf 208}, 19 (2013)
  [arXiv:1212.5226 [astro-ph.CO]].

\bibitem{Ade:2015lrj} 
  P.~A.~R.~Ade {\it et al.} [Planck Collaboration],
  arXiv:1502.02114 [astro-ph.CO].

\bibitem{Ade:2015xua} 
  P.~A.~R.~Ade {\it et al.} [Planck Collaboration],
  arXiv:1502.01589 [astro-ph.CO].

\bibitem{Planck:2006aa} 
  J.~Tauber {\it et al.} [Planck Collaboration],
  astro-ph/0604069.

\bibitem{Shiraishi:2013vha} 
  M.~Shiraishi,
  JCAP {\bf 1311}, 006 (2013)
  [arXiv:1308.2531 [astro-ph.CO]].

\bibitem{Shiraishi:2013kxa} 
  M.~Shiraishi, A.~Ricciardone and S.~Saga,
  JCAP {\bf 1311}, 051 (2013)
  [arXiv:1308.6769 [astro-ph.CO]].

\bibitem{Ade:2015ava} 
  P.~A.~R.~Ade {\it et al.} [Planck Collaboration],
  arXiv:1502.01592 [astro-ph.CO].

\bibitem{Shiraishi:2010sm} 
  M.~Shiraishi, S.~Yokoyama, D.~Nitta, K.~Ichiki and K.~Takahashi,
  Phys.\ Rev.\ D {\bf 82}, 103505 (2010)
  [arXiv:1003.2096 [astro-ph.CO]].

\bibitem{Shiraishi:2010kd} 
  M.~Shiraishi, D.~Nitta, S.~Yokoyama, K.~Ichiki and K.~Takahashi,
  Prog.\ Theor.\ Phys.\  {\bf 125}, 795 (2011)
  [arXiv:1012.1079 [astro-ph.CO]].

\bibitem{Cook:2013xea} 
  J.~L.~Cook and L.~Sorbo,
  JCAP {\bf 1311}, 047 (2013)
  [arXiv:1307.7077 [astro-ph.CO]].

\bibitem{Kamionkowski:2010rb} 
  M.~Kamionkowski and T.~Souradeep,
  Phys.\ Rev.\ D {\bf 83}, 027301 (2011)
  [arXiv:1010.4304 [astro-ph.CO]].

\bibitem{Shiraishi:2011st} 
  M.~Shiraishi, D.~Nitta and S.~Yokoyama,
  Prog.\ Theor.\ Phys.\  {\bf 126}, 937 (2011)
  [arXiv:1108.0175 [astro-ph.CO]].

\bibitem{Maldacena:2011nz} 
  J.~M.~Maldacena and G.~L.~Pimentel,
  JHEP {\bf 1109}, 045 (2011)
  [arXiv:1104.2846 [hep-th]].

\bibitem{Soda:2011am} 
  J.~Soda, H.~Kodama and M.~Nozawa,
  JHEP {\bf 1108}, 067 (2011)
  [arXiv:1106.3228 [hep-th]].
  
\bibitem{Shiraishi:2012sn} 
  M.~Shiraishi,
  JCAP {\bf 1206}, 015 (2012)
  [arXiv:1202.2847 [astro-ph.CO]].

\bibitem{Shiraishi:2014roa} 
  M.~Shiraishi, M.~Liguori and J.~R.~Fergusson,
  JCAP {\bf 1405}, 008 (2014)
  [arXiv:1403.4222 [astro-ph.CO]].

\bibitem{Shiraishi:2014ila} 
  M.~Shiraishi, M.~Liguori and J.~R.~Fergusson,
  JCAP {\bf 1501}, no. 01, 007 (2015)
  [arXiv:1409.0265 [astro-ph.CO]].

\bibitem{Babich:2004yc} 
  D.~Babich and M.~Zaldarriaga,
  Phys.\ Rev.\ D {\bf 70}, 083005 (2004)
  [astro-ph/0408455].

\bibitem{Yadav:2007ny} 
  A.~P.~S.~Yadav, E.~Komatsu, B.~D.~Wandelt, M.~Liguori, F.~K.~Hansen and S.~Matarrese,
  Astrophys.\ J.\  {\bf 678}, 578 (2008)
  [arXiv:0711.4933 [astro-ph]].

\bibitem{Yadav:2007rk} 
  A.~P.~S.~Yadav, E.~Komatsu and B.~D.~Wandelt,
  Astrophys.\ J.\  {\bf 664}, 680 (2007)
  [astro-ph/0701921].
  
\bibitem{merzbacher}  
  E.~Merzbacher, ``Quantum  mechanics,''
  Wiley, 3rd edition (1997), 672 p
 
\end{thebibliography}
\end{document}